\documentclass[12pt,a4paper,twoside]{book}
\usepackage[textheight=0.72\paperheight,textwidth=0.705\paperwidth,inner=2.5cm,bindingoffset=0.5cm]{geometry}

\usepackage[T1]{fontenc}
\usepackage[utf8]{inputenc}
\usepackage[english]{babel}

\usepackage{graphicx}
\usepackage{amsfonts}
\usepackage{amssymb}
\usepackage{amsmath}
\usepackage{mathtools}
\usepackage{slashed}
\usepackage{palatino}
\usepackage[notext,classicReIm]{kpfonts}
\usepackage{mathpazo}
\usepackage{microtype}
\usepackage{titlesec}
\usepackage{titletoc}
\usepackage{tocloft}
\usepackage[labelsep=period,tableposition=top,font=small,labelfont=bf]{caption}
\usepackage{tabularx}
\usepackage{booktabs}
\usepackage{csquotes}
\usepackage{enumitem}
\usepackage{pdflscape}
\usepackage{afterpage}
\usepackage{pdfpages}
\usepackage{hyperref}
\usepackage{scrextend}
\usepackage{chngcntr}
\usepackage{tikz}
\usepackage{feynmp}

\usepackage[backend=biber,citestyle=numeric-comp,maxbibnames=5,sorting=none, sortcites=false,doi=false,url=false,abbreviate=false]{biblatex}
\renewcommand{\newunitpunct}{\addcomma\space}

\renewbibmacro{in:}{}
\renewbibmacro{volume+number+eid}{
  \printfield{volume}
  \setunit{\addcomma\space no.}
  \printfield{number}
  \setunit{\addcomma\space}
  \printfield{eid}}
\DeclareFieldFormat[article,inbook,incollection,inproceedings,thesis,unpublished]{title}{#1\isdot}
\DeclareFieldFormat{pages}{#1}
\DeclareNameAlias{sortname}{first-last}
\AtEveryBibitem{\ifentrytype{inproceedings}{\clearfield{year}}{}}
\AtEveryBibitem{\ifentrytype{unpublished}{\clearfield{year}}{}}
\AtEveryBibitem{\ifentrytype{online}{\clearfield{year}}{}}
\DefineBibliographyExtras{english}{%
  \let\finalandcomma\empty
  \let\finalandsemicolon\empty
}

\newbool{titlefirst}

\newbibmacro*{titlefirst}{%
  \ifboolexpr{
    test {\iffieldundef{title}}
    and
    test {\iffieldundef{subtitle}}
  }
    {}
    {\printtext[title]{%
       \bf%
       \printfield[titlecase]{title}%
       \setunit{\subtitlepunct}%
       \printfield[titlecase]{subtitle}}%
     \newunit}%
  \printfield{titleaddon}}

\AtEveryBibitem{
  \ifbool{titlefirst}{
    \renewbibmacro*{begentry}{%
      \usebibmacro{titlefirst}\newline\nopunct\newunit}
    \renewbibmacro*{title}{} %
  }
}

\DeclareBibliographyCategory{contribution}

\hypersetup{
pdfauthor={Petja Paakkinen},
pdftitle={New constraints for nuclear parton distribution functions from hadron--nucleus collision processes}
pdfsubject={PhD thesis}
}

\def\fig#1{Figure~\ref{fig:#1}}

\def\eq#1{Eq.~(\ref{eq:#1})}
\def\eqs#1#2{Eqs.~(\ref{eq:#1})~and~(\ref{eq:#2})}
\def\tab#1{Table~\ref{tab:#1}}
\def\chap#1{Chapter~\ref{chap:#1}}
\def\sec#1{Section~\ref{sec:#1}}
\def\secs#1#2{Sections~\ref{sec:#1}~and~\ref{sec:#2}}

\DeclareMathOperator{\tr}{Tr}
\DeclareMathOperator{\su}{SU}
\DeclareMathOperator{\usym}{U}

\newcommand{\iseq}{\overset{\text{\raisebox{-0.25ex}{IS}}}{=}}
\newcommand{\cceq}{\overset{\text{\raisebox{-0.25ex}{CC}}}{=}}
\newcommand{\lleq}{\overset{\text{\raisebox{-0.25ex}{LL}}}{=}}
\newcommand{\trfeq}{\overset{\text{\raisebox{-0.25ex}{TRF}}}{=}}
\newcommand{\breiteq}{\overset{\text{\raisebox{-0.25ex}{Breit}}}{=}}

\newcommand{\smallfrac}[2]{\frac{\text{\small$#1$}}{\text{\small$#2$}}}

\renewcommand{\D}[1]{\mathrm{D^{#1}}}
\newcommand{\ptave}{p_{\rm T}^\text{ave}}

\newcommand{\nf}{n_f}
\newcommand{\tf}{T_\mathrm{F}}
\newcommand{\cf}{C_\mathrm{F}}
\newcommand{\ca}{C_\mathrm{A}}
\newcommand{\gs}{g_\mathrm{s}}
\newcommand{\alphas}{\alpha_\mathrm{s}}
\newcommand{\alphaem}{\alpha_\mathrm{em}}

\newcommand{\mh}{M}

\newcommand{\eqsq}{e_q^2}
\newcommand{\halfeqsq}{\frac{\eqsq}{2}}

\newcommand{\msbar}{\overline{\text{MS}}}

\newcommand{\fsp}{\slashed{p}}
\newcommand{\fspup}{\slashed{P}}

\newcommand{\fsn}{\slashed{n}}

\newcommand{\e}{\mathrm{e}}
\renewcommand{\i}{\mathrm{i}}
\renewcommand{\d}[1]{\mathrm{d} #1}

\newcommand{\parderiv}[2]{\frac{\partial #1}{\partial #2}}

\newcommand{\mconv}[3]{#1 \otimes #2 \left( #3 \right)}

\newcommand{\bigo}[1]{{\cal O}\left(#1\right)}

\newcommand{\pdotn}{{p \cdot n}}

\newcommand{\vect}[1]{\boldsymbol{#1}}

\newcommand{\kt}[1][]{k_{#1 \perp}}

\newcommand{\ktv}[1][]{\vect{k}_{#1 \perp}}
\newcommand{\ktvsq}[1][]{\ktv[#1]^2}

\newcommand{\mean}[1]{\langle{#1}\rangle}

\newcommand{\set}[1]{\{{#1}\}}

\newcommand{\pij}{P_{ij}}
\newcommand{\pqq}{P_{qq}}
\newcommand{\pqg}{P_{qg}}
\newcommand{\pgq}{P_{gq}}
\newcommand{\pgg}{P_{gg}}

\begin{document}

\frontmatter

\includepdf[pages=1-2]{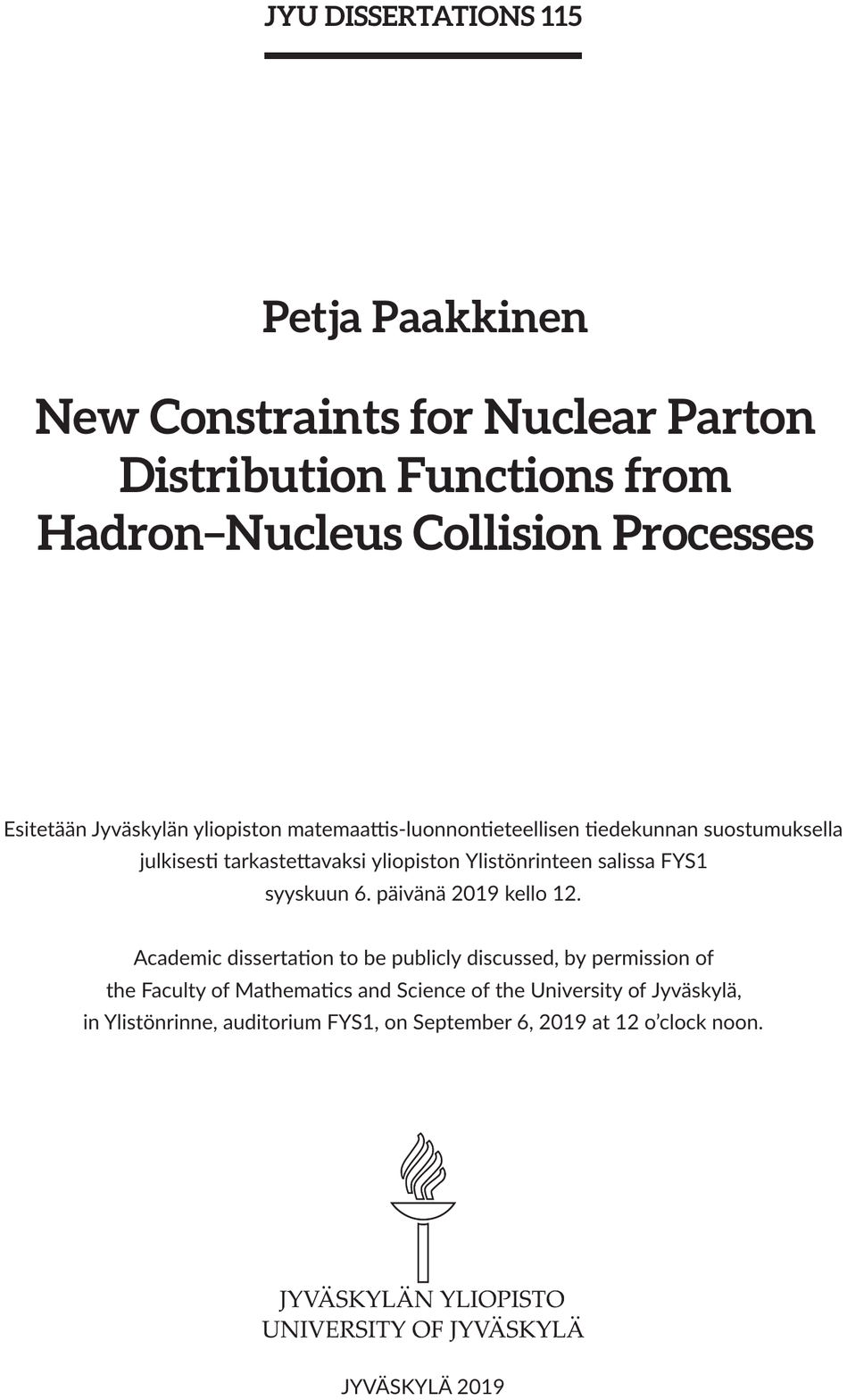}
\pagestyle{empty}

\phantom{.}
\vfill
\noindent\textbf{\Large Abstract}

\smallskip\bigskip\noindent Paakkinen, Petja\\
New constraints for nuclear parton distribution functions from hadron--nucleus collision processes\\
Jyväskylä: University of Jyväskylä, 2019\\
(JYU Dissertations\\
ISSN 2489-9003, 115)\\
ISBN 978-951-39-7828-0

\bigskip\noindent This work studies collinearly factorizable nuclear parton distribution functions (nPDFs) in perturbative Quantum Chromodynamics (QCD) at next-to-leading order in the light of hadron--nucleus collision data which have not been included in nPDF analyses previously. The aim is at setting new constraints on the nuclear modifications of the gluon distribution and on the flavour separation of quark nuclear modifications. The introductory part provides an outline of the theoretical framework of QCD collinear factorization and the used statistical methods and relates the work presented here to other similar contemporary analyses.

As a result, a new set of nPDFs, EPPS16, is presented, including for the first time electroweak-boson and dijet production data from CERN-LHC proton--lead collisions and allowing a full flavour separation in the fit. The flavour separation is constrained with Drell--Yan dilepton-production data from fixed target pion--nucleus experiments and neutrino--nu\-cleus deep-inelastic scattering data, which are shown to give evidence for the similarity of the $u$ and $d$ valence-quark nuclear modifications. For studying the gluon degrees of freedom, collider data are essential and in the EPPS16 analysis new constraints are derived from the dijet production at the LHC.

Possible further constraints for the gluons are investigated in terms of the LHC data on nuclear modification ratios of dijet and D-meson production. Using a non-quadratically improved Hessian reweighting method, these measurements are found to put stringent constraints on the gluon modifications in the lead nucleus, reaching smaller values of the nucleon momentum fraction than previously accessible. A study on the future prospects of constraining nPDFs within a multi-observable approach with the BNL-RHIC is also given.
\vfill

\clearpage
\phantom{.}
\vfill
\begin{description}[labelsep=0.25em,labelwidth=2.5cm,leftmargin=4.0cm,itemindent=0.0cm]
  \item[Author]
    Petja Paakkinen\\
    Department of Physics\\
    University of Jyväskylä\\
    Finland
  \item[Supervisors]
    Doc.~Hannu Paukkunen\\
    Department of Physics\\
    University of Jyväskylä\\
    Finland

  \vspace{\itemsep}
    Prof.~Kari J.~Eskola\\
    Department of Physics\\
    University of Jyväskylä\\
    Finland
  \item[Reviewers]
    Dr.~Urs A.~Wiedemann\\
    Theoretical Physics Department\\
    CERN\\
    Switzerland

  \vspace{\itemsep}
    Prof.~N\'estor Armesto\\
    Departamento de F\'isica de Part\'iculas and IGFAE\\
    Universidade de Santiago de Compostela\\
    Spain
  \item[Opponent]
    Prof.~Juan Rojo\\
    Department of Physics and Astronomy\\
    Vrije Universiteit Amsterdam\\
    The Netherlands
\end{description}
\vfill

\clearpage
\pagestyle{plain}

\chapter*{Preface}

\vspace{-0.355cm}
The research presented in this thesis has been carried out at the University of Jyväskylä during the period from November 2015 to August 2019. In addition to the University of Jyväskylä, the work has been funded by the Magnus Ehrnrooth Foundation and during the very final stages of the work also by the Academy of Finland, Project 308301. The Academy of Finland, Project 297058, is also ac\-knowl\-edged for the generous travel funding which has enabled the author to present the results of this work at various scientific meetings around the world.

My foremost thanks go to my supervisors Doc.\ Hannu Paukkunen and Prof.\ Kari J.\ Eskola. It was in Kari's Particle Physics course where I first got introduced to the concept of parton distribution functions. His enthusiasm on the topic was, least to say, contagious, so that when I was asked whether I wanted to work on the subject, I did not need to think twice. I am thankful for his guidance and support ever since. I have come to know Hannu as one of the leading younger researchers in the field. I feel grateful for the encouraging and collaborative manner with which he has overseen my progress. I wish to thank also my collaborators Prof.\ Carlos A.\ Salgado, Dr.\ Ilkka Helenius, Prof.\ John Lajoie and Dr.\ Joseph D.\ Osborn for their valuable contributions to the articles presented in this thesis. I thank Dr.\ Urs A.\ Wiedemann and Prof.\ N\'estor Armesto for reviewing the manuscript and Prof.~Juan Rojo for promising to act as the opponent at the public examination.

Many colleagues and friends, all of whom I cannot possibly name here, have had a less direct influence on the completion of this thesis. I feel I should thank the whole Jyväskylä QCD theory group, past and present, for the inspiring research environment that we have. The work would have been far less fun without the people at the office YFL 353 a.k.a.\ Holvi. Special thanks go to the latest generation of young aspiring scientists H.~Hänninen, L.~Jokiniemi, M.~Kuha, T.~Löytäinen and P.~Pirinen. I also wish to thank Matias, Valtteri, Pekka and Juha, with whom I spent most of my undergraduate years, for the long-lasting friendship.

I thank my father, whose book collection gave me some of the first inspirations to study theoretical physics, my mother, who always encouraged me to study mathematical subjects, and my whole family for their unconditional support on all of my endeavours. Finally, with the utmost gratitude, I wish to thank Sanna for all her love.

\bigskip\noindent Jyväskylä, August 2019 \hfill Petja Paakkinen

\addtocategory{contribution}{Paakkinen:2016wxk,Eskola:2016oht,Eskola:2019dui,Helenius:2019lop,Eskola:2019bgf}
\nocite{Paakkinen:2016wxk,Eskola:2016oht,Eskola:2019dui,Helenius:2019lop,Eskola:2019bgf}
\defbibnote{descr}{This thesis consists of an introductory part and of the following publications:}
\defbibnote{authorcontr}{The author performed all the calculations and wrote the first draft for the article~\cite{Paakkinen:2016wxk}. For the article~\cite{Eskola:2016oht}, the author produced look-up tables for fast calculations of the cross-section ratios discussed in the article~\cite{Paakkinen:2016wxk} and participated in writing of the manuscript. The author performed all the calculations presented in the article~\cite{Eskola:2019dui} except those for the 7 TeV proton--proton inclusive jet cross sections and wrote the first version of the manuscript and handled the revision process. For the articles~\cite{Helenius:2019lop} and \cite{Eskola:2019bgf}, the author performed the reweighting analysis using the tools developed for the article~\cite{Eskola:2019dui} and participated in the preparation of the manuscripts.}
\booltrue{titlefirst}
\printbibliography[category=contribution,title=List of Publications,prenote=descr,postnote=authorcontr]
\boolfalse{titlefirst}

\cleardoublepage

\setcounter{tocdepth}{2}
\tableofcontents

\mainmatter

\chapter{Introduction}
\label{chap:intro}

The Standard Model of particle physics describes our present-day best knowledge of the fundamental particles of Nature and their electromagnetic, weak and strong interactions. It is a renormalizable quantum field theory with a local $\usym(1)\times\su(2)\times\su(3)$ gauge symmetry. The $\usym(1)\times\su(2)$ symmetry, spontaneously broken through the Higgs mechanism, gives rise to the electroweak interactions, while the unbroken $\su(3)$ symmetry dictates, in a theory called Quantum Chromodynamics (QCD), the strong interaction between particles carrying an $\su(3)$ colour charge. What makes the strong interaction different from electroweak phenomena is the property of confinement: the strong force binds coloured particles, quarks and gluons, into colourless hadrons. We thus never observe freely propagating quarks and gluons, only the hadrons they constitute.

This poses a difficulty in the theoretical description of the QCD phenomena, as the asymptotic states are not the fundamental degrees of freedom of the theory. Fortunately, scattering processes involving a large momentum transfer factorize~\cite{Collins:1989gx}, i.e.\ the cross sections of these hard processes can be obtained by convoluting the scattering probabilities of the fundamental particles with long-distance functions describing their distributions in the involved hadrons. This makes it possible to study the distributions of partons, particles inside hadrons, by measurements of hard-process cross sections. These long-distance functions are called \emph{parton distribution functions} (PDFs).

The PDFs are universal, independent of the scattering process, and hence distributions extracted from one process can be used to make predictions of another. It is not, however, possible to determine the PDFs of all different parton flavours independently from a single observable and instead large sets of data from different measurements are needed for their reliable extraction. This has led to the development of the field of PDF global analyses with ever-increasing precision in the obtained PDFs~\cite{Gao:2017yyd}.

This thesis deals with PDFs of a particular kind, the nuclear PDFs (nPDFs), describing the partonic content of nucleons bound in nuclei. Even 20 years after the pioneering works~\cite{Frankfurt:1990xz,Eskola:1992zb,Eskola:1998iy,Eskola:1998df}, the nPDFs carry large uncertainties. Until very recently, the possible asymmetry in nuclear modifications of different valence and sea-quark flavours has not been considered in the nPDF global analyses. Also the nuclear modifications of gluons, for which direct constraints have been scarce, have remained poorly known. These open problems are addressed in this thesis. In particular, we will discuss the impact of new constraints from hadron--nucleus collision processes which have not been used in nPDF global analyses previously. These include CERN-LHC proton--lead measurements of electroweak bosons, dijets and inclusive $\D0$-production, but also older pion--nucleus Drell--Yan measurements. Further, by considering prospects at present and upcoming experiments, we will try to pave the way towards a better understanding of the PDF nuclear modifications in the future. The work presented in this thesis is performed at the level of next-to-leading order perturbative QCD.

The introductory part is organized as follows. \chap{pdfsincollfact} introduces the the\-o\-ret\-i\-cal framework of the thesis, the collinear factorization of QCD. The discussion here is rather minimal, with the aim at presenting the relevant concepts, but avoiding any unnecessary calculational details. In \chap{globalanalysis}, the used statistical methods are presented. Emphasis is given to the treatment of correlated uncertainties, which will become increasingly important in the nPDF fits with precision data from the LHC becoming available. The novel physics results are discussed and compared to the results of earlier analyses in \chap{nuclearpdfs} and summarized in \chap{concl}.

\chapter{Parton distributions in collinear factorization}
\label{chap:pdfsincollfact}

Quantum Chromodynamics, the theory of strong interactions within the Standard Model, is characterized by its Lagrangian, defined in terms of the quark and gluon fields $\psi_i$ and $A^a_\mu$ as~\cite{Muta:1987mz,Ellis:1991qj}
\begin{equation} \label{eq:lagrangian}
    {\cal L} = \sum_i\bar{\psi}_i(\i\slashed{D} - m_i)\psi_i - \frac{1}{4}F^a_{\mu\nu}F^{a,\mu\nu},
\end{equation}
where the sum goes over the quark flavours with masses $m_i$ and
\begin{gather}
    D_\mu = \partial_\mu - \i \gs A^a_\mu t^a, \\
    F^a_{\mu\nu} = \partial_\mu A^a_\nu - \partial_\nu A^a_\mu + \gs f^{abc} A^b_\mu A^c_\nu
\end{gather}
are the covariant derivative and the gluon field strength tensor, respectively. Here, $t^a$ are the $\su(3)$ generators in the fundamental representation and $f^{abc}$ the structure constants. The Lagrangian is invariant in local $\su(3)$ gauge transformations, which in the case of an infinitesimal shift $\theta$ can be written as
\begin{equation}
    \psi \rightarrow \psi + \i \theta^a t^a \psi, \quad A^a_\mu \rightarrow A^a_\mu + \frac{1}{\gs} \partial_\mu \theta^a + f^{abc} A^b_\mu \theta^c,
\end{equation}
leading to conservation of the $\su(3)$ colour charge.

The strength of the QCD interactions is set by the coupling $\gs$. In renormalizing the theory, this bare coupling must be traded with the running coupling, usually expressed in terms of $\alphas(Q^2) = \gs^2(Q^2)/4\piup$, which depends on the interaction scale $Q^2$ as
\begin{equation}
  Q^2\parderiv{\alphas}{Q^2} = \beta(\alphas),
\end{equation}
with a negative beta function, $\beta < 0$. At low energies the coupling is large, permitting colour confinement, but towards higher scales the coupling gets weaker, asymptotically approaching zero. This phenomenon is called \emph{asymptotic freedom}~\cite{Gross:1973id,Politzer:1974fr} and it allows one to use perturbation theory to calculate high-momentum-transfer cross sections in QCD.

\section{Deep inelastic scattering in parton model}

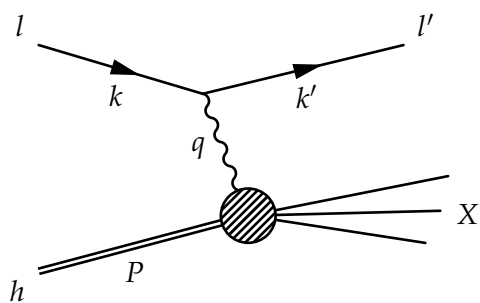
\begin{figure}[b]
    \centering
    \small
    \bigskip
    \begin{fmffile}{diagrams/dis}
        \begin{fmfgraph*}(60,30)
            \fmfleft{ip,il}
            \fmfright{o1,o2,o3,o4,o5,o6,o7,ol}
            \fmf{fermion,label=$k$,tension=1.5}{il,vl}
            \fmf{fermion,label=$k'$}{vl,ol}
            \fmf{double,label=$P$,tension=2.5}{ip,vp}
            \fmf{photon,label=$q$}{vl,vp}
            \fmfblob{.12w}{vp}
            \fmf{plain}{vp,o2}
            \fmf{plain}{vp,o3}
            \fmf{plain}{vp,o4}
            \fmflabel{$l$}{il}
            \fmflabel{$l'$}{ol}
            \fmflabel{$h$}{ip}
            \fmflabel{$X$}{o3}
        \end{fmfgraph*}
    \end{fmffile}
    \bigskip
    \caption{Deep inelastic lepton--hadron scattering.}
    \label{fig:dis}
\end{figure}

The easiest way to study experimentally the inner workings of hadrons is by deep inelastic scattering (DIS). In this process, illustrated in \fig{dis}, a lepton $l$ with high energy $E$ scatters off a hadron $h$, which then breaks apart into an inclusive final state $X$ with a large invariant mass $W \gg \mh$, where $\mh$ is the hadron mass. In the target rest frame (TRF), the square of the four-momentum transfer from the lepton to the hadron $q \trfeq (E-E',\vect{k}-\vect{k}')$, where $\vect{k}$, $\vect{k}'$ are the three-momenta of the initial and final state leptons, is given in terms of the energy $E'$ and scattering angle $\theta$ of the final state lepton $l'$ by
\begin{equation}
  Q^2 \coloneqq -q^2 \trfeq 2EE'(1-\cos\theta).
\end{equation}
The other relevant kinematical quantities for this process, the Bjorken $x$ and the inelasticity $y$, are defined as
\begin{equation}
  x \coloneqq \frac{Q^2}{2P \cdot q} \trfeq \frac{Q^2}{2\mh(E-E')}, \qquad y \coloneqq \frac{P \cdot q}{P \cdot k} \trfeq 1 - \frac{E'}{E^{\phantom{\prime}}},
\end{equation}
where $P \trfeq (M,0,0,0)$ is the four momentum of the hadron.
In these Lorentz invariant variables, the unpolarized double-differential cross section can be expressed as
\begin{equation}
    \frac{\d{\sigma}}{\d{Q^2}\d{x}} = \frac{4\piup\alphaem^2}{Q^4}\,\frac{y^2}{2Q^2}\,L_{\mu\nu}(k,k')\,W^{\mu\nu}(P,q),
\end{equation}
where $\alphaem = e^2/4\piup$ is the electromagnetic fine-structure constant and $L_{\mu\nu}$, $W^{\mu\nu}$ refer to the leptonic and hadronic tensors, respectively.

In a neutral-current electromagnetic scattering mediated by a virtual photon, the leptonic tensor is simply
\begin{equation}
    L_{\mu\nu}(k,k') = 2\,(k^{\phantom{\prime}}_\mu k'_\nu + k'_\mu k^{\phantom{\prime}}_\nu - k \cdot k' g^{\phantom{\prime}}_{\mu\nu}).
\end{equation}
and, by conservation of current and the leptonic tensor being real and symmetric, the hadronic tensor can be expressed as~\cite{Collins:2011zzd}
\begin{equation}
  \begin{split}
    W^{\mu\nu}(P,q) = &-\left( g^{\mu\nu} - \frac{q^\mu q^\nu}{q^2} \right) F_1(x,Q^2) \\&+ \frac{1}{P \cdot q}\left( P^\mu - \frac{P \cdot q}{q^2} q^\mu \right)\left( P^\nu - \frac{P \cdot q}{q^2} q^\nu \right) F_2(x,Q^2),
  \end{split}
\end{equation}
where the structure functions $F_{1,2}$ encode our ignorance of the hadron structure.
In these terms, the cross section reads
\begin{equation} \label{eq:structfxsec}
    \frac{\d{\sigma}}{\d{Q^2}\d{x}} = \frac{4\piup\alphaem^2}{Q^4}\,\frac{1}{x}\left\{ xy^2 F_1(x,Q^2) + \left(1-y-x^2y^2\frac{M^2}{Q^2}\right) F_2(x,Q^2) \right\}.
\end{equation}

\subsubsection{Parton model}

Now, let us consider the DIS in a frame where the hadron is moving very fast, e.g.\ the Breit frame, where assuming $Q^2 \gg M^2$, we can take
\begin{equation}
  P \breiteq (Q/2x,0,0,Q/2x), \qquad q \breiteq (0,0,0,-Q), \qquad Q = \sqrt{Q^2}.
\end{equation}
In such a frame the hadron is Lorentz contracted and the interaction times of its constituents are strongly dilated. During the short phase when the collision with the lepton takes place the hadron is thus ``frozen'' and the lepton can scatter incoherently from the individual partons. This description is the basis of the ``naive'' parton model~\cite{Feynman:1969ej,Bjorken:1969ja} giving the leading behaviour of the DIS cross section. In this picture, the partons move collinearly with the parent hadron and we can define
\begin{description}[labelsep=0.25em,labelindent=0.5em]
  \item[$f_i(\xi) =$] the probability density of finding a parton $i$ within the hadron carrying a fraction $\xi$ of the hadrons momentum.
\end{description}
In more formal terms the PDFs can be defined as operator expectation values, see Refs.~\cite{Collins:2011zzd,Kovarik:2019xvh}.

\begin{figure}[tb]
    \centering
    \small
    \begin{tikzpicture}[scale=2.5]
        \node at (0,0.5) {
            \begin{fmffile}{diagrams/handbag}
                \begin{fmfgraph*}(45,30)
                    \fmfleft{i1,i2}
                    \fmfright{o1,o2}
                    \fmflabel{$P$}{i1}
                    \fmflabel{$q$}{i2}
                    \fmf{photon,tension=5}{i2,vl}
                    \fmf{photon,tension=5}{vr,o2}
                    \fmf{phantom}{i1,vl,vr,o1}
                    \fmffreeze
                    \fmf{double,tension=2}{i1,vg1}
                    \fmf{fermion,label=$p$,label.side=left}{vg1,vl}
                    \fmf{fermion,label={$p'$\hspace{5mm}},label.side=left}{vl,vr}
                    \fmf{fermion}{vr,vg2}
                    \fmf{double,tension=2}{vg2,o1}
                    \fmf{phantom,tension=0}{vg1,vg2}
                \end{fmfgraph*}
            \end{fmffile}
        };
        \draw[dashed, thick] (0.01,-0.2) -- (0.01,1.2);
        \draw[fill=black!10, thick] (0,0.2) ellipse (0.8 and 0.2);
        \node at (0.0,0.2) {$f_q$};
    \end{tikzpicture}
    \bigskip
    \caption{Parton-model picture of the hadronic tensor. Labels refer to the four-momenta of the particles.}
    \label{fig:born}
\end{figure}
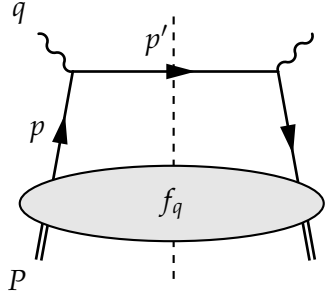

The photon couples only to electrically charged particles and hence at the leading order the hadronic tensor takes the form
\begin{equation}
    W^{\mu\nu}(P,q) = \sum_{i=q,\bar{q}} \int\frac{\d{\xi}}{\xi}\,f_i(\xi)\,\hat{W}_i^{\mu\nu}(p,q) + \bigo{\frac{1}{Q^2}},
    \label{eq:hadrontensor}
\end{equation}
where $p = \xi P \breiteq (\xi Q/2x,0,0,\xi Q/2x)$ and $\hat{W}_i^{\mu\nu}$ denotes a so-called partonic tensor. This can be pictorically represented as a ``handbag'' diagram, given in \fig{born}, where the left-hand side of the cut represents the scattering amplitude and the right-hand side its complex conjugate. The antiquark contribution is obtained simply by changing the direction of the fermion line. At low scales, additional ``higher-twist'' contributions, essentially originating from multi-parton interactions, to the simple parton model picture, denoted by $\bigo{1/Q^2}$ in \eq{hadrontensor} can become important. At the clearly perturbative scales $Q^2 \gg M^2$ these should be negligible and we do not discuss them further here. In this leading order (LO), or ``Born'', approximation the quark-initiated partonic tensor is
\begin{equation}
    \hat{W}_{q,\text{Born}}^{\mu\nu}(p,q) = \frac{x}{2Q^2}\,\halfeqsq\,\tr[\fsp\gamma^\nu\fsn\gamma^\mu]\,\delta(\xi-x),
    \label{eq:quarktensor}
\end{equation}
where
\begin{equation}
  n = q + xP \breiteq (Q/2,0,0,-Q/2), \qquad n^2 = 0, \label{eq:nvector}
\end{equation}
and $\eqsq$ is the square of quark fractional charge.
The delta function in \eq{quarktensor} arises from integrating over the final state quark momentum $p'$ and shows us that, to leading order perturbative accuracy, the Bjorken $x$ measures the momentum fraction of the parton.

Now, using \eqs{hadrontensor}{quarktensor}, the differential cross section takes the form
\begin{equation}
    \left(\frac{\d{\sigma}}{\d{Q^2}\d{x}}\right)_\text{LO} = \sum_q\eqsq\,f_q(x)\,\left(\frac{\d{\hat{\sigma}}}{\d{Q^2}\d{x}}\right)_\text{Born}, \label{eq:pmdisxsec}
\end{equation}
where the sum is understood to be over both quarks and antiquarks and
\begin{equation}
    \left(\frac{\d{\hat{\sigma}}}{\d{Q^2}\d{x}}\right)_\text{Born} = \frac{4\piup\alphaem^2}{Q^4}\left\{ \frac{y^2}{2} + \left(1-y-x^2y^2\frac{M^2}{Q^2}\right) \right\},
\end{equation}
or equivalently, if expressed in terms of the structure functions, \eq{structfxsec}, we have
\begin{equation}
    2xF_1(x) = F_2(x) = x\sum_q\eqsq\,f_q(x).
\end{equation}
That is, in the naive parton model, the structure functions depend only on $x$ and not on the scale $Q^2$. This phenomenon, called \emph{Bjorken scaling}, is however broken by radiative corrections, as we will discuss next.

\section{DGLAP evolution}

The leading-order DIS cross section found in the previous section is subject to various radiative and virtual corrections at higher orders in perturbation theory. For massless partons, these corrections include collinear and soft divergences. The soft and final-state collinear divergences cancel at the level of summing over different contributions, but for the initial-state collinear divergences this cancellation is not complete. One can, however, resum these initial state divergences into the definitions of the PDFs, leading to the Dokshitzer--Gribov--Lipatov--Altarelli--Parisi (DGLAP) evolution of the parton densities~\cite{Dokshitzer:1977sg,Gribov:1972ri,Lipatov:1974qm,Altarelli:1977zs}. In the following, we present the general idea of how this is done. For more thorough discussions and calculational details the reader is directed to Refs.~\cite{Dokshitzer:1978hw,Altarelli:1981ax,Dokshitzer:1991wu,Paukkunen:2009ks,Paakkinen:2015}.

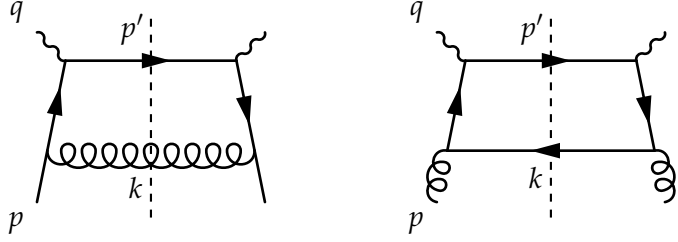
\begin{figure}
  \centering
  \begin{fmffile}{diagrams/realemission}
    \small
    \begin{tikzpicture}[scale=2.66]
      \node at (0,0.5) {
      \begin{fmfgraph*}(37.5,22.5)
        \fmfleft{i1,i2}
        \fmfright{o1,o2}
        \fmflabel{$p$}{i1}
        \fmflabel{$q$}{i2}
        \fmf{photon,tension=5}{i2,vl}
        \fmf{photon,tension=5}{vr,o2}
        \fmf{phantom}{i1,vl,vr,o1}
        \fmffreeze
        \fmf{plain,tension=1.75}{i1,vg1}
        \fmf{fermion}{vg1,vl}
        \fmf{fermion,label={$p'$\hspace{5mm}},label.side=left}{vl,vr}
        \fmf{fermion}{vr,vg2}
        \fmf{plain,tension=1.75}{vg2,o1}
        \fmf{gluon,tension=0,label={$k$\hspace{4mm}},label.side=right,label.dist=3.5mm}{vg1,vg2}
      \end{fmfgraph*}
      };
      \draw[dashed, thick] (0,0) -- (0,1);
    \end{tikzpicture}
  \end{fmffile}
  \qquad
  \begin{fmffile}{diagrams/initialgluon}
    \small
    \begin{tikzpicture}[scale=2.66]
      \node at (0,0.5) {
      \begin{fmfgraph*}(37.5,22.5)
        \fmfleft{i1,i2}
        \fmfright{o1,o2}
        \fmflabel{$p$}{i1}
        \fmflabel{$q$}{i2}
        \fmf{photon,tension=5}{i2,vl}
        \fmf{photon,tension=5}{vr,o2}
        \fmf{phantom}{i1,vl,vr,o1}
        \fmffreeze
        \fmf{gluon,tension=1.75}{vg1,i1}
        \fmf{gluon,tension=1.75}{o1,vg2}
        \fmf{fermion}{vg1,vl}
        \fmf{fermion,label={$p'$\hspace{5mm}},label.side=left}{vl,vr}
        \fmf{fermion}{vr,vg2}
        \fmf{fermion,tension=0,label={$k$\hspace{4mm}},label.side=left}{vg2,vg1}
      \end{fmfgraph*}
      };
      \draw[dashed, thick] (0,0) -- (0,1);
    \end{tikzpicture}
  \end{fmffile}
  \medskip
  \caption{Ladder diagrams at order $\alphas$. Left: Real-gluon emission from the initial-state quark. Right: Initial-state gluon splitting into a quark--antiquark pair.}
  \label{fig:ladder}
\end{figure}

For the problem at hand, it is convenient to use the light-cone gauge \cite{Bassetto:1984dq,Leibbrandt:1987qv}, where a gauge-fixing term ${\cal L}_\text{gauge-fixing} = - \frac{1}{2\lambda}\,(n^\mu A^a_\mu)^2$, with the limit $\lambda \rightarrow 0$ understood, is added to the Lagrangian in \eq{lagrangian} and the gauge vector $n$ is set to be that in \eq{nvector}. In this gauge, the only non-cancelling divergent contributions come from the ``ladder''-type diagrams shown in \fig{ladder} (here as well, also the contributions obtained by reversing the fermion line need to be summed). These diagrams contain fermion propagators with denominators $(p-k)^2$, which diverge at the limit where $k$ is collinear to $p$. Decomposing the momentum $k$ \`a la Sudakov~\cite{Sudakov:1954sw},
\begin{equation}
    k^\mu = (1-z) p^\mu + \frac{\ \ktvsq}{1-z} \frac{n^\mu}{2\pdotn} + \kt^\mu,
\end{equation}
where $\kt \breiteq (0,\ktv,0)$ is the component of momentum $k$ transverse to $p$ and $n$, we find that the contribution from the ladder diagram in \fig{ladder} (left) to the quark tensor is
\begin{equation}
    \hat{W}^{\mu\nu}_{q,\text{Ladder}}
    = \frac{x}{2Q^2} \halfeqsq \tr[\fsp\gamma^\nu\fsn\gamma^\mu] \,\frac{\alphas}{2\piup} \int_x^1\frac{\d{z}}{z}\, \cf \left(\frac{\,1+z^2}{1-z}\right)\delta\left(\xi-\frac{x}{z}\right)\,\int_0^{Q^2}\frac{\d{\ktvsq}}{\ktvsq} + \ldots,
    \label{eq:quarkladder}
\end{equation}
with a colour factor $\cf = 4/3$ and where ``\ldots'' refers to terms that do not contain a collinear divergence. The divergence thus appears as a logarithmic $\ktv$-integral, which we here regulate with a simple cut-off $\ktvsq \geq m^2$, whereby the integral becomes
\begin{equation}
    \int_{m^2}^{Q^2}\frac{\d{\ktvsq}}{\ktvsq} = \log\left(\frac{Q^2}{m^2}\right).
\end{equation}
\eq{quarkladder} still contains a soft divergence at the limit $z \rightarrow 1$, but this cancels when we take into account the quark self-energy (SE) correction to the quark field-strength renormalization.

Combining the real and virtual corrections, we find the total divergent $\bigo{\alphas}$ contribution to the hadronic tensor to be
\begin{equation}
    W^{\mu\nu}_\text{quark,Ladder+SE} \lleq \frac{x}{4Q^2}\tr[\fspup\gamma^\nu\fsn\gamma^\mu] \sum_q\eqsq\,\frac{\alphas}{2\piup}\log\left(\frac{Q^2}{m^2}\right)\,\int_x^1 \frac{\d{z}}{z}\,\pqq(z)\,f_q\left(\frac{x}{z}\right),
\end{equation}
where
\begin{equation}
  \pqq(z) = \cf\left[ \frac{1+z^2}{(1-z)_+} + \frac{3}{2}\delta(1-z) \right]
\end{equation}
is the leading-order \emph{Altarelli--Parisi splitting function} for the quark-to-quark transition~\cite{Altarelli:1977zs}, with the plus distribution defined in terms of an integral equation
\begin{equation}
  \int_0^1 \d{z}\,\frac{f(z)}{(1-z)_+} = \int_0^1 \d{z}\,\frac{f(z) - f(1)}{(1-z)}
\end{equation}
Similarly, the divergent contribution from initial-state gluons given by the ladder diagram in \fig{ladder} (right) can be expressed as
\begin{equation}
    W^{\mu\nu}_\text{gluon,Ladder} \lleq \frac{x}{4Q^2}\tr[\fspup\gamma^\nu\fsn\gamma^\mu] \sum_q\eqsq\,\frac{\alphas}{2\piup}\log\left(\frac{Q^2}{m^2}\right)\,\int_x^1 \frac{\d{z}}{z}\,\pqg(z)\,f_g\left(\frac{x}{z}\right),
\end{equation}
where we have the leading-order gluon-to-quark splitting function
\begin{equation}
    \pqg(z) = \tf\left((1-z)^2 + z^2\right), \qquad \tf = \frac{1}{2}.
\end{equation}

In the above equations, we have denoted by ``LL'' that we are only considering the \emph{leading logarithmic} contributions to the hadronic tensor. There are also further contributions at order $\alphas$ (see e.g.\ Ref.~\cite{Altarelli:1979ub}), but as these are of non-divergent nature, they are not important for the present discussion.
Now, summing all the $\bigo{\alphas}$ leading logarithmic terms with the leading order expression, we have
\begin{align} \label{eq:nloleadinglog}
    \left(\frac{\d{\sigma}}{\d{Q^2}\d{x}}\right)_\text{NLO} \lleq \sum_q\eqsq\,\bigg\{ \mconv{\left[ 1 + \frac{\alphas}{2\piup}\log\left(\frac{Q^2}{m^2}\right)\pqq \right]&}{f_q}{x} \\+\; \mconv{\frac{\alphas}{2\piup}\log\left(\frac{Q^2}{m^2}\right)\pqg&}{f_g}{x} \bigg\}\left(\frac{\d{\hat{\sigma}}}{\d{Q^2}\d{x}}\right)_\text{Born}, \notag
\end{align}
where the symbol $\otimes$ above denotes a multiplicative convolution, defined as
\begin{equation}
   \mconv{h}{f}{x} = \int_x^1 \frac{\d{z}}{z}\,h(z)\,f\left(\frac{x}{z}\right), \qquad \mconv{1}{f}{x} = \int_x^1 \frac{\d{z}}{z}\,\delta(1-z)\,f\left(\frac{x}{z}\right) = f(x).
\end{equation}
We notice that the result in \eq{nloleadinglog} is nothing but the parton-model cross section in \eq{pmdisxsec} with the parton distribution $f_q(x)$ replaced with the term in curly braces. The collinear divergences thus \emph{factorize} from the partonic process.

At this point, as we have seen that the collinear divergences occur when an internal quark gets on-shell and is thus allowed to propagate a long distance before the scattering with the virtual photon, it appears natural to redefine the PDFs as to include these long-distance effects. But before we do so, we have to note that similar collinear divergences can appear at \emph{all} orders of perturbation theory, thus potentially spoiling this simple picture. The crucial thing here is that these divergent contributions exponentiate, and the DIS cross section can be written, in the leading-logarithm accuracy, as
\begin{equation} \label{eq:lldis}
    \frac{\d{\sigma}}{\d{Q^2}\d{x}} \lleq \sum_{q} e_q^2 \left(\begin{matrix}
        1 & 0
    \end{matrix}\right) \mconv{\exp\!\left[\frac{\alphas}{2\piup}\log\left(\frac{Q^2}{m^2}\right)\left(\begin{matrix}
        \pqq & \pqg \\ \pgq & \pgg
    \end{matrix}\right)\right]\!}{\!\left(\begin{matrix}
        f_q \\ f_g
    \end{matrix}\right)}{x}\left(\frac{\d{\hat{\sigma}}}{\d{Q^2}\d{x}}\right)_\text{\!Born},
\end{equation}
where the exponential convolution should be understood as
\begin{equation}
  \mconv{\exp\!\left[\mathcal{P}\right]}{f}{x} = \sum_n \frac{1}{n!} \underbrace{\mathcal{P} \otimes \cdots \otimes \mathcal{P}}_{n\ \text{times}} \otimes f \left( x \right)
\end{equation}
and where we now also have the leading-order quark-to-gluon and gluon-to-gluon splitting functions
\begin{align}
    \pgq(z) &= \cf\left(\frac{1+(1-z)^2}{z}\right), \\
    \pgg(z) &= 2\,\ca\left(\frac{1-z}{z} + \frac{z}{(1-z)_+} + z(1-z)\right) + \left(\frac{11}{6}\ca - \frac{2}{3}\nf\tf\right)\delta(1-z), \notag
\end{align}
where $\ca = 3$ and $\nf$ is the number of active quark flavours.
Hence it makes sense to define \emph{scale-dependent} parton distribution functions as
\begin{equation} \label{eq:pdfdef}
    \left(\begin{matrix}
        f_q(x,Q^2) \\ f_g(x,Q^2)
    \end{matrix}\right)
    \coloneqq \mconv{\exp\!\left[\frac{\alphas}{2\piup}\log\left(\frac{Q^2}{m^2}\right)\left(\begin{matrix}
        \pqq & \pqg \\ \pgq & \pgg
    \end{matrix}\right)\right]\!}{\!\left(\begin{matrix}
        f_q \\ f_g
    \end{matrix}\right)}{x},
\end{equation}
from where, by taking the $Q^2$ derivative, we find the \emph{Dokshitzer--Gribov--Lipatov--Altarelli--Parisi \emph{(DGLAP)} evolution equations}~\cite{Dokshitzer:1977sg,Gribov:1972ri,Lipatov:1974qm,Altarelli:1977zs},
\begin{equation} \label{eq:dglap}
    Q^2\parderiv{}{Q^2}\left(\begin{matrix}
        f_q(x,Q^2) \\ f_g(x,Q^2)
    \end{matrix}\right)
    \lleq \mconv{\frac{\alphas}{2\piup}\left(\begin{matrix}
        \pqq & \pqg \\ \pgq & \pgg
    \end{matrix}\right)}{\left(\,\begin{matrix}
        f_q(Q^2) \\ f_g(Q^2)
    \end{matrix}\,\right)}{x}.
\end{equation}
Now, substituting the definition in \eq{pdfdef} back to \eq{lldis}, the physical predictions become independent of the collinear regulator and in this \emph{QCD-improved} parton model, the full, \emph{finite}, leading-order $+$ leading-logarithm DIS cross section reads
\begin{equation} \label{eq:qcdimproveddisxsec}
    \left(\frac{\d{\sigma}}{\d{Q^2}\d{x}}\right)_{\text{LO}+\text{LL}} = \sum_{q} e_q^2\,f_q(x,Q^2)\left(\frac{\d{\hat{\sigma}}}{\d{Q^2}\d{x}}\right)_\text{Born},
\end{equation}
with the PDF scale evolution governed by \eq{dglap}.

\section{Factorization schemes and scales}

In the discussion above, we have only considered the leading logarithmic contributions. At higher orders, $\alphas^{n+1} \log^n(Q^2/m^2)$, etc., both the partonic cross sections after the extraction of divergences (or \emph{coefficient functions}) and the splitting functions get perturbative corrections~\cite{Altarelli:1979ub,Furmanski:1980cm,Curci:1980uw}. Moreover, the definition given in \eq{pdfdef} is not unique, leading to \emph{scheme dependence} of the PDFs and of the splitting and coefficient functions~\cite{Brock:1993sz}.
Any physical predictions are still independent of the scheme to the perturbative order to which they have been calculated.
To elaborate this more, let us write here the full NLO expression of the structure function $F_2$ as
\begin{equation}
  \begin{split}
    F_2^{\rm NLO} = x\sum_q\eqsq\,\bigg\{ \mconv{\left[ 1 + \frac{\alphas}{2\piup}\log\left(\frac{Q^2}{m^2}\right)\pqq + R_q \right]&}{f_q}{x} \\+\; \mconv{\left[ \frac{\alphas}{2\piup}\log\left(\frac{Q^2}{m^2}\right)\pqg + R_g \right]&}{f_g}{x} \bigg\}
  \end{split}
\end{equation}
where $R_{q,g}$ denote the remainder parts including all the non-divergent $\bigo{\alphas}$ terms which we neglected in the discussion leading to \eq{nloleadinglog}. While we must include the large logarithms to the redefinitions of the PDFs, nothing prevents us from including also some of the finite parts. Defining now
\begin{equation}
  f_i(x,Q^2) \coloneqq \sum_{j}\mconv{\left[ \delta_{ij} + \frac{\alphas}{2\piup}\log\left(\frac{Q^2}{m^2}\right)\pij + f_{ij}^{\rm scheme} \right]}{f_j}{x} + \bigo{\alphas^2},
\end{equation}
we can write
\begin{equation}
  F_2^{\rm NLO} = x\sum_q\eqsq\,\big\{ \mconv{\underbrace{\big[ 1 + R_q - f_{qq}^{\rm scheme} \big]}_{\phantom{C_q^{\rm scheme}}\ \eqqcolon C_q^{\rm scheme}}}{f_q(Q^2)}{x} + \mconv{\underbrace{\big[ R_g - f_{qg}^{\rm scheme} \big]}_{\phantom{C_g^{\rm scheme}}\ \eqqcolon C_g^{\rm scheme}}}{f_g(Q^2)}{x} \big\},
\end{equation}
where $C_{q,g}^{\rm scheme}$ are now the NLO coefficient functions in the chosen scheme.

In a similar fashion, one can also choose to define the PDFs at some \emph{factorization scale} $Q_f$ different from $Q$, including the remaining $\log(Q^2/Q_f^2)$ terms in the coefficient functions,
\begin{equation}
  \begin{split}
    F_2^{\rm NLO} = x\sum_q\eqsq\,\bigg\{ \mconv{\overbrace{\bigg[ 1 + \frac{\alphas}{2\piup}\log\bigg(\frac{Q^2}{Q_f^2}\bigg)\pqq + R_q - f_{qq}^{\rm scheme} \bigg]}^{\phantom{C_q^{\rm scheme}(Q^2/Q_f^2)}\ \eqqcolon C_q^{\rm scheme}(Q^2/Q_f^2)}&}{f_q(Q_f^2)}{x} \\+\; \mconv{\underbrace{\bigg[ \frac{\alphas}{2\piup}\log\bigg(\frac{Q^2}{Q_f^2}\bigg)\pqg + R_g - f_{qg}^{\rm scheme} \bigg]}_{\phantom{C_g^{\rm scheme}(Q^2/Q_f^2)}\ \eqqcolon C_g^{\rm scheme}(Q^2/Q_f^2)}&}{f_g(Q_f^2)}{x} \bigg\}.
  \end{split}
\end{equation}
Again the different scale choices are formally equivalent up to corrections of one order of $\alphas$ higher. For this property, it is possible to estimate the uncertainties arising from the termination of the perturbative expansion by calculating so-called scale uncertainties through variations of the indefinite scales of the process.

While for the purpose of demonstrating the appearance of collinear divergences and their resummation in the discussion above it was useful to work in four spacetime dimensions and use cut-off regulators, it is more common in practical calculations to use dimensional regularization~\cite{tHooft:1972tcz}, which does not break any symmetries of the theory. In the dimensional regularization the spacetime is continued to $4 - 2\varepsilon$ dimensions and the collinear divergences now appear as poles at $\varepsilon = 0$. The NLO $F_2$ structure function takes in this case the form
\begin{equation}
  \begin{split}
    F_2^{\rm NLO} = x\sum_q\eqsq\,\bigg\{ \mconv{\left[ 1 + \frac{\alphas}{2\piup}\left(-\frac{1}{\hat{\varepsilon}} + \log\left(\frac{Q^2}{\mu^2}\right)\right)\pqq + R_q \right]&}{f_q}{x} \\+\; \mconv{\left[ \frac{\alphas}{2\piup}\left(-\frac{1}{\hat{\varepsilon}} + \log\left(\frac{Q^2}{\mu^2}\right)\right)\pqg + R_g \right]&}{f_g}{x} \bigg\},
  \end{split}
\end{equation}
where $\mu^2$ is an arbitrary scale needed in the dimensional regularization in order to keep the coupling dimensionless, and $1/\hat{\varepsilon} = 1/\varepsilon - \gamma_{\rm E} + \log(4\piup)$ with $\gamma_{\rm E}$ being the Euler--Mascheroni constant. This suggests the use of
\begin{equation}
  f_i(x,Q_f^2) \coloneqq \sum_{j}\mconv{\left[ \delta_{ij} + \frac{\alphas}{2\piup}\left(-\frac{1}{\hat{\varepsilon}} + \log\left(\frac{Q_f^2}{\mu^2}\right)\right)\pij \right]}{f_j}{x} + \bigo{\alphas^2},
\end{equation}
defining the modified minimal subtraction $\msbar$ scheme~\cite{Bardeen:1978yd,Furmanski:1981cw}, which is also the scheme employed in this thesis. The structure function can be expressed as
\begin{equation}
    F_2^{\rm NLO} = x\sum_q e_q^2\,\bigg\{ \mconv{C_q^{\msbar}(Q^2/Q_f^2)}{f_q(Q_f^2)}{x} + \mconv{C_g^{\msbar}(Q^2/Q_f^2)}{f_g(Q_f^2)}{x} \bigg\},
\end{equation}
with the $\msbar$ coefficient functions $C_{q,g}^{\msbar}$ available e.g.\ in Ref.~\cite{Furmanski:1981cw}.
To be exact, in the global PDF analysis presented in this thesis, we take the DIS and other partonic cross sections to NLO accuracy, and evolve the PDFs according to DGLAP equations using NLO splitting functions~\cite{Furmanski:1980cm,Curci:1980uw}.

\section{Heavy-quark PDFs}
\label{sec:hq}

So far we have treated all partons as massless, but for heavy quarks, particularly charm and bottom, with their masses in the GeV range, this is not always justifiable~\cite{Thorne:2008xf}.
When the energy of the process is not high enough to produce heavy quarks, they should simply not contribute to the cross section.
Above the mass threshold, the heavy-quark production becomes possible, in DIS through the partonic processes like the one shown in \fig{ladder} (right).
Here, the heavy-quark mass $m_{\rm H}$ regulates the $\ktv$-integrals and thus the cross section remains finite. At very high scales $Q \gg m_{\rm H}$, however, the resulting logarithms $\log(Q^2/m_{\rm H}^2)$ become very large and their resummation into heavy-quark PDFs becomes necessary. How to interpolate between the extremes of high ($Q \gg m_{\rm H}$) and low ($Q \sim m_{\rm H}$) scales is, again, scheme dependent.

In the simplest \emph{zero-mass variable flavour number scheme} (ZM-VFNS), one treats the heavy quark above the threshold as a massless active parton, using the same massless coefficient functions as for the $\nf$ light quarks with now $\nf+1$ flavours participating in the DGLAP evolution. This, however, ignores the mass effects important at scales $Q \sim m_{\rm H}$. To take account of the mass effects, the simplest way is that of a \emph{fixed flavour number scheme} (FFNS), where one keeps the number of flavours in the evolution fixed to $\nf$ and uses massive coefficient functions for the heavy quark, but this approach loses its validity at the high scales $Q \gg m_{\rm H}$.
Combining the above two approaches with validity extended to all scales, in a \emph{general-mass variable flavour number scheme} (GM-VFNS), one switches to the $\nf+1$ evolution and uses the massive coefficient functions subtracted with terms that prevent double counting. These subtraction terms depend on which mass terms one chooses to include in the heavy-quark coefficient functions and thus there is not one, but many different GM-VFN schemes.

The scheme utilized in the articles of this thesis is that of the simplified Aivazis--Collins--Olness--Tung (SACOT)~\cite{Collins:1998rz,Kramer:2000hn}. In this scheme, one uses the $\msbar$ coefficient functions together with the heavy-quark PDFs, $C_{q_{\rm H}} = C_q^{\msbar}$. Below a transition scale $Q_t \sim m_{\rm H}$ a fixed-flavour prescription is used, with the $F_2$ structure function expressible as
\begin{align}
    F_2^{\rm NLO} \overset{Q^2 < Q_t^2}{=} x&\sum_{q_\ell} e_{q_\ell}^2\,\bigg\{ \mconv{C_q^{\msbar}(Q^2/Q_f^2)}{f_{q_\ell}(Q_f^2)}{x} + \mconv{C_g^{\msbar}(Q^2/Q_f^2)}{f_g(Q_f^2)}{x} \bigg\} \notag \\
    \qquad+ x&\sum_{q_{\rm H}} e_{q_{\rm H}}^2\,\mconv{C_{g \rightarrow q_{\rm H}}^{\rm FFNS}(m_{\rm H}^2/Q^2)}{f_g(Q_f^2)}{\chi},
\end{align}
where $\chi = x(1+4m_{\rm H}^2/Q^2)$ is the rescaling variable which accounts for the energy needed in heavy-quark pair production, and $\nf$ flavours are taken in the evolution. The fixed-flavour coefficient function $C_{g \rightarrow q_{\rm H}}^{\rm FFNS}$ can be found in Ref.~\cite{Gluck:1979aw}. Above the transition scale, the structure function can be writen as~\cite{Tung:2001mv}
\begin{align}
    F_2^{\rm NLO} \overset{Q^2 > Q_t^2}{=} x&\sum_{q_\ell} e_{q_\ell}^2\,\bigg\{ \mconv{C_q^{\msbar}(Q^2/Q_f^2)}{f_{q_\ell}(Q_f^2)}{x} + \mconv{C_g^{\msbar}(Q^2/Q_f^2)}{f_g(Q_f^2)}{x} \bigg\} \notag \\
    \qquad+ x&\sum_{q_{\rm H}} e_{q_{\rm H}}^2\,\bigg\{ \mconv{\bigg[C_{g \rightarrow q_{\rm H}}^{\rm FFNS}(m_{\rm H}^2/Q^2) - \frac{\alphas}{2\piup}\log(Q_f^2/m_{\rm H}^2)\pqg\bigg]}{f_g(Q_f^2)}{\chi} \notag \\&\qquad\qquad+\; \mconv{C_q^{\msbar}(Q^2/Q_f^2)}{f_{q_{\rm H}}(Q_f^2)}{\chi} \bigg\},
\end{align}
with now $\nf+1$ flavours in the scale evolution. In the above expressions, the sums should again be understood to go over both quarks and antiquarks, in the first sum for the light (massless) flavours and in the second for the heavy-quark flavour.

\section{Sum rules and symmetry relations}
\label{sec:sumrule}

Due to the conservation of flavour in QCD interactions, we have the following sum rules for the proton PDFs
\begin{equation}
  \int_0^1 \d{x}\,u^{p}_{\rm V}(x,Q^2) = 2, \qquad \int_0^1 \d{x}\,d^{p}_{\rm V}(x,Q^2) = 1,
\end{equation}
where the valence distributions are defined as $q_{\rm V} = q - \bar{q}$ and where we have introduced the shorthand $q(x,Q^2) = f_q(x,Q^2)$, $g(x,Q^2) = f_g(x,Q^2)$. Similarly, the conservation of momentum requires the sum rule
\begin{equation}
  \int_0^1 \d{x}\,x \sum_i f_i(x,Q^2) = 1.
\end{equation}

Due to the degeneracy in the $u$ and $d$ quark masses, QCD has an approximative symmetry called isospin. This symmetry relates the PDFs of the proton and neutron with
\begin{equation}
  u^{p} \iseq d^{n}, \qquad d^{p} \iseq u^{n}, \qquad \bar{u}^{p} \iseq \bar{d}^{n}, \qquad \bar{d}^{p} \iseq \bar{u}^{n}
  \label{eq:isospinrel}
\end{equation}
and $f_i^p \iseq f_i^n$ for $i \neq u, d$. One can also use charge conjugation (CC), which is an exact symmetry of QCD, to relate the PDFs of proton and antiproton,
\begin{equation}
  q^{p} \cceq \bar{q}^{\bar{p}}, \qquad \bar{q}^{p} \cceq q^{\bar{p}}, \qquad g^{p} \cceq g^{\bar{p}},
\end{equation}
or, using both symmetries, the PDFs of charged pions,
\begin{equation}
  \begin{split}
    u^{\pi^{+}} \iseq d^{\pi^{-}} \cceq \bar{d}^{\pi^{+}} \iseq \bar{u}^{\pi^{-}}, \qquad d^{\pi^{+}} \iseq u^{\pi^{-}} \cceq \bar{u}^{\pi^{+}} \iseq \bar{d}^{\pi^{-}},\\
    q^{\pi^{+}} \iseq q^{\pi^{-}} \cceq \bar{q}^{\pi^{+}} \iseq \bar{q}^{\pi^{-}}\ \text{for}\ q \neq u,d, \qquad g^{\pi^{-}} \cceq g^{\pi^{+}}.
  \end{split}
  \label{eq:pionsymrel}
\end{equation}
The isospin symmetry relations for nucleons, \eq{isospinrel}, are essential for the discussion in \chap{nuclearpdfs}, assumed by practically all nPDF analyses. We also need to employ the charged pion relations, \eq{pionsymrel}, when discussing the results of the article~\cite{Paakkinen:2016wxk} in \sec{piady}.

\section{Factorization in hadron--hadron collisions}

The same perturbative approach which we have discussed in previous sections in the case of DIS also applies to hadron--hadron collision processes. This is stated formally in the factorization theorem which says that, order by order in perturbation theory, the collinear logarithms arising in hard-process calculations can be resummed into scale dependent long-distance functions in such a way that the full cross section becomes finite~\cite{Collins:1989gx}. Importantly, the structure of collinear divergences is the same in DIS and hadron--hadron processes, leading to \emph{universality} of the PDFs.

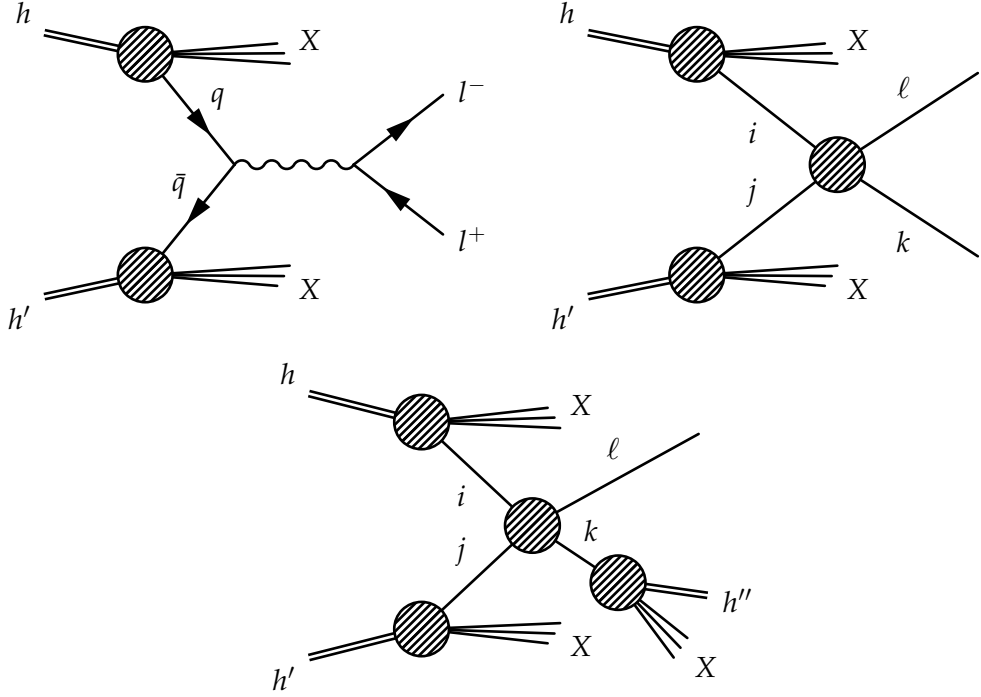
\begin{figure}
    \centering
    \small
    \bigskip
    \begin{fmffile}{diagrams/dy}
    	\begin{fmfgraph*}(60,35)
      	\fmfleft{i1,i2}
      	\fmfright{o1,o2,o3,o4,m1,m2,m3,m4,m5,o5,o6,o7,o8}
      	\fmf{double,tension=5}{i1,p1}
      	\fmf{double,tension=5}{i2,p2}
        \fmflabel{$h'$}{i1}
        \fmflabel{$h$}{i2}
      	\fmfblob{.12w}{p1}
      	\fmfblob{.12w}{p2}
      	\fmf{plain}{p1,x1}
      	\fmf{plain}{p1,x2}
      	\fmf{plain}{p1,x3}
      	\fmf{plain}{p2,x4}
      	\fmf{plain}{p2,x5}
      	\fmf{plain}{p2,x6}
      	\fmf{phantom}{x1,o1}
      	\fmf{phantom}{x2,o2}
      	\fmf{phantom}{x3,o3}
      	\fmf{phantom}{x4,o6}
      	\fmf{phantom}{x5,o7}
      	\fmf{phantom}{x6,o8}
      	\fmflabel{$X$}{x3}
      	\fmflabel{$X$}{x4}
      	\fmf{fermion,label=$q$}{p2,v1}
      	\fmf{fermion,label=$\bar{q}$}{v1,p1}
      	\fmf{fermion}{o4,v2}
      	\fmf{fermion}{v2,o5}
      	\fmflabel{$l^+$}{o4}
      	\fmflabel{$l^-$}{o5}
      	\fmf{photon,tension=1.5}{v1,v2}
    	\end{fmfgraph*}
    \end{fmffile}
    \qquad
    \begin{fmffile}{diagrams/dijet}
    	\begin{fmfgraph*}(60,35)
      	\fmfleft{i1,i2}
      	\fmfright{o1,o2,o3,o4,m1,m2,m3,m4,m5,o5,o6,o7,o8}
      	\fmf{double,tension=5}{i1,p1}
      	\fmf{double,tension=5}{i2,p2}
        \fmflabel{$h'$}{i1}
        \fmflabel{$h$}{i2}
      	\fmfblob{.12w}{p1}
      	\fmfblob{.12w}{p2}
      	\fmf{plain}{p1,x1}
      	\fmf{plain}{p1,x2}
      	\fmf{plain}{p1,x3}
      	\fmf{plain}{p2,x4}
      	\fmf{plain}{p2,x5}
      	\fmf{plain}{p2,x6}
      	\fmf{phantom}{x1,o1}
      	\fmf{phantom}{x2,o2}
      	\fmf{phantom}{x3,o3}
      	\fmf{phantom}{x4,o6}
      	\fmf{phantom}{x5,o7}
      	\fmf{phantom}{x6,o8}
      	\fmflabel{$X$}{x3}
      	\fmflabel{$X$}{x4}
      	\fmf{plain,label=$i$}{p2,v1}
      	\fmf{plain,label=$j$}{v1,p1}
      	\fmf{plain,label=$k$}{o3,v1}
      	\fmf{plain,label=$\ell$}{v1,o6}
      	\fmfblob{.12w}{v1}
    	\end{fmfgraph*}
    \end{fmffile}
    \\ \vspace{1.25cm}
    \begin{fmffile}{diagrams/inclhadr}
    	\begin{fmfgraph*}(60,35)
      	\fmfleft{i1,i2}
      	\fmfright{o1,o2,o3,o4,m1,m2,m3,m4,m5,o5,o6,o7,o8}
      	\fmf{double,tension=5}{i1,p1}
      	\fmf{double,tension=5}{i2,p2}
        \fmflabel{$h'$}{i1}
        \fmflabel{$h$}{i2}
      	\fmfblob{.12w}{p1}
      	\fmfblob{.12w}{p2}
      	\fmf{plain}{p1,x1}
      	\fmf{plain}{p1,x2}
      	\fmf{plain}{p1,x3}
      	\fmf{plain}{p2,x4}
      	\fmf{plain}{p2,x5}
      	\fmf{plain}{p2,x6}
      	\fmf{phantom}{x1,o1}
      	\fmf{phantom}{x2,o2}
      	\fmf{phantom}{x3,o3}
      	\fmf{phantom}{x4,o6}
      	\fmf{phantom}{x5,o7}
      	\fmf{phantom}{x6,o8}
      	\fmflabel{$X$}{x3}
      	\fmflabel{$X$}{x4}
      	\fmf{plain,label=$i$,tension=1.5}{p2,v1}
      	\fmf{plain,label=$j$,tension=1.5,label.side=right}{v1,p1}
      	\fmf{phantom}{o3,v1}
      	\fmf{plain,label=$\ell$}{v1,o6}
        \fmffreeze
      	\fmf{plain,label=$k$,tension=2.5}{v2,v1}
      	\fmf{double}{o4,v2}
      	\fmf{plain}{o1,v2}
      	\fmf{plain}{o2,v2}
      	\fmf{phantom}{o1,v3}
      	\fmf{phantom}{o2,v3}
      	\fmf{plain,tension=0}{v3,v2}
      	\fmflabel{$h''$}{o4}
      	\fmflabel{$X$}{v3}
      	\fmfblob{.12w}{v1}
      	\fmfblob{.12w}{v2}
    	\end{fmfgraph*}
    \end{fmffile}
    \bigskip
    \caption{Drell--Yan dilepton pair (upper left), dijet (upper right) and inclusive hadron (bottom) production in hadron--hadron collision at leading order of perturbation theory.}
    \label{fig:dydijetinclhadron}
\end{figure}

The relevant processes for this thesis are illustrated in \fig{dydijetinclhadron}. In the work presented in this thesis, various publicly available codes have been used in calculating them at the NLO level. In first of these processes, Drell--Yan (DY) dilepton production, $h + h' \rightarrow l^-l^+ + X$, the leading-order process happens through an annihilation of a quark and antiquark originating from the colliding hadrons $h$ and $h'$, as shown in \fig{dydijetinclhadron} (upper left). In more general terms, the cross section factorizes, schematically
\begin{equation}
  \sigma^{h + h'}_\text{DY} = \sum_{i,j=q,\bar{q},g}\, f_i^h \otimes f_j^{h'} \otimes \hat{\sigma}^{ij \rightarrow l^-l^+ + X},
\end{equation}
where there are now two PDFs, $f_i^h$ and $f_j^{h'}$, convoluted with the perturbatively calculable pieces. The production of massive electroweak (EW) gauge bosons proceeds in a similar way.
For practical applications, the MCFM program~\cite{Campbell:2015qma} was used in calculating the NLO pion--nucleus DY cross sections in the articles~\cite{Paakkinen:2016wxk} and~\cite{Eskola:2016oht}, and for the EW-boson cross sections in the article~\cite{Eskola:2016oht}.

It is also possible to consider the production of various hadronic final states, such as production of a dijet system, $h + h' \rightarrow \text{jet} + \text{jet} + X$. In this process at leading order, initial-state partons $i,j$ undergo a scattering into final-state partons $\ell, k$, which are observed as high-$p_{\rm T}$ hadronic jets in the detector, as shown in \fig{dydijetinclhadron} (upper right). At higher orders, this simple parton-to-jet correspondence is lost, and the jets are defined in terms of jet algorithms.
Formally still, the perturbative part of the cross section can be expressed in terms of a measurement function $F_\text{dijet}$ that defines the dijet,
\begin{equation}
  \sigma^{h + h'}_\text{dijet} = \sum_{i,j=q,\bar{q},g}\, f_i^h \otimes f_j^{h'} \otimes \hat{\sigma}^{ij}[F_\text{dijet}].
\end{equation}
For the calculation of dijet cross sections, the article~\cite{Eskola:2019dui} utilized the NLOJet++ code~\cite{Nagy:2003tz}, while the MEKS program~\cite{Gao:2012he} was used in the jet calculations of the articles~\cite{Eskola:2016oht} and~\cite{Helenius:2019lop}.

Instead of measuring jets, one can alternatively consider final states inclusive in a hadron species $h''$,
\begin{equation}
  \sigma^{h + h' \rightarrow h'' + X} = \sum_{i,j,k=q,\bar{q},g}\, f_i^h \otimes f_j^{h'} \otimes \hat{\sigma}^{ij \rightarrow k + X} \otimes D_k^{h''},
\end{equation}
illustrated in \fig{dydijetinclhadron} (bottom).
In such processes also the final state collinear logarithms need to be resummed, this time into \emph{fragmentation functions} $D_k^{h''}(z,Q^2)$, which give the probability for finding a final state hadron $h$, which has fragmented off from a hard parton $i$, carrying a fraction $z$ of the partons momentum. The inclusive pion-production cross sections considered in the article~\cite{Eskola:2016oht} were calculated with the INCNLO code~\cite{Aversa:1988vb}. The calculations for heavy-flavoured mesons are much more involved~\cite{Kniehl:2005mk} with various mass schemes again applicable, similarly to what was discussed in \sec{hq}. In the article~\cite{Eskola:2019bgf}, a recently developed variant of the SACOT scheme~\cite{Helenius:2018uul} was used with the zero-mass contributions obtained from the INCNLO~\cite{Aversa:1988vb} and the massive contributions from the MNR~\cite{Mangano:1991jk} codes.

\chapter{Global analysis and uncertainty estimation}
\label{chap:globalanalysis}

As discussed in the previous chapter, the PDFs describe long-range physics and cannot be calculated perturbatively from first principles. The common approach for obtaining them is then to use the means of statistical inference: By performing a ``global analysis'' on multiple observables sensitive to the PDFs, one aims to deduce the partonic structure from the measured hard-process data. This is in principle an infinite-dimensional optimization problem, as there is no a priori knowledge of the functional form. However, we do not have an infinite amount of perfectly precise data from which the PDFs could be obtained by inversion. For this reason, the PDFs need to be parametrized in a way or another, be it some suitably chosen functional form or a neural network~\cite{Forte:2002fg}.

\begin{figure}
  \centering
  \vspace{-0.1cm}
  \includegraphics[width=\textwidth]{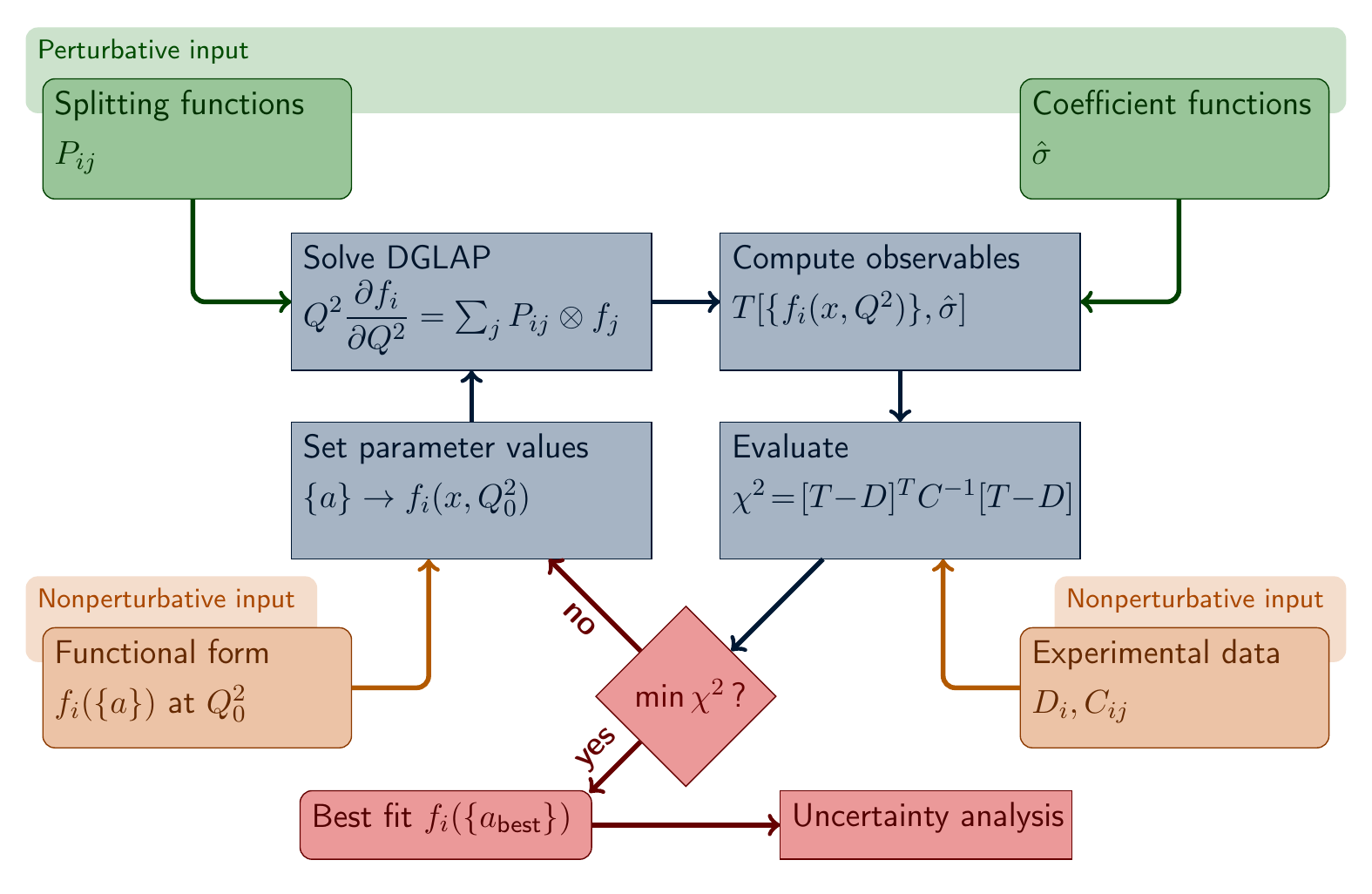}
  \caption{Flow of the $\chi^2$ minimization in a PDF global analysis. Figure from Ref.~\cite{Paakkinen:2018zbs}.}
  \label{fig:ga}
\end{figure}

Once the parametrization form is decided upon, one then needs to find the range of parameter values the data would support. For this, one defines a goodness-of-fit function $\chi^2$, the minimum of which corresponds to the best-fit values of the parameters. The various steps needed in the $\chi^2$ minimization are illustrated in \fig{ga}. One begins by setting a suitable first guess for the parameter values, which give the PDFs at a chosen parametrization scale $Q^2_0$. Using the DGLAP equations, these PDFs are then evolved to higher scales and convoluted with the coefficient functions to obtain theoretical predictions. To reduce the time required by the fitting, fast methods for performing these convolutions are needed~\cite{Carli:2010rw,Wobisch:2011ij,Bertone:2014zva}, such as the use of look-up tables as explained in the Section 3.3 of the article~\cite{Eskola:2016oht}. Comparing these predictions with the measured values, one calculates the $\chi^2$ figure-of-merit value for the chosen parameters. This procedure is then repeated for different sets of parameter values, until the minimum of $\chi^2$ is reached.

In addition to the functional form, the obtained result depends on various other inputs. The most obvious of these is which data one chooses to use. In principle, one would like to include as much data as possible to have the best constraining power, but care must be taken to only include measurements where one can trust the theoretical description of the process to avoid possible biases. For example, one should only include processes which are clearly in the perturbative regime to be able to neglect power corrections, but the exact value of minimum $Q^2$ to allow is somewhat arbitrary and different cuts are used by different groups, see~\tab{npdfs} for conventions in nPDF fits.

The results of minimization also depend on the level or perturbative accuracy in the used splitting and coefficient functions. It is hard to quantify the size of these theoretical uncertainties and they are usually neglected in reporting PDF errors, although work towards taking these uncertainties into account in global analyses is ongoing~\cite{Cacciari:2011ze,Harland-Lang:2018bxd,AbdulKhalek:2019bux}. Therefore, one usually only propagates the experimental uncertainties into the uncertainties of the PDFs and the subsequent predictions. \sec{hessianuncert} discusses how this is done in the Hessian formalism~\cite{Pumplin:2001ct} applied in this thesis work.

\section{Statistical basis of global analysis}
\label{sec:uncorruncert}

In this and the following section, we show how the $\chi^2$-function minimization arises as a maximum-likelihood estimator of the parameters. The viewpoint taken here is that of frequentist probability theory, for a Bayesian equivalent we refer the reader to Ref.~\cite{Kovarik:2019xvh}.

Due to experimental uncertainties, each measured value $D_i$ of any observable differs from its true value $T_i$ by some error $\delta_i$,
\begin{equation}
  D_i = T_i + \delta_i.
  \label{eq:dtaerr}
\end{equation}
Let us first assume that these errors are uncorrelated between the measurements, $\delta_i = \delta^\text{uncorr.}_i$ with $\mean{(\delta^\text{uncorr.}_i - \mean{\delta^\text{uncorr.}_i})(\delta^\text{uncorr.}_j - \mean{\delta^\text{uncorr.}_j})} = 0$ for $i \neq j$, and follow a Gaussian distribution with a zero mean, $\mean{\delta^\text{uncorr.}_i} = 0$, and a variance $\mean{(\delta^\text{uncorr.}_i - \mean{\delta^\text{uncorr.}_i})^2} = (\sigma^\text{uncorr.}_i)^2$. The probability density for each $\delta^\text{uncorr.}_i$ thus reads
\begin{equation}
  P(\delta^\text{uncorr.}_i) = \smallfrac{1}{\sqrt{2\piup}\sigma^\text{uncorr.}_i} \e^{-(\delta^\text{uncorr.}_i)^2/2(\sigma^\text{uncorr.}_i)^2}.
  \label{eq:uncerrprob}
\end{equation}
Since the errors are independent, the joint probability of a set of errors $\set{\delta^\text{uncorr.}_i}$ is simply
\begin{equation}
  P(\set{\delta^\text{uncorr.}_i}) = \prod_i P(\delta^\text{uncorr.}_i).
\end{equation}

By changing variables to $D_i$ according to \eq{dtaerr}, we can construct the joint probability for obtaining a set of $N_{\rm data}$ mutually independent measurements $\set{D_i}$ for given $\set{T_i}$,
\begin{equation}
  \begin{split}
    &P(\set{D_i}|\set{T_i}) = \prod_i \int\d{\delta^\text{uncorr.}_i} \delta(D_i - T_i - \delta^\text{uncorr.}_i) P(\delta^\text{uncorr.}_i)
    \\&\quad= \smallfrac{1}{(2\piup)^{N_{\rm data}/2}\prod_i\sigma^\text{uncorr.}_i} \e^{-\frac{1}{2}\sum_i(D_i - T_i)^2/(\sigma^\text{uncorr.}_i)^2}.
  \end{split}
\end{equation}
In PDF fits the true values $\set{T_i}$ are of course not known, but neglecting the theoretical uncertainties, one can trade these with the pQCD predictions with PDFs given by a set of parameters $\set{a}$, $T_i = T_i(\set{a})$. The \emph{likelihood} for a certain set of values of $\set{a}$ is then related to the probability of obtaining $\set{D_i}$ for given $\set{a}$ as
\begin{equation}
  L(\set{a}) \coloneqq P(\set{D_i}|\set{a}) = \smallfrac{1}{(2\piup)^{N_{\rm data}/2}\prod_i\sigma^\text{uncorr.}_i} \e^{-\frac{1}{2}\sum_i(D_i - T_i(\set{a}))^2/(\sigma^\text{uncorr.}_i)^2}.
\end{equation}
In the global fit, we wish to find the parameter values which maximize this likelihood function.

The parameter values $\set{a}$ which give the maximal likelihood also minimize the $\chi^2$ function
\begin{equation}
    \chi^2(\set{a}) \coloneqq -2\log L(\set{a}) + \text{const.} = \sum_i\left(\frac{D_i - T_i(\set{a})}{\sigma^\text{uncorr.}_i}\right)^2,
  \label{eq:uncorrchisq}
\end{equation}
which just shows that in the case of Gaussian errors, the maximum-likelihood and least-squares estimators are the same~\cite{Tanabashi:2018oca}.
Note that the $\chi^2$ function defined above essentially compares observed data fluctuations $D_i - T_i$ to expected ones $\sigma^\text{uncorr.}_i$ and in the limit of perfect theoretical description of the data we should obtain $\chi^2 \approx N_{\rm d.o.f.} = N_{\rm data} - N_{\rm par.}$, the number of degrees of freedom, where $N_{\rm par.}$ is the number of free parameters. Thus, on one hand, a value much higher than this would then tell that the fit does not describe the data well and, on the other hand, a significantly smaller value would be a signal of possible overfitting. In this sense, the $\chi^2$ is a goodness-of-fit function. A similar interpretation cannot be given for the value of the likelihood function at its maximum due to the way it is normalized.

In deriving \eq{uncorrchisq} we have assumed that the errors have a Gaussian distribution. This is an assumption that we often make in lack of better knowledge. In fact, the measured quantities are often cross sections, which should not go negative, but with the Gaussian distribution, we are assuming a nonvanishing probability for the measured value to be less than zero. However, when uncertainties are small, any corrections to \eq{uncorrchisq} should be small and its usage perfectly valid.

\section{Fitting to data with correlated uncertainties}
\label{sec:corruncert}

Let us now discuss the treatment of data with correlated uncertainties. We take these to be additive, leaving the treatment of multiplicative uncertainties to \sec{multcorruncert}. The total measurement error can then be decomposed as
\begin{equation}
  \delta_i = \delta_i^\text{uncorr.} + \delta_i^\text{corr.},
  \label{eq:errsum}
\end{equation}
where $\delta_i^\text{uncorr.}$ is the uncorrelated error distributed according to \eq{uncerrprob} and $\delta_i^\text{corr.} = \sum_k \beta_i^k \lambda_k$ sums the errors from independent systematical sources $\lambda_k$. \secs{marginalization}{profiling} discuss two ways of treating the $\lambda_k$ in formulating the $\chi^2$ function, ``marginalization'' and ``profiling''. In the case of additive Gaussian uncertanties these methods give identical results~\cite{Stump:2001gu}.

\subsection{Covariance matrix from marginalization}
\label{sec:marginalization}

We take here the $\lambda_k$ to be Gaussian distributed random variables with zero mean and normalized such that
\begin{equation}
  P(\lambda_k) = \smallfrac{1}{\sqrt{2\piup}} \e^{-\lambda_k^2/2}.
  \label{eq:problambdas}
\end{equation}
This way, $\beta_i^k$ can be interpreted as the response of the $i$th data point on a one standard deviation shift in the $k$th experimental systematic source of error. While the $\delta_i^\text{corr.}$ defined this way are correlated amongst themselves, the $\lambda_k$ are taken to be independent and hence
\begin{equation}
  P(\set{\delta^\text{uncorr.}_i},\set{\lambda_k}) = \prod_i P(\delta^\text{uncorr.}_i) \prod_k P(\lambda_k).
\end{equation}
Again, we can trade the $\delta^\text{uncorr.}_i$ with $D_i$ using \eqs{dtaerr}{errsum} to obtain
\begin{align}
    &P(\set{D_i},\set{\lambda_k}|\set{a}) \label{eq:probdatalambdas}
    \\&\quad= \smallfrac{1}{(2\piup)^{N_{\rm data}/2}\prod_i\sigma^\text{uncorr.}_i(2\piup)^{N_{\rm syst.}/2}} \e^{-\frac{1}{2}\sum_i(D_i - T_i(\set{a}) - \sum_k \beta_i^k \lambda_k)^2/(\sigma^\text{uncorr.}_i)^2 -\frac{1}{2}\sum_k\lambda_k^2}, \notag
\end{align}
where $N_{\rm syst.}$ is the number of systematical sources.
We can integrate over the $\set{\lambda_k}$ in \eq{probdatalambdas} to get the \emph{marginal probability distribution} for the data points,
\begin{align}
    &P(\set{D_i}|\set{a}) = \int\prod_k\d{\lambda_k}P(\set{D_i},\set{\lambda_k}|\set{a}) \notag
    \\&\quad= \smallfrac{1}{(2\piup)^{N_{\rm data}/2}\prod_i\sigma^\text{uncorr.}_i} \e^{-\frac{1}{2}\sum_i(D_i - T_i)^2/(\sigma^\text{uncorr.}_i)^2} \label{eq:probdataint}
    \\&\quad\phantom{=}\quad\times\smallfrac{1}{(2\piup)^{N_{\rm syst.}/2}}  \int\prod_k\d{\lambda_k}\e^{-\frac{1}{2}\sum_{k,\ell}\lambda_k\overbrace{\scriptstyle{\left(\sum_i\frac{\beta_i^k\beta_i^\ell}{(\sigma^\text{uncorr.}_i)^2}+\delta^{k\ell}\right)}}^{\phantom{A^{k\ell}}\ \eqqcolon A^{k\ell}}\lambda_\ell + \sum_k\sum_i(D_i - T_i)\frac{\beta_i^k}{(\sigma^\text{uncorr.}_i)^2}\lambda_k}, \notag
\end{align}
where we dropped the explicit $\set{a}$ dependence of $T_i$ for simplicity. The matrix $A$, with components defined above, is symmetric and positive definite, whereby the Gaussian integral in \eq{probdataint} can be performed. This yields
\begin{equation}
  \begin{split}
    P(\set{D_i}|\set{a}) &= \smallfrac{1}{(2\piup)^{N_{\rm data}/2}\prod_i\sigma^\text{uncorr.}_i} \e^{-\frac{1}{2}\sum_i(D_i - T_i)^2/(\sigma^\text{uncorr.}_i)^2}
    \\&\phantom{=}\quad\times\smallfrac{1}{\sqrt{\det A}} \e^{-\frac{1}{2}\sum_{i,j}(D_i-T_i)\left[\sum_{k,\ell}\frac{\beta_i^k}{(\sigma^\text{uncorr.}_i)^2}(A^{-1})^{k\ell}\frac{\beta_j^\ell}{(\sigma^\text{uncorr.}_j)^2}\right](D_j-T_j)}
  \end{split}
  \label{eq:probdata}
\end{equation}

The likelihood function is then defined similarly as with the uncorrelated uncertainties in \sec{uncorruncert},
\begin{equation}
  L(\set{a}) \coloneqq P(\set{D_i}|\set{a}) = \smallfrac{1}{(2\piup)^{N_{\rm data}/2}\prod_i\sigma^\text{uncorr.}_i}\smallfrac{1}{\sqrt{\det A}} \e^{-\frac{1}{2}\chi^2(\set{a})},
\end{equation}
where now
\begin{align}
    \chi^2(\set{a}) &= \sum_{i,j}(D_i-T_i)\!\bigg[\!\underbrace{\frac{\delta_{ij}}{(\sigma^\text{uncorr.}_i)^2} -\sum_{k,h}\frac{\beta_i^k}{(\sigma^\text{uncorr.}_i)^2}(A^{-1})^{k h}\frac{\beta_j^h}{(\sigma^\text{uncorr.}_j)^2}}_{\phantom{C^{-1}_{ij}}\,\eqqcolon C^{-1}_{ij}}\!\bigg]\!(D_j-T_j) \notag
    \\&= [D - T]^T C^{-1} [D - T]. \label{eq:chi2covmat}
\end{align}
The matrix $C^{-1}$ defined above is simply the inverse of the covariance matrix of the data, which is given by
\begin{equation}
  C_{ij} = \mean{(D_i - \mean{D_i})(D_j - \mean{D_j})} = (\sigma^\text{uncorr.}_i)^2\delta_{ij} + \sum_k \beta_i^k\beta_j^k,
\end{equation}
as can be easily shown by taking the matrix product
\begin{equation}
  \sum_n C_{in} C^{-1}_{nj} \!=\! \delta_{ij} + \sum_k \frac{\beta_i^k\beta_j^k}{(\sigma^\text{uncorr.}_i)^2} - \sum_{h,\ell} \bigg[\underbrace{\beta_i^h + \sum_{n,k}\frac{\beta_i^k\beta_n^k\beta_n^h}{(\sigma^\text{uncorr.}_i)^2}}_{\phantom{\sum_k\beta_i^k A^{kh}}\,= \sum_k\beta_i^k A^{kh}}\bigg](A^{-1})^{ h\ell}\frac{\beta_j^\ell}{(\sigma^\text{uncorr.}_j)^2} \!=\! \delta_{ij}.
\end{equation}
\eq{chi2covmat} is the standard covariance-matrix formulation of the $\chi^2$ function. It reduces to the uncorrelated form \eq{uncorrchisq} in the limit where $\beta_i^k \rightarrow 0$ for all $i,k$.

\subsection{Nuisance parameter profiling}
\label{sec:profiling}

Another way to treat the correlated uncertainties is to take the systematic shifts $\lambda_k$ to be free parameters of our statistical model. As these are not parameters of primary interest, they are called ``nuisance parameters''.
Since parameters are not allowed to have probabilities in the frequentist approach that we have adopted, \eq{problambdas} does not apply directly here. Rather, we should understand each of the nuisance parameters to be constrained by some systematical statistic $\hat{\lambda}_k$ distributed by
\begin{equation}
  P(\hat{\lambda}_k|\lambda_k) = \smallfrac{1}{\sqrt{2\piup}} \e^{-(\hat{\lambda}_k - \lambda_k)^2/2},
\end{equation}
and having an experimental value $\hat{\lambda}_k = 0$.
The likelihood function for the full set of parameters then reads
\begin{equation}
  \begin{split}
    L(\set{a},\set{\lambda_k}) &\coloneqq P(\set{D_i},\set{\hat{\lambda}_k = 0}|\set{a},\set{\lambda_k}) \\&= \smallfrac{1}{(2\piup)^{(N_{\rm data}+N_{\rm syst.})/2}\prod_i\sigma^\text{uncorr.}_i} \e^{-\frac{1}{2}\chi^2(\set{a},\set{\lambda_k})},
  \end{split}
  \label{eq:likelihoodlambdas}
\end{equation}
where the $\chi^2$ function in this case is defined as
\begin{equation}
  \chi^2(\set{a},\set{\lambda_k}) = \sum_i\frac{(D_i - T_i(\set{a}) - \sum_k \beta_i^k \lambda_k)^2}{(\sigma^\text{uncorr.}_i)^2} + \sum_k\lambda_k^2.
  \label{eq:chi2lambdas}
\end{equation}
As \eq{chi2lambdas} is quadratic in $\lambda_k$ we can find the minimum analytically. Setting the first derivatives to zero,
\begin{equation}
  \left.\parderiv{\chi^2}{\lambda_k}\right|_{\set{\lambda_\ell}=\set{\lambda^\text{min}_\ell}}\! = -2\beta_i^k\sum_i\frac{D_i - T_i - \sum_\ell \beta_i^\ell \lambda^\text{min}_\ell}{(\sigma^\text{uncorr.}_i)^2} + 2\lambda^\text{min}_k = 0 \quad \text{for all $k$},
\end{equation}
we find
\begin{equation}
  \sum_\ell \bigg(\underbrace{\sum_i\frac{\beta_i^k\beta_i^\ell}{(\sigma^\text{uncorr.}_i)^2}+\delta^{k\ell}}_{\phantom{A^{k\ell}}\ = A^{k\ell}}\bigg)\lambda^\text{min}_\ell = \sum_i(D_i-T_i)\frac{\beta_i^k}{(\sigma^\text{uncorr.}_i)^2},
  \label{eq:alambdaprod}
\end{equation}
where the matrix $A$ is the same which we encountered in \eq{probdataint}. Performing a matrix multiplication with its inverse
to \eq{alambdaprod} gives
\begin{equation}
  \lambda_h^\text{min} = \sum_k(A^{-1})^{hk}\sum_\ell A^{k\ell}\lambda^\text{min}_\ell = \sum_i(D_i-T_i)\sum_k(A^{-1})^{hk}\frac{\beta_i^k}{(\sigma^\text{uncorr.}_i)^2}.
\end{equation}

The obtained values can be substituted back to \eq{likelihoodlambdas}, giving us a \emph{profile likelihood}, which is a function of $\set{a}$ only. At the minimum of \eq{chi2lambdas} we have
\begin{align}
    &\sum_i\frac{(D_i - T_i - \sum_k \beta_i^k \lambda^\text{min}_k)^2}{(\sigma^\text{uncorr.}_i)^2}
    \\&\quad= \sum_{i,j}(D_i-T_i)\bigg[\frac{\delta_{ij}}{(\sigma^\text{uncorr.}_i)^2} -2\overbrace{\sum_{k,h}\frac{\beta_i^k}{(\sigma^\text{uncorr.}_i)^2}(A^{-1})^{k h}\frac{\beta_j^h}{(\sigma^\text{uncorr.}_j)^2}}^{\phantom{B^{(1)}_{ij}}\ \eqqcolon B^{(1)}_{ij}} \notag
    \\&\quad\phantom{=}\qquad+\underbrace{\sum_{k,\ell,m,h}\frac{\beta_i^k}{(\sigma^\text{uncorr.}_i)^2}(A^{-1})^{k \ell}\sum_n\frac{\beta_n^\ell\beta_n^m}{(\sigma^\text{uncorr.}_n)^2}(A^{-1})^{m h}\frac{\beta_j^h}{(\sigma^\text{uncorr.}_j)^2}}_{\phantom{B^{(2)}_{ij}}\ \eqqcolon B^{(2)}_{ij}}\bigg](D_j-T_j) \notag
\end{align}
and
\begin{equation}
  \sum_k(\lambda^\text{min}_k)^2 = \sum_{i,j}(D_i-T_i)\bigg[\underbrace{\sum_{k,\ell,h}\frac{\beta_i^k}{(\sigma^\text{uncorr.}_i)^2}(A^{-1})^{k \ell}(A^{-1})^{\ell h}\frac{\beta_j^h}{(\sigma^\text{uncorr.}_j)^2}}_{\phantom{B^{(3)}_{ij}}\ \eqqcolon B^{(3)}_{ij}}\bigg](D_j-T_j).
\end{equation}
Here we notice that
\begin{align}
    B^{(2)}_{ij} + B^{(3)}_{ij} &= \!\sum_{k,\ell,m,h}\frac{\beta_i^k}{(\sigma^\text{uncorr.}_i)^2}(A^{-1})^{k \ell}\bigg(\!\underbrace{\sum_n\frac{\beta_n^\ell\beta_n^m}{(\sigma^\text{uncorr.}_n)^2}+\delta^{\ell m}}_{\phantom{A^{k\ell}}\ = A^{\ell m}}\!\bigg)(A^{-1})^{m h}\frac{\beta_j^h}{(\sigma^\text{uncorr.}_j)^2} \notag
    \\&= \sum_{k,h}\frac{\beta_i^k}{(\sigma^\text{uncorr.}_i)^2}(A^{-1})^{k h}\frac{\beta_j^h}{(\sigma^\text{uncorr.}_j)^2} = B^{(1)}_{ij}
\end{align}
and hence
\begin{align}
    \min_{\set{\lambda_k}}\chi^2 &= \sum_{i,j}(D_i-T_i)\!\bigg[\!\underbrace{\frac{\delta_{ij}}{(\sigma^\text{uncorr.}_i)^2} -\sum_{k,h}\frac{\beta_i^k}{(\sigma^\text{uncorr.}_i)^2}(A^{-1})^{k h}\frac{\beta_j^h}{(\sigma^\text{uncorr.}_j)^2}}_{\phantom{C^{-1}_{ij}}\,= C^{-1}_{ij}}\!\bigg]\!(D_j-T_j) \notag
    \\&= [D - T]^T C^{-1} [D - T].
\end{align}
This shows that the covariance-matrix and nuisance-parameter formulations of the $\chi^2$ function are equivalent and either one can be used to treat the correlated uncertainties.

The nuisance-parameter approach facilitates an easy way for a graphical data-to-theory comparison in situations where simply adding quadratically the correlated and uncorrelated uncertainties would exaggerate the uncertainties. By defining
\begin{equation}
  D_i^\text{shifted} \coloneqq D_i  - \sum_k \beta_i^k \lambda_k^\text{min},
  \label{eq:datashifted}
\end{equation}
we may write
\begin{equation}
  \min_{\set{\lambda_k}}\chi^2 = \sum_i\frac{(D_i^\text{shifted} - T_i)^2}{(\sigma^\text{uncorr.}_i)^2} + \sum_k(\lambda_k^\text{min})^2.
\end{equation}
That is, if we shift the data according to \eq{datashifted}, the remaining differences between data and theory should be from the uncorrelated uncertainties, point by point. This method was used for example in the article~\cite{Eskola:2019dui} for presenting the inclusive jet data.

\subsection{Normalization uncertainties}
\label{sec:multcorruncert}

Until now we have taken the considered uncertainties to be of additive nature, i.e.\ each of the errors simply adds on the difference between the measured and true value, irrespective of what these values are. However, some uncertainties are known to be multiplicative in the sense that their magnitudes depend on the measured (or true) value. Luminosity uncertainties are good examples of such: the errors they pose on the measured cross sections are proportional to the (expected) number of events. Experiments often give these uncertainties in terms of normalization uncertainties, where each measured data point is subject to a mutual, fully correlated, percentual uncertainty, but also more complicated situations are possible. These uncertainties need to be treated correctly to avoid possible biases, as we will discuss next.

\subsubsection{d'Agostini bias}

Since the normalization uncertainty is a property of the data, it might appear natural to take it into account in the $\chi^2$ by introducing a normalization factor $f_N$ multiplying the data points and assuming a Gaussian uncertainty $\sigma^\text{norm.}$ for it, and therefore write
\begin{equation}
  \chi^2(\set{a},f_N) = \sum_i\left(\frac{f_N D_i - T_i(\set{a})}{\sigma^\text{uncorr.}_i}\right)^2 + \left(\frac{f_N - 1}{\sigma^\text{norm.}}\right)^2,
  \label{eq:biasedchisq}
\end{equation}
as was done e.g.\ in Ref.~\cite{Stump:2001gu} and also in the article~\cite{Eskola:2016oht} of this thesis. However, it can be shown that this formulation is subjective to so-called d'Agostini bias~\cite{DAgostini:1993arp}.

Following the example given in Ref.~\cite{DAgostini:1993arp}, let us assume that we have taken $N_{\rm data}$ measurements $\set{D_i}$ of a single observable quantity and that these data points share a common normalization uncertainty $\sigma^\text{norm.}$. We would then like to find the best estimate for the true value $T$ from which the measured values derive. For simplicity, let us also assume that the data points all have identical uncorrelated statistical uncertainties with variances $(\sigma^\text{uncorr.})^2$. The $\chi^2$ function of \eq{biasedchisq} then becomes
\begin{equation}
  \chi^2(T,f_N) = \frac{1}{(\sigma^\text{uncorr.})^2}\sum_i(f_N D_i - T)^2 + \left(\frac{f_N - 1}{\sigma^\text{norm.}}\right)^2.
  \label{eq:biasedsimplechisq}
\end{equation}
This is easily minimized with respect to both $T$ and $f_N$. We find
\begin{equation}
    T^{\rm min} = f_N^{\rm min} \hat{D}, \qquad
    f_N^{\rm min} = \frac{1}{1 + N_{\rm data}\frac{(\sigma^\text{norm.})^2}{(\sigma^\text{uncorr.})^2}\sigma_D^2},
\end{equation}
where
\begin{equation}
    \hat{D} = \frac{1}{N_{\rm data}} \sum_i D_i, \qquad
    \sigma_D^2 = \frac{1}{N_{\rm data}}\sum_i D_i^2 - \left(\frac{1}{N_{\rm data}}\sum_i D_i\right)^2
\end{equation}
are the sample mean and the sample variance of the data, respectively.
Now, as we have not introduced a statistical model, but taken the $\chi^2$ function as given, it is not clear how $\sigma_D^2$
is related to the uncorrelated error.
However, if we assume the \emph{true} normalization to be simply unity, one can then show that
\begin{equation}
  \mean{\sigma_D^2} = \frac{N_{\rm data} - 1}{N_{\rm data}} (\sigma^\text{uncorr.})^2
\end{equation}
and hence the expected value for the optimal normalization following from \eq{biasedsimplechisq} is
\begin{equation}
  \mean{f_N^{\rm min}} = \frac{1}{1 + (N_{\rm data} - 1)(\sigma^\text{norm.})^2}.
\end{equation}
This is clearly biased, as it tends towards zero as we increase the number of measurements. One can see why this happens also in a more general case by looking at \eq{biasedchisq}. By making both $f_N$ and $\set{T_i}$ smaller, also the difference in the numerator of the first term in \eq{biasedchisq} diminishes. As there is no similar compensation in the denominator, such a decrease in the normalization is favoured in the fit, whether that be truly statistically motivated or not. This can cause a significant bias in the found $\set{T_i}$ and thus also in the fitted parameters.

In real world PDF fits, such as in the article~\cite{Eskola:2016oht}, the bias is typically not as severe as in the above simple case. Here we assumed that the quantity of interest $T$ was completely free in the fit, but in a typical global fit the parameters are constrained by multiple independent data sets and limited also by the sum rules. Still, in article~\cite{Helenius:2019lop} of this thesis we encountered a case where this bias had an effect on the results and an unbiased method was called for.

\subsubsection{Unbiased fitting}

Let us assume, in a general setting, that each of the measured values $D_i$ deviates from the true value $T_i$ by a common normalization factor $f_N$ plus an individual, uncorrelated error $\delta_i^\text{uncorr.}$ such that
\begin{equation}
  D_i = f_N T_i + \delta_i^\text{uncorr.}
\end{equation}
and, treating $f_N$ as a nuisance parameter, the measured normalization deviates from the true one by
  $\hat{f}_N = f_N + \delta^\text{norm.}$.
Taking all uncertainties to be Gaussian distributed and independent, with \eq{uncerrprob} and
\begin{equation}
  P(\delta^\text{norm.}) = \smallfrac{1}{\sqrt{2\piup}\sigma^\text{norm.}} \e^{-(\delta^\text{norm.})^2/2(\sigma^\text{norm.})^2},
  \label{eq:normerrprob}
\end{equation}
and taking the experimental value $\hat{f}_N = 1$, we have
\begin{align}
    &P(\set{D_i},\hat{f}_N = 1|\set{T_i},f_N) \\&\quad= \smallfrac{1}{(2\piup)^{(N_{\rm data}+1)/2}\sigma^\text{norm.}\prod_i\sigma^\text{uncorr.}_i} \e^{-\frac{1}{2}\sum_i(D_i - f_N T_i)^2/(\sigma^\text{uncorr.}_i)^2 -\frac{1}{2}(f_N-1)^2/(\sigma^\text{norm.})^2}. \notag
\end{align}
In this case, the likelihood function takes the form
\begin{equation}
  L(\set{a},f_N) = \smallfrac{1}{(2\piup)^{(N_{\rm data}+1)/2}\sigma^\text{norm.}\prod_i\sigma^\text{uncorr.}_i} \e^{-\frac{1}{2}\chi^2(\set{a},f_N)},
\end{equation}
maximized at the minimum of
\begin{equation}
  \chi^2(\set{a},f_N) = \sum_i\left(\frac{D_i - f_N T_i(\set{a})}{\sigma^\text{uncorr.}_i}\right)^2 + \left(\frac{f_N - 1}{\sigma^\text{norm.}}\right)^2,
  \label{eq:unbiasedchisq}
\end{equation}
which only differs from \eq{biasedchisq} in so that $f_N$ multiplies the theory values, not the data. Note, on the contrary, that minimization of \eq{biasedchisq} does \emph{not} follow directly as a maximum-likelihood estimator from assuming $f_N D_i = T_i + \delta_i^\text{uncorr.}$ as in this case the likelihood function would have $f_N$ in its normalization.
Now, in the simple scenario discussed previously, one finds
\begin{equation}
    T^{\rm min} = \hat{D}, \qquad
    f_N^{\rm min} = 1,
\end{equation}
as should be the case when the data cannot provide additional information on the normalization. \eq{unbiasedchisq} is thus free from the d'Agostini bias. We note that there is also another way to treat the multiplicative uncertainties, called $t_0$ method, which is free also from a ``non-decoupling bias'', see Ref.~\cite{Ball:2009qv}.

\section{Uncertainty estimation in Hessian method}
\label{sec:hessianuncert}

In a global analysis, one aims at finding the best estimate for the PDFs based on available data and, importantly, determining the uncertainties in the results and communicating these in a way that allows to propagate the uncertainties into predictions made with the obtained PDFs. A common way to do this is the Hessian method~\cite{Pumplin:2001ct}. Having found the values $\set{a_i^\text{min}}$ which minimize the $\chi^2$, we can approximate the behaviour around the minimum by
\begin{equation}
  \chi^2(\set{a_i}) \approx \chi^2_0 + \sum_{i,j}\,(a_i - a_i^\text{min})\,H_{ij}\,(a_j - a_j^\text{min}), \label{eq:chisqexp}
\end{equation}
where $\chi^2_0 = \chi^2(\set{a_i^\text{min}})$ is the value at the minimum and $H_{ij} = \frac{1}{2} \partial^2 \chi^2 / \partial a_i \partial a_j|_{\!\set{a_i^\text{min}}}$ are the elements of the Hessian matrix, which is symmetric and must be positive definite, for otherwise we would not be at the minimum. Due to these properties, the Hessian matrix has a complete set of orthonormal eigenvectors $\vec{v}^{(k)}$ with positive eigenvalues $\varepsilon_k$,
\begin{gather}
  \sum_j\,H_{ij}\,v^{(k)}_j = \varepsilon_k\,v^{(k)}_i, \\
  \sum_i\,v^{(k)}_i\,v^{(\ell)}_i = \delta_{k\ell}, \qquad \sum_k\,v^{(k)}_i\,v^{(k)}_j = \delta_{ij}.
\end{gather}
By defining new parameters
\begin{equation}
  z_k = \sum_i \sqrt{\varepsilon_k}\,v^{(k)}_i\,(a_i - a_i^\text{min})
\end{equation}
the Hessian matrix can be diagonalized and the Equation~\eqref{eq:chisqexp} written as
\begin{equation}
  \chi^2 \approx \chi^2_0 + \sum_{k}\,z_k^2. \label{eq:chisqquadrappr}
\end{equation}

This facilitates an easy way to propagate the uncertainties. Let us assume that we associate each of the new parameters $z_k$ with an uncertainty $\Delta z_k$. Since the parameters in this basis are uncorrelated up to non-quardratic corrections, the related uncertainty in any quantity $y_i$ can be written in this approximation by the standard law of error propagation as
\begin{equation}
  (\Delta y_i)^2 = \sum_k \left(\frac{\partial y_i}{\partial z_k} \Delta z_k\right)^2.
  \label{eq:observerror}
\end{equation}
It then becomes a question of how large variations $\Delta z_k$ one should allow. These can be related in the quadratic approximation to a global tolerance $\Delta\chi^2$ in the growth of the $\chi^2$ from its minimum simply as $\Delta z_k = \sqrt{\Delta\chi^2}$.
In presence of ideal Gaussian statistics one could further derive values of $\Delta\chi^2$ corresponding to exact confidence regions in the parameters~\cite{Tanabashi:2018oca}.
However, for non-quadratic $\chi^2$ functions
using such pre-determined $\Delta\chi^2$ values can give only approximate coverage of the true parameter values~\cite{Cousins:1994yw}. Using a $\Delta\chi^2$ larger than some idealized value has also been motivated by conflicts between data sets~\cite{Pumplin:2001ct} and parametrization uncertainties~\cite{Pumplin:2009bb}. In fact, it has become more common to obtain the error tolerances by requiring that all the data sets remain in agreement within some confidence criterion under variations in each of the parameter directions, either separately~\cite{Martin:2009iq}, or on average as in the article~\cite{Eskola:2016oht}. This method is described in detail in the article~\cite{Eskola:2016oht} and thus will not be discussed further here.

\begin{figure}
  \vspace{-0.1cm}
  \includegraphics[width=\textwidth]{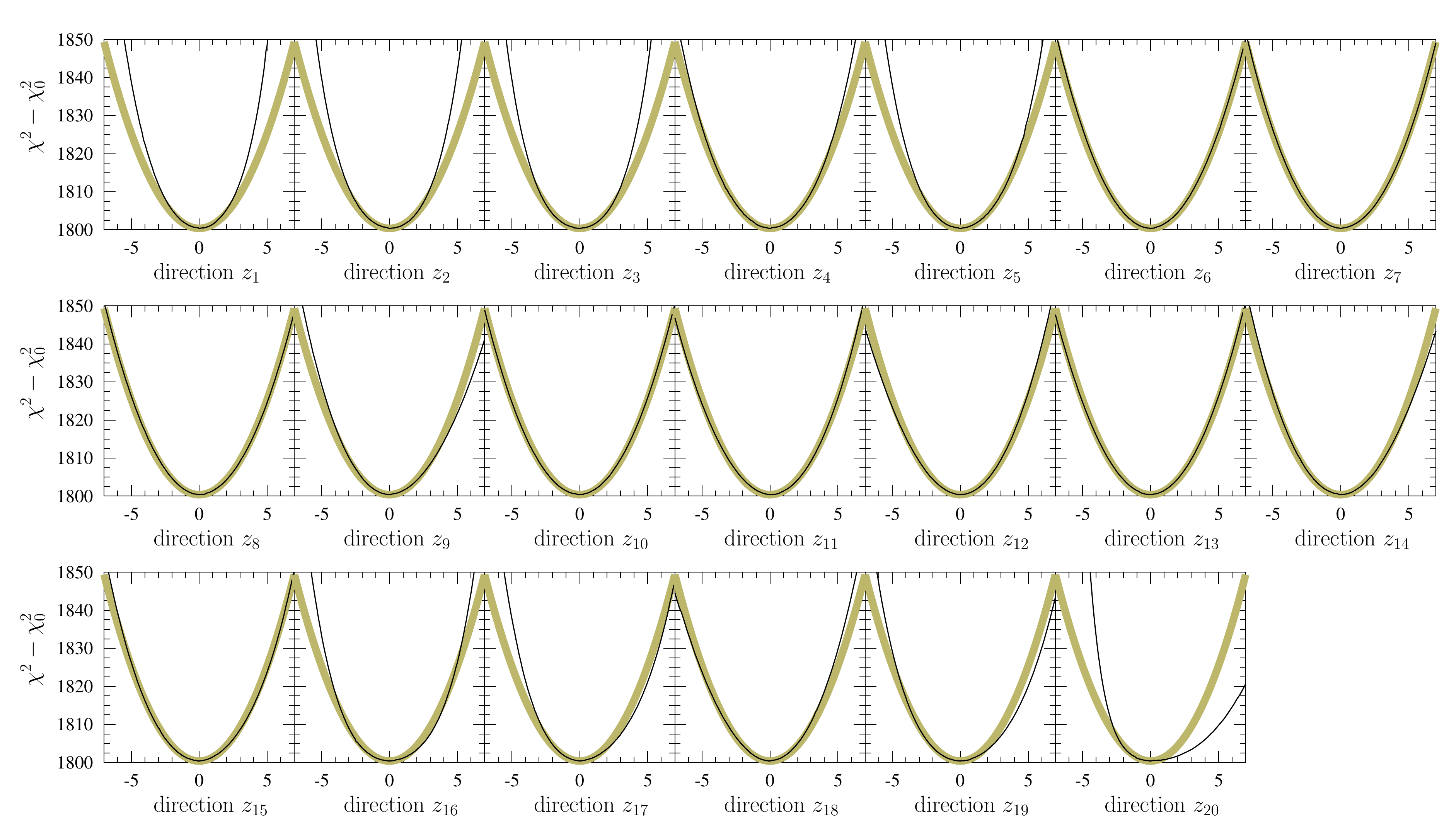}
  \caption{Growth of the $\chi^2$ function in each of the eigendirections of the EPPS16 analysis. Black lines show the true dependence, whereas yellow lines represent the ideal quadratic behaviour. Figure from article~\cite{Eskola:2016oht}.}
  \label{fig:quadratictest}
\end{figure}

\fig{quadratictest} shows the shape of the $\chi^2$ function around the minimum in the EPPS16 analysis~\cite{Eskola:2016oht}. The quadratic approximation is typically very good, with only few eigendirections showing clear cubic
or quartic
components. To take into account such deviations from the ideal behaviour, one defines $\Delta z_k = (\delta z^+_k - \delta z^-_k) / 2$, where $\delta z^\pm_k$ are the values of $z_k$ where $\chi^2$ has grown from its minimum by $\Delta\chi^2$. To simplify the expressions, it is useful to define PDF error sets $S^\pm_i$ obtained with parameter values
\begin{equation}
	z_k[S^\pm_i] =
  \delta_{ki}\,\delta z^\pm_i.
  \label{eq:errorsets}
\end{equation}
The derivative in \eq{observerror} can then be approximated with
\begin{equation}
	\frac{\partial y_i}{\partial z_k} = \frac{y_i[S^+_k] - y_i[S^-_k]}{2\,\Delta z_k}, \label{eq:derivapprox}
\end{equation}
whereby the errors in PDFs or related observables can be calculated simply by using
\begin{equation}
  (\Delta y_i)^2 = \frac{1}{4}\sum_k \left(y_i[S^+_k] - y_i[S^-_k]\right)^2.
	\label{eq:sympresc}
\end{equation}
It is also possible to extend this expression into an asymmetric error prescription~\cite{Lai:2010vv}
\begin{equation}
	(\delta y_i^\pm)^2 = \sum_k \left[ \substack{\max \\ \min} \left\{ y_i[S^+_k]-y_i[S_0],y_i[S^-_k]-y_i[S_0],0 \right\} \right]^2,
	\label{eq:asympresc}
\end{equation}
where $S_0$ is the central set with $z_k[S_0] = 0$ for all $k$.

\section{Hessian PDF reweighting}
\label{sec:hessianrw}

Using the Hessian uncertainty estimation, it is also possible to estimate the impact of a new data set on the PDFs~\cite{Paukkunen:2013grz,Paukkunen:2014zia,Schmidt:2018hvu,Eskola:2019dui}. Assume that
\begin{equation}
  \chi^2_\text{old}(\set{z_k}) \approx \chi^2_0 + \sum_{k}\,z_k^2 \label{eq:chisqoldquadrappr}
\end{equation}
is the $\chi^2$ function of a PDF global analysis. To add a new data set $\set{D_i^\text{new}}$ to the analysis, we can simply write
\begin{equation}
	\chi^2_\text{new} = \chi^2_\text{old} + \chi^2_\text{new\;data}, \label{eq:chisqnew}
\end{equation}
where
\begin{equation}
	\chi^2_\text{new\;data}(\set{z_k}) = \sum_{i,j}\,(T_i^\text{new}(\set{z_k}) - D_i^\text{new})\,C^{-1}_{ij}\,(T_j^\text{new}(\set{z_k}) - D_j^\text{new}). \label{eq:chisqnewdata}
\end{equation}

By using \eq{derivapprox}, where we take $\Delta z_k = \sqrt{\Delta\chi^2}$ in accordance with the quad\-ratic approximation in \eq{chisqoldquadrappr}, we can estimate the parameter dependence of \emph{any} PDF-dependent quantity with a linear function
\begin{equation}
	y_i(\set{z_k}) \approx y_i[S_0] + \sum_k \frac{\partial y_i}{\partial z_k} z_k, \qquad \frac{\partial y_i}{\partial z_k} = \frac{y_i[S_k^+] - y_i[S_k^-]}{2 \sqrt{\Delta\chi^2}}. \label{eq:ylinappr}
\end{equation}
Applying this approximation to $\set{T_i^\text{new}}$, we find that $\chi^2_\text{new}$ is a quadratic function of $\set{z_k}$ and can be minimized analytically. The new minimum is found at~\cite{Paukkunen:2014zia}
\begin{equation}
  z_k^{\rm min} = \sum_\ell {h}^{-1}_{k\ell}\left[\sum_{i,j}\frac{\partial T_i^\text{new}}{\partial z_\ell}\,C^{-1}_{ij}\,(D_j^\text{new} - T_j^\text{new}[S_0])\right],
\end{equation}
where
\begin{equation}
  \frac{\partial T_i^\text{new}}{\partial z_k} = \frac{T_i^\text{new}[S_k^+] - T_i^\text{new}[S_k^-]}{2 \sqrt{\Delta\chi^2}}
\end{equation}
and
\begin{equation}
  {h}_{k\ell} = \delta_{k\ell} + \sum_{i,j}\frac{\partial T_{i\phantom{j}}^\text{new}}{\partial z_k}\,C^{-1}_{ij}\,\frac{\partial T_j^\text{new}}{\partial z_\ell} \label{eq:newhessian}
\end{equation}
is the new Hessian matrix in
\begin{equation}
	\chi^2_\text{new}(\set{z_k}) \approx \chi^2_\text{new}(\set{z^\text{min}}) + \sum_{k\ell}\,(z_k - z_k^\text{min})\,{h}_{k\ell}\,(z_\ell - z_\ell^\text{min}).
\end{equation}

Now, updated central predictions for related quantities can be obtained simply by substituting the found $z_k^{\rm min}$ to the linear approximation in \eq{ylinappr}. For example, the new best-fit PDFs are a simple weighted sum of the original ones
\begin{equation}
  f_i[S_0^\text{new}] \approx f_i[S_0] + \sum_k \frac{z_k^{\rm min}}{2 \sqrt{\Delta\chi^2}} ( f_i[S_k^+] - f_i[S_k^-] ),
\end{equation}
that is, the PDFs are \emph{reweighted} in the process. Similarly, one can diagonalize the new Hessian matrix in \eq{newhessian} and find in these new eigendirections the parameter values corresponding to the tolerance $\Delta\chi^2$ to obtain the new error sets and then use \eq{ylinappr} to propagate the updated uncertainties into the observables.
It should be emphasized that the obtained results are only approximative of those of a full global fit, limited by the approximations made and also restricted by all the assumptions that were made in the original analysis, such as the functional forms assumed.

\subsubsection{Including higher-order terms}

\begin{figure}
  \centering
  \includegraphics[width=0.51\textwidth]{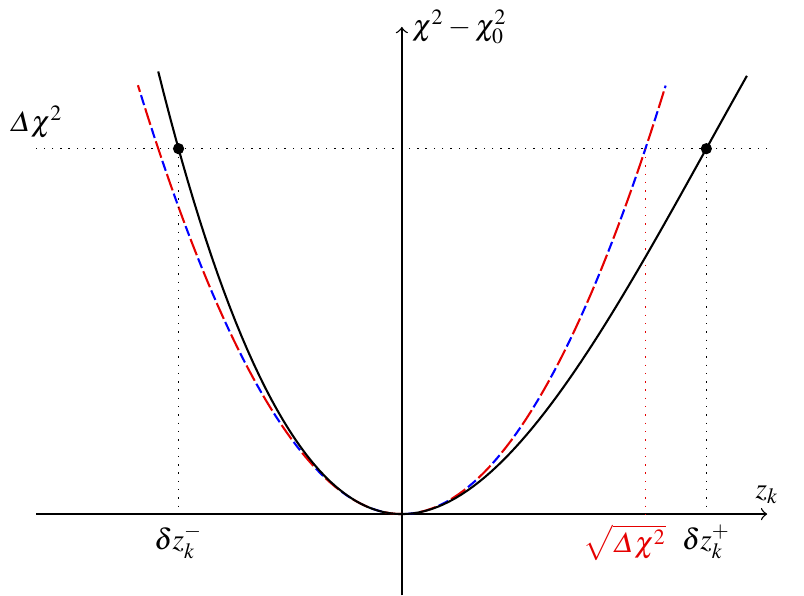}
  \hspace{-0.035\textwidth}
  \includegraphics[width=0.51\textwidth]{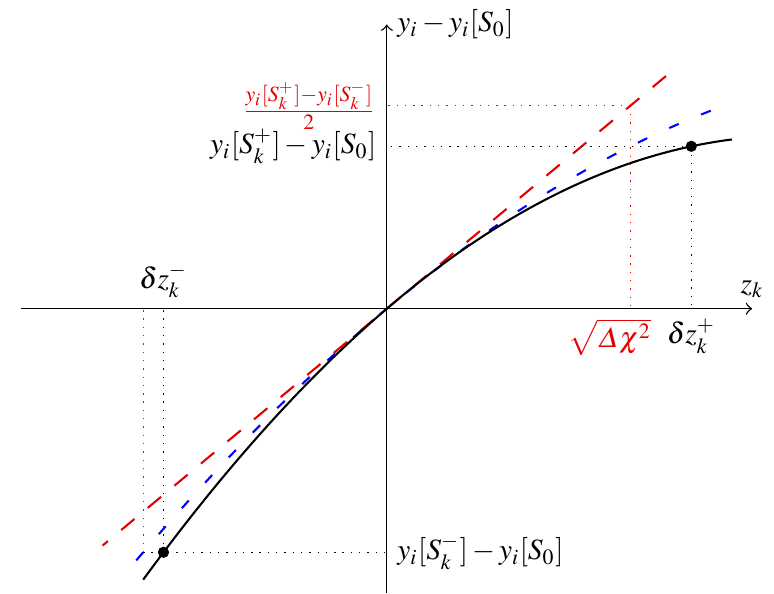}
  \caption{Different approximations of the $\chi^2$ function (left) and the PDF-dependent quantities (right) in the reweighting. Figure from article~\cite{Eskola:2019dui}.}
  \label{fig:rwapprox}
\end{figure}

As discussed at length in the article~\cite{Eskola:2019dui}, the Hessian reweighting with the quadratic approximation of $\chi^2_\text{old}$ and a linear approximation in the predictions $y_i$, shown as dashed red lines in \fig{rwapprox}, can be extended to include also higher-order terms.
Simply by using only the PDF central and error sets, one can extend \eq{ylinappr} to include also quadratic terms, shown with blue dashed lines in \fig{rwapprox}, as derived in Ref.~\cite{Paukkunen:2014zia}.
However, if additional information on the original fit is provided, one can also include cubic terms in the approximation of the original $\chi^2$ function,
\begin{equation}
	\chi^2_\text{old} \approx \chi^2_0 + \sum_{k} (a_k z_k^2 + b_k z_k^3), \label{eq:chisqnonquadrappr}
\end{equation}
with
\begin{equation}
	a_k = \frac{\Delta\chi^2}{\delta z^+_k - \delta z^-_k}\left(\frac{\delta z^+_k}{(\delta z^-_k)^2} - \frac{\delta z^-_k}{(\delta z^+_k)^2}\right), \quad
	b_k = \frac{\Delta\chi^2}{\delta z^+_k - \delta z^-_k}\left(\frac{1}{(\delta z^+_k)^2} - \frac{1}{(\delta z^-_k)^2}\right).
\end{equation}
where $\delta z^\pm_k$ are the parameter values determining the error sets in \eq{errorsets}. Then, approximating the PDF-dependent quantities with a quadratic function,
\begin{equation}
	y_i \approx y_i[S_0] + \sum_k (d_{ik} z_k + e_{ik} z_k^2), \label{eq:ynonquadrappr}
\end{equation}
the coefficients then read
\begin{equation}
  \begin{split}
  	d_{ik} &= \frac{1}{\delta z^+_k - \delta z^-_k}\bigg[-\frac{\delta z^-_k}{\delta z^+_k}\left(y_i[S_k^+] - y_i[S_0]\right) + \frac{\delta z^+_k}{\delta z^-_k}\left(y_i[S_k^-] - y_i[S_0]\right)\bigg], \\
  	e_{ik} &= \frac{1}{\delta z^+_k - \delta z^-_k}\bigg[\frac{1}{\delta z^+_k}\left(y_i[S_k^+] - y_i[S_0]\right) - \frac{1}{\delta z^-_k}\left(y_i[S_k^-] - y_i[S_0]\right)\bigg].
  \end{split}
\end{equation}
This approximation is shown as solid black lines in \fig{rwapprox}. These additions can help improve the accuracy of the method, especially in situations when uncertainties are large.
On a downside, in including these terms the simple quadratic form of $\chi^2_\text{new}$ is lost and the minimization needs to be done numerically.
\looseness=1

\chapter{Nuclear modifications of partonic structure}
\label{chap:nuclearpdfs}

As a first approximation, one could think of a nucleus as a loosely bound ensemble of nucleons. There is, however, ample experimental evidence that this simple picture is too crude to explain hard-scattering phenomena and that the partonic structure of the nucleons in nuclei is modified in a nontrivial way. Already from early DIS measurements on deuteron targets it was known that the \emph{Fermi motion} of the bound nucleons increases the probability of finding a parton with a large momentum fraction $x_N$ with respect to the average nucleon momentum. What came as a surprise in DIS experiments with heavy nuclei was that the quark distributions in bound nucleons are suppressed compared to those of a free proton for $0.3 \lesssim x_N \lesssim 0.8$. This phenomenon carries the name \emph{EMC effect} due to its first observation by the European Muon Collaboration (EMC)~\cite{Aubert:1983xm}. Later experiments also revealed an enhancement in the parton content at $0.03 \lesssim x_N \lesssim 0.3$ and a suppression again at $x_N \lesssim 0.03$, nowadays known as \emph{antishadowing} and \emph{shadowing}, respectively.

Over the years, a plethora of models to explain the nuclear effects have appeared, see Refs.~\cite{Arneodo:1992wf,Geesaman:1995yd,Piller:1999wx,Norton:2003cb,Armesto:2006ph} for reviews. The approach taken in nPDF analyses is, however, rather different. By parametrizing the nPDFs with suitably flexible functions and determining their parameters through a global analysis as described in \chap{globalanalysis}, one aims to get rid of any model dependence and to obtain a fully data-driven estimate of the nuclear modifications of parton distributions. From these, one can then make model-independent predictions for, e.g., production rates of hard probes of the Quark Gluon Plasma in ultrarelativistic heavy-ion collisions~\cite{Mangano:2004wq} or for ultra-high energy scattering cross sections in neutrino telescopes~\cite{Bertone:2018dse} and importantly also quantify the bias in free-proton PDFs caused by using nuclear data in their fits~\cite{Ball:2018twp}.

The PDFs of different nuclei are, in principle, independent quantities and should be determined from the data nucleus by nucleus, but the present data are far from sufficient to do so reliably for any single nucleus other than perhaps lead. Therefore, the mass-number dependence is parametrized in the nPDF fits.
It is conventional to decompose the PDFs of an average nucleon $f_i^A$ in a nucleus with a mass number $A$ and an atomic number $Z$ as
\begin{equation}
  f_i^A(x_N,Q^2) = \frac{Z}{A} f_i^{p/A}(x_N,Q^2) + \frac{A-Z}{A} f_i^{n/A}(x_N,Q^2),
\end{equation}
where $f_i^{p/A}$ is the PDF of a proton bound in a nucleus and $f_i^{n/A}$ the PDF of a bound neutron, with the latter obtained from the first by the approximative isospin symmetry according to \eq{isospinrel}. With this, one disentangles the isospin effects from other nuclear modifications.

\section{Nuclear PDF parametrizations}

\begin{figure}
  \centering
  \includegraphics[width=0.6\textwidth]{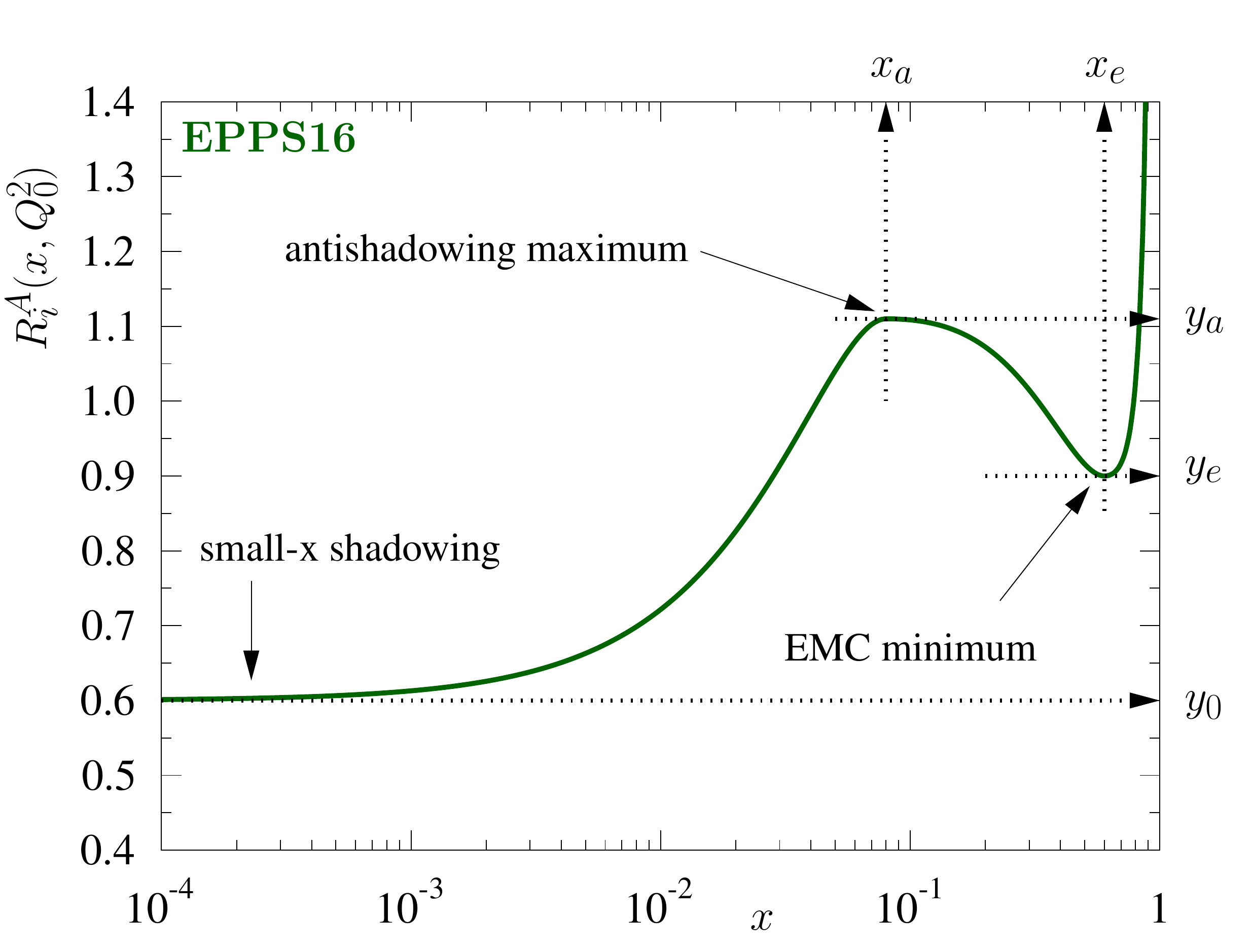}
  \caption{The functional form of the PDF nuclear modifications used in the EPPS16 analysis. Figure from article~\cite{Eskola:2016oht}.}
  \label{fig:fitform}
\end{figure}

By far the most common way to parametrize the nPDFs is through nuclear modification functions $R_i^A$, such that at the parametrization scale $Q^2_0$ the PDFs of a proton bound in a nucleus are defined as
\begin{equation}
  f_i^{p/A}(x_N,Q^2_0) = R_i^A(x_N,Q^2_0) f_i^p(x_N,Q^2_0),
\end{equation}
where $f_i^p$ are the PDFs of the free proton. This approach has been adopted by e.g.\ the EPS09~\cite{Eskola:2009uj}, DSSZ~\cite{deFlorian:2011fp}, KA15~\cite{Khanpour:2016pph} and EPPS16~\cite{Eskola:2016oht} analyses. An illustration of the $R_i^A$ parametrization of the EPPS16 analysis is given in \fig{fitform}. The functional form follows the pattern anticipated by the nuclear effects discussed in the beginning of the chapter, with free parameters controlling the amount of shadowing, antishadowing and EMC-effect and the location of the extremum of the latter two.

The nCTEQ15~\cite{Kovarik:2015cma} and nNNPDF1.0~\cite{AbdulKhalek:2019mzd} analyses have taken a different approach and parameterized the bound nucleon PDFs $f_i^{p/A}$ directly, in the case of nCTEQ15 by making the PDF fit function parameters an $A$-dependent and in nNNPDF1.0 by using a common neural network to parametrize all the nuclei with $A$ as an input to the network. Yet another approach was taken in the nDS analysis~\cite{deFlorian:2003qf}, where the nPDFs were defined as a convolution
\begin{equation}
  f_i^{p/A}(x_N,Q^2_0) = \int_{x_N}^A \frac{\d{y}}{y} W_i^A(y) f_i^p(x_N/y,Q^2_0)
\end{equation}
with suitably parametrized weight functions $W_i^A(y)$. Note that the integration range goes up to $A$ allowing $x_N > 1$. This is perfectly valid, as in the nuclear environment individual partons can borrow momentum from different nucleons. The parton distributions in this region are, however, expected to be very small and therefore, and in lack of constraining data, most of the analyses have simply assumed $f_i^{p/A}(x_N>1) = 0$. With this assumption, $f_i^{p/A}$ follow the same sum rules and evolution equations as the free-proton PDFs and the different nPDF parametrization approaches are practically equivalent.

\tab{npdfs} summarizes the most recent nPDF global analyses. We list here the perturbative order of the analysis, the included data types, the minimum scale at which data is included and the total number of data points. Further indicated are the number of free parameters, the approach in error analysis, the used free-proton PDFs and heavy-quark mass scheme and the amount of detail in flavour separation. We will discuss the similarities and differences of these analyses further in the following sections.

\afterpage{\begin{landscape}
  \begin{table}
    \centering
    \caption{nPDF releases from the past ten years. Table adapted from Ref.~\cite{Paukkunen:2017bbm}.}
    \small
    \newcolumntype{Y}{>{\centering\arraybackslash}X}
    \begin{tabularx}{207.5mm}{|c|Y|Y|Y|Y|Y|Y|}
      \hline
      & EPS09 & DSSZ & KA15 & nCTEQ15 & EPPS16 & nNNPDF1.0 \\
      \hline
      \hline
      Order in $\alpha_s$ & LO \& NLO & NLO & NNLO & NLO & NLO & NNLO \\
      $l$A/$l$d NC DIS & \checkmark & \checkmark & \checkmark & \checkmark & \checkmark & \checkmark \\
      pA/pd DY & \checkmark & \checkmark & \checkmark & \checkmark & \checkmark & \\
      RHIC dAu/pp $\pi$ & \checkmark & \checkmark & & \checkmark & \checkmark & \\
      $\nu$A DIS & & \checkmark & & & \checkmark & \\
      $\pi$A DY & & & & & \checkmark & \\
      LHC pPb W, Z & & & & & \checkmark & \\
      LHC pPb jets & & & & & \checkmark & \\
      & & & & & & \\
      $Q$ cut in DIS & 1.3 GeV & 1 GeV & 1 GeV & 2 GeV & 1.3 GeV & 1.87 GeV \\
      Data points & 929 & 1579 & 1479 & 708 & 1811 & 451 \\
      Free parameters & 15 & 25 & 16 & 16 & 20 & $\sim$183\phantom{$\sim$} \\
      Error analysis & Hessian & Hessian & Hessian & Hessian & Hessian & Monte Carlo \\
      Error tolerance $\Delta\chi^2$ & 50 & 30 & not given & 35 & 52 & not applicable \\
      Free proton PDFs & CTEQ6.1 & MSTW2008 & JR09 & CTEQ6M-like & CT14 & NNPDF3.1 \\
      HQ treatment & ZM-VFNS & GM-VFNS & ZM-VFNS & GM-VFNS & GM-VFNS & GM-VFNS \\
      Flavour separation & no & no & no & valence & full & no \\
      Reference & \cite{Eskola:2009uj} & \cite{deFlorian:2011fp} & \cite{Khanpour:2016pph} & \cite{Kovarik:2015cma} & \cite{Eskola:2016oht} & \cite{AbdulKhalek:2019mzd} \\
      \hline
    \end{tabularx}
    \label{tab:npdfs}
  \end{table}
\end{landscape}}

\section{Resolving flavour asymmetry}
\label{sec:flavourasym}

The bulk of the data used in the nPDF global analyses consists of electromagnetic neutral-current DIS measurements. At large $x_N$, where contributions from sea quarks can be neglected, the per-nucleon structure function $F_2^A$ at leading order reads
\begin{equation}
  F_2^A \approx \frac{5}{18} x_N \left[ \left(u_\mathrm{V}^{p/A} + d_\mathrm{V}^{p/A}\right) + \frac{3}{5} \left(\frac{2Z}{A} - 1\right) \left(u_\mathrm{V}^{p/A} - d_\mathrm{V}^{p/A}\right) \right].
  \label{eq:f2a}
\end{equation}
For isoscalar nuclei the $\frac{2Z}{A} - 1$ prefactoring the valence PDF difference is exactly zero and even for neutron-rich isotopes such as $^{208}{\rm Pb}$ it is approximately only $-0.2$. Hence $F_2^A$ at large $x_N$ is always predominantly sensitive to the sum of the valence quarks making it difficult to constrain the flavour separation. The same happens for the sea quarks at small $x_N$.

Moreover, the published structure functions are often \emph{isoscalarized}, i.e.\ reported in terms of
$F_2^{\text{isoscalar-}A} = \beta F_2^A$,
where the factor
\begin{equation}
  \beta = \frac{A}{2} \left(1+\frac{F_2^n}{F_2^p}\right)\bigg/\left(Z+(A-Z)\frac{F_2^n}{F_2^p}\right),
\end{equation}
with the ratio of neutron and proton structure functions $F_2^n/F_2^p$ suitably paramet\-rized, is applied to facilitate an easy comparison with the deuteron structure function, such that $F_2^{\text{isoscalar-}A}/F_2^{\rm D}$ would be unity if there were no nuclear modifications beyond isospin effects.
Unfortunately, this ``correction'' makes the extraction of flavour separation even more challenging.
For the above reasons, most of the nPDF analyses (cf. \tab{npdfs}) have made
simplifying assumptions to fix the flavour dependence of the valence quarks, and separately for the sea quarks. The first exception from this rule was the nCTEQ15 analysis, where the valence quarks were parametrized independently. However, in lack of constraining data, the uncertainties on the flavour separation remained large.

\subsection{Pion--nucleus Drell--Yan as a novel probe}
\label{sec:piady}

It was suggested in Ref.~\cite{Dutta:2010pg} that by studying pion-induced fixed-target DY data, one would get additional information on the flavour separation in the EMC region. In particular, the cross section ratios
\begin{equation}
R^{-}_{A_1/A_2}(x_2) = \frac{\frac{1}{A_1} \mathrm{d}\sigma^{\pi^- + A_1}_\text{DY} / \mathrm{d}x_2}{\frac{1}{A_2} \mathrm{d}\sigma^{\pi^- + A_2}_\text{DY} / \mathrm{d}x_2}, \qquad
    R^{+/-}_A(x_2) = \frac{\mathrm{d}\sigma^{\pi^+ + A}_\text{DY} / \mathrm{d}x_2}{\mathrm{d}\sigma^{\pi^- + A}_\text{DY} / \mathrm{d}x_2}
\end{equation}
were advocated. These are differential in $x_2 = \frac{M}{\sqrt{s}} \mathrm{e}^{-y}$, where $\sqrt{s}$ is the pion--nucleon center-of-mass energy, $M$ is the invariant mass of the lepton pair and $y$ its rapidity in the center-of-mass frame, probing at leading order the momentum fraction of the parton from the nucleus, $x_2 \approx x_N$. Employing the isospin and charge-conjugation relations of \eq{pionsymrel} and assuming that we are in a kinematic region where the pion sea quarks can be neglected, the leading order expressions for these ratios can be written as
\begin{equation}
R^{-}_{A_1/A_2}(x_2) \approx \frac{4u^{A_1}(x_2) + \bar{d}^{A_1}(x_2)}{4u^{A_2}(x_2) + \bar{d}^{A_2}(x_2)}, \qquad
  R^{+/-}_A(x_2) \approx \frac{4\bar{u}^A(x_2) + d^A(x_2)}{4u^A(x_2) + \bar{d}^A(x_2)}.
  \label{eq:piadyapprox}
\end{equation}
This shows that the remaining pion valence PDFs, which are not well known, cancel in these ratios, making them potential probes of the PDF nuclear modifications. Due to having valence anti-quarks in the pions, these ratios probe different flavour combinations than the DIS structure functions. In particular in the region $x_2 \gtrsim 0.1$, we have
\begin{equation}
  R^{-}_{A/\text{D}} \approx \frac{u_{\rm V}^{p/A} + d_{\rm V}^{p/A}}{u_{\rm V}^{p} + d_{\rm V}^{p}} + \left(\frac{2Z}{A} - 1\right)\frac{u_{\rm V}^{p/A} - d_{\rm V}^{p/A}}{u_{\rm V}^{p} + d_{\rm V}^{p}},
  \label{eq:piadyisospin}
\end{equation}
where we notice a factor $5/3$ increase in the sensitivity to the flavour separation compared to that in the $F_2^A$ in \eq{f2a}.

\begin{figure}[p]
  \centering
  \includegraphics[width=0.745\textwidth]{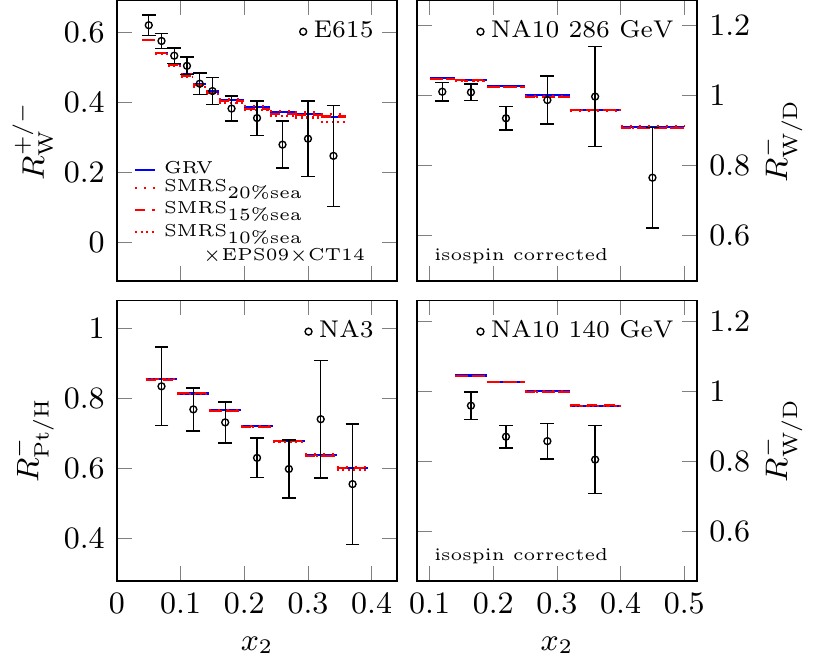}
  \caption{The cancellation of pion PDFs in pion-induced Drell--Yan cross-section ratios at next-to-leading order. Figure from article~\cite{Paakkinen:2016wxk}.}
  \label{fig:piadypipdf}
\end{figure}

Article~\cite{Paakkinen:2016wxk} discusses in detail the applicability and prospects of using the existing measurements of these observables in the nPDF global analysis. An important check was to make sure that the cancellation of pion degrees of freedom in the ratios would work also beyond leading order. This is shown in \fig{piadypipdf}, where NLO calculations performed with the public MCFM code~\cite{Campbell:2015qma} using pion PDFs from GRV~\cite{Gluck:1991ey} and SMRS~\cite{Sutton:1991ay} analyses are compared with measurements from the NA3~\cite{Badier:1981ci}, NA10~\cite{Bordalo:1987cs} and E615~\cite{Heinrich:1989cp} experiments. The NA10 data have been published with a similar isospin correction applied to them as discussed above and this had to be taken into account in the calculations. As can be seen from the figure, the cancellation of the pion PDFs is extremely good, and hence these ratios are insensitive to the rather poorly known pion structure.

\begin{figure}[p]
  \centering
  \includegraphics[width=0.745\textwidth]{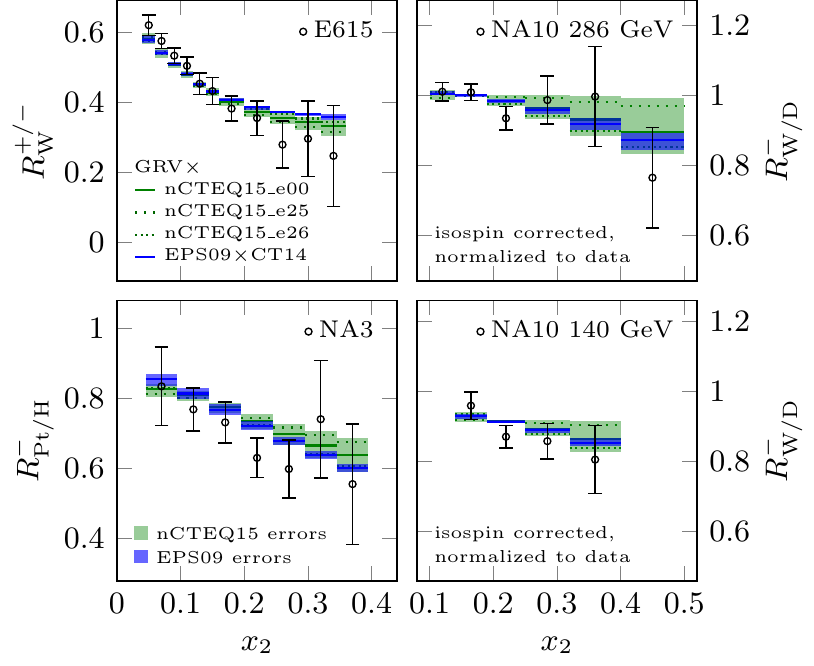}
  \caption{A comparison of measured pion-induced Drell--Yan cross-section ratios with predictions using the nCTEQ15 and EPS09 nuclear PDFs. Figure from article~\cite{Paakkinen:2016wxk}.}
  \label{fig:piadynpdf}
\end{figure}

\fig{piadynpdf} compares these data with NLO calculations using the nCTEQ15 and EPS09 nuclear PDFs. The NA10 data have a large normalization uncertainty, which was treated by normalizing predictions from each PDF set according to the optimal normalization found with \eq{biasedchisq} (with the PDF parameters kept fixed, there is no danger of d'Agostini bias in this case). The overall agreement between data and theory is rather good, which shows that these data can be used in a nPDF global fit.

Since valence flavour separation was allowed in the nCTEQ15 analysis, the related uncertainty bands in \fig{piadynpdf} are larger than those in EPS09. To study this in more detail, we show also the results with the nCTEQ15 error sets 25 and 26, which have the largest and smallest flavour asymmetry, respectively. The error set 25 of nCTEQ15 shows a flatter $x_2$ dependence than that in the NA3 and NA10 data, while predictions with nCTEQ15 error set 26 and EPS09 central set are in perfect agreement with the measurements. This hints towards similarity of valence-quark nuclear modifications, but as the experimental uncertainties are large, more stringent constraints are clearly needed.

\subsection{Global analysis with full flavour separation}
\label{sec:eppsflavoursep}

\begin{figure}
  \centering
  \includegraphics[width=0.92\textwidth]{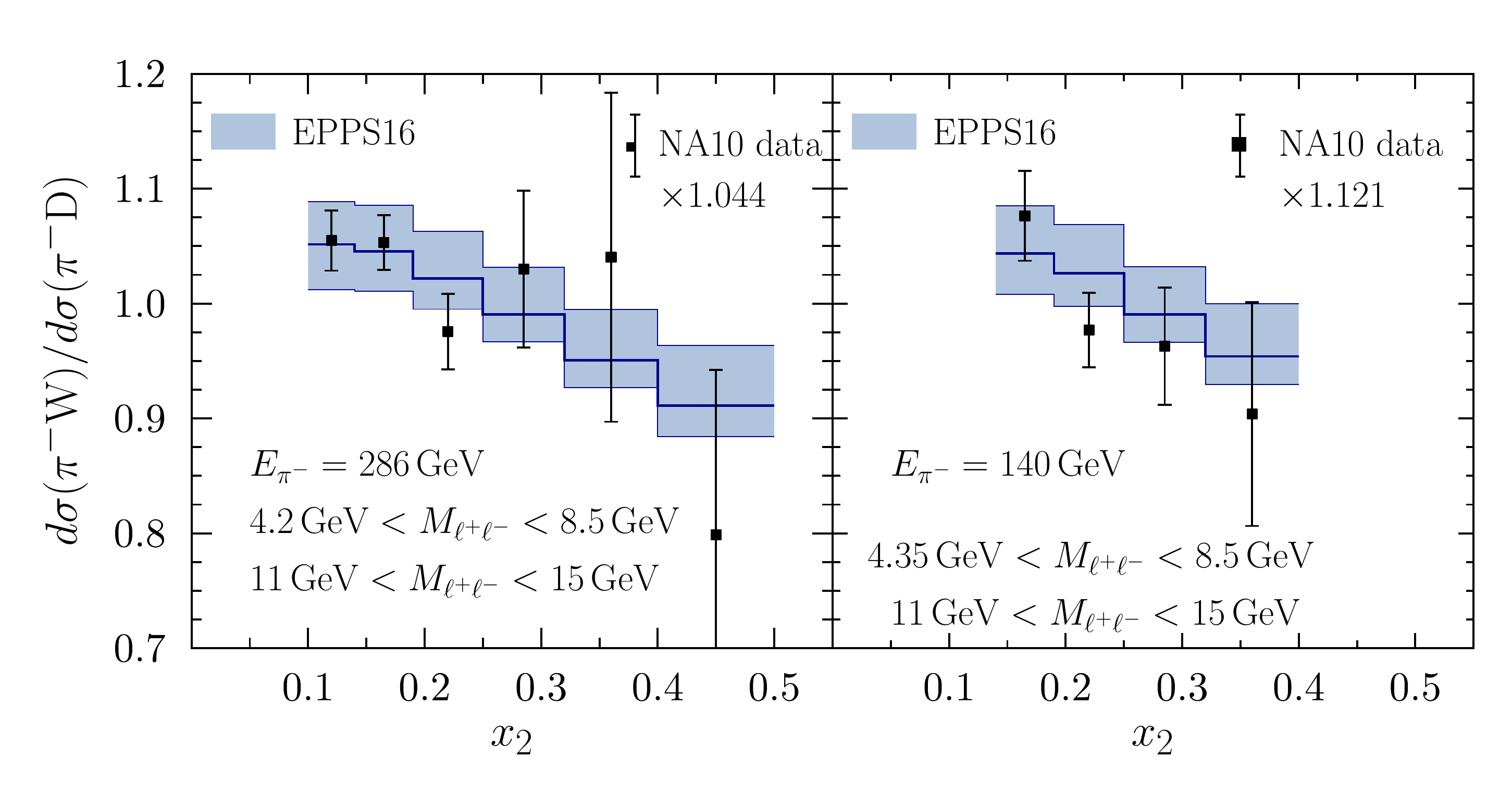}
  \caption{NA10 pion--tungsten DY data compared with the EPPS16 fit results. Figure from article~\cite{Eskola:2016oht}.}
  \label{fig:eppspi}
\end{figure}

In article~\cite{Eskola:2016oht}, we provided the first nPDF global analysis with full flavour separation, EPPS16, using the above-mentioned pion--nucleus DY data and other observables to constrain the valence and sea quark asymmetries. A very good fit to the pion DY data was found, shown as an example for NA10 data in \fig{eppspi}, but due to large data uncertainties the impact in the fit was somewhat limited.
Also new in this analysis, by using published isoscalar-correction factors $\beta$ of the charged-lepton DIS data, the non-isoscalarized ratios $F_2^A/F_2^{\rm D}$ were obtained from the ``corrected'' ones, gaining enhanced sensitivity to the flavour separation compared to a case where the fit would be simply considered to be done on isoscalar targets.

\begin{figure}
  \centering
  \includegraphics[width=0.49\textwidth]{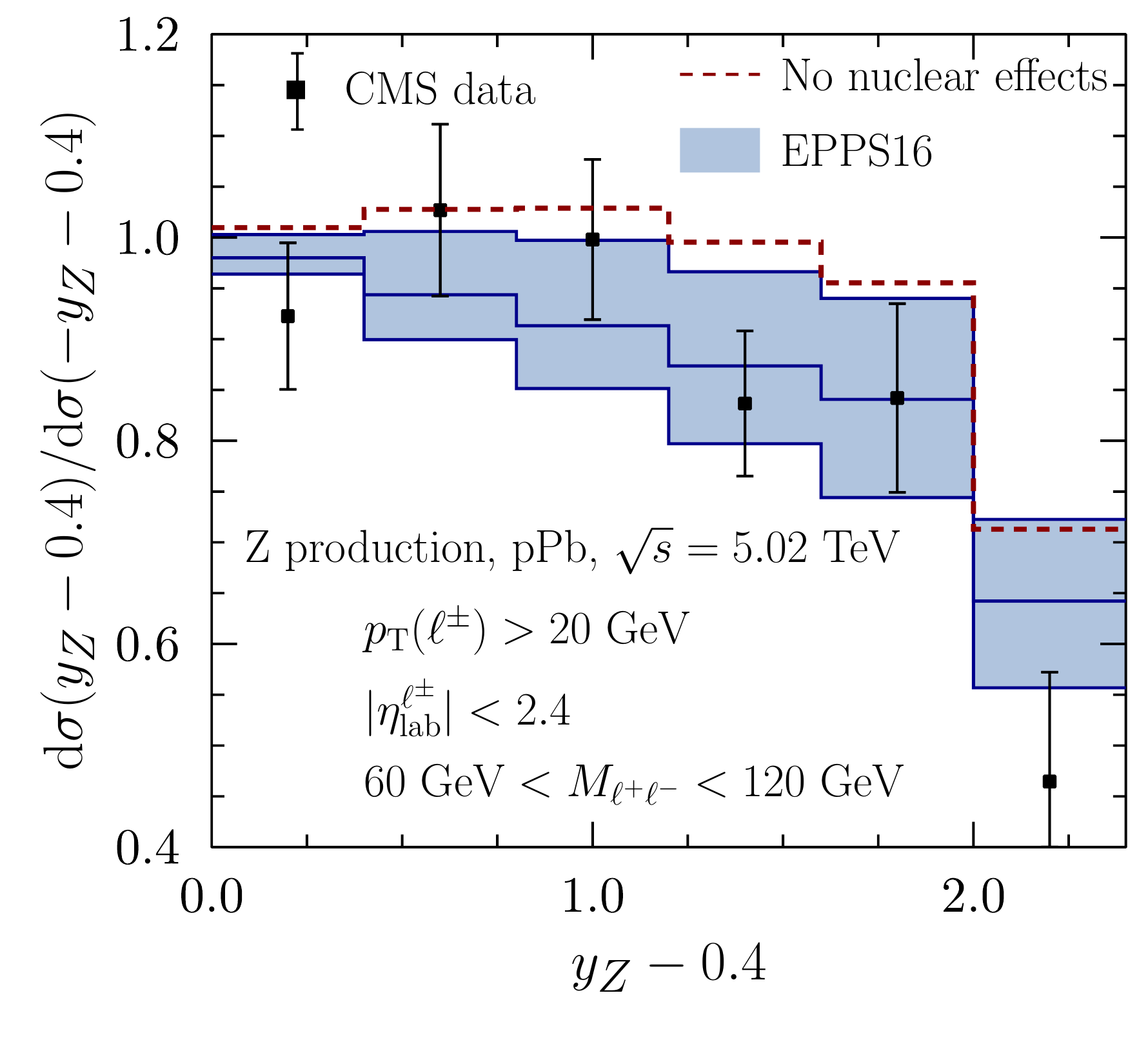}
  \includegraphics[width=0.49\textwidth]{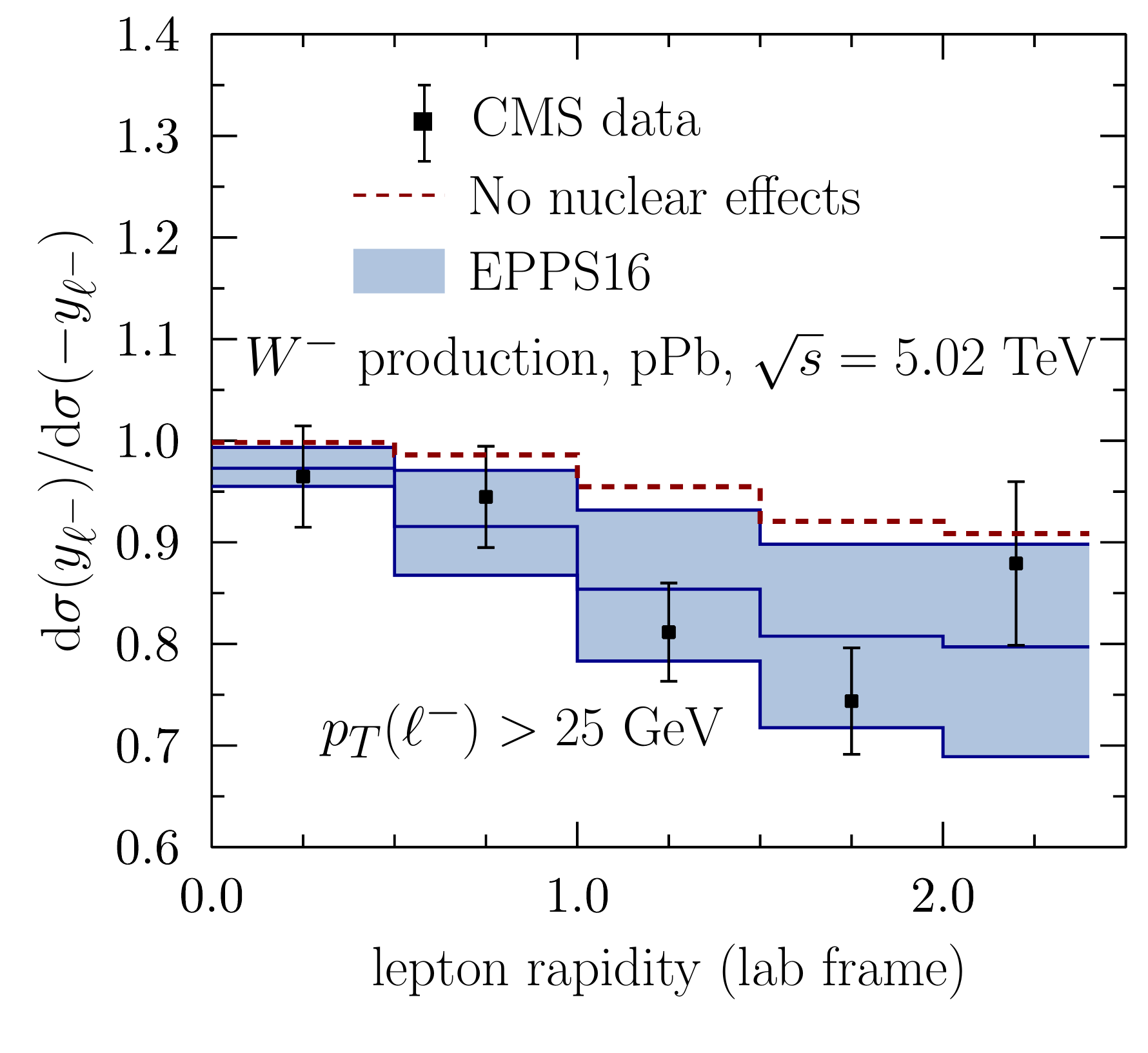}
  \caption{CMS proton--lead $Z$ and $W^-$ data compared with the EPPS16 fit results. Figure from article~\cite{Eskola:2016oht}.}
  \label{fig:ewbosons}
\end{figure}

The EPPS16 analysis was the first to use data from LHC proton--lead collisions, including W and Z production data from CMS and ATLAS experiments~\cite{Khachatryan:2015hha,Khachatryan:2015pzs,Aad:2015gta}. In principle, these observables are sensitive to different flavour combinations than the neutral-current DIS and DY data and could help constrain the flavour separation. However, as at the time no proton--proton baseline measurements were available, these data were added in the analysis as forward-to-backwards ratios, where the differential cross sections at forward rapidities $\d{\sigma}(y)$ are divided with those at backward rapidities $\d{\sigma}(-y)$ to reduce the uncertainties arising from the applied free-proton PDFs.
\fig{ewbosons} shows these observables for the $Z$ and $W^-$ production with a comparison of the CMS data and EPPS16 fit results. As can be seen from the figure, the agreement is very good and supports nuclear modifications of the PDFs, namely shadowing, in the region $x \lesssim 0.1$ probed by the data. However, because of low statistics, the data did not give strong constraints. Moreover, as these data probe the PDFs at high $Q^2$, any small differences at the parametrization scale are likely to be hidden under a large flavour-symmetric sea component generated through $g \rightarrow q\bar{q}$ splittings in the scale evolution, hindering the potential constraints for flavour separation. This, and also direct contribution from quark--gluon scattering at NLO make these data sensitive to also gluon nuclear modifications, discussed in more detail in the context of the EPPS16 analysis in \sec{eppsgluonconstr}.

\begin{figure}
  \centering
  \includegraphics[width=\textwidth]{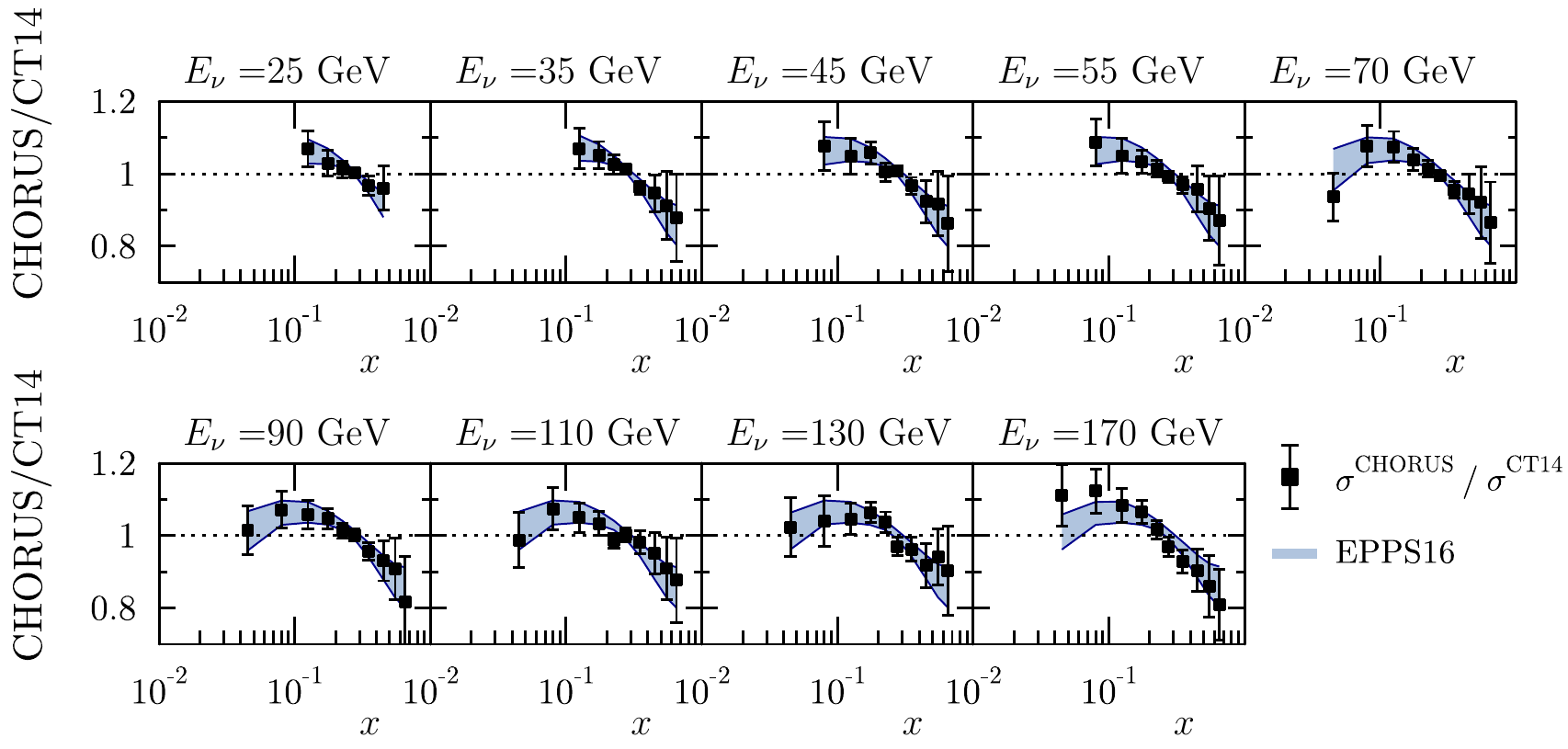}
  \caption{Normalized CHORUS neutrino DIS data compared with the results of the EPPS16 analysis. Figure from article~\cite{Eskola:2016oht}.}
  \label{fig:chorus}
\end{figure}

More stringent constraints were obtained from CHORUS neutrino and antineutrino DIS data~\cite{Onengut:2005kv}. These data were included already in the DSSZ analysis~\cite{deFlorian:2011fp}, but as no flavour separation was allowed in the fit, their constraining potential was not fully utilized. In EPPS16, these data were self-normalized at each beam-energy bin according to a procedure introduced in Ref.~\cite{Paukkunen:2013grz} to deal with data normalization uncertainty, propagating the data correlations, given in terms of the systematic shifts discussed in \sec{corruncert}, consistently to the normalized cross sections. \fig{chorus} shows the normalized neutrino-beam data, again in comparison with the EPPS16 results, divided with predictions with no nuclear effects in the PDFs to ease the interpretation. These data had a large impact in the fit, giving the $u$ and $d$ valence quark modifications a similar shape.

\begin{figure}
  \centering
  \vspace{-0.05cm}
  \includegraphics[width=0.5\textwidth]{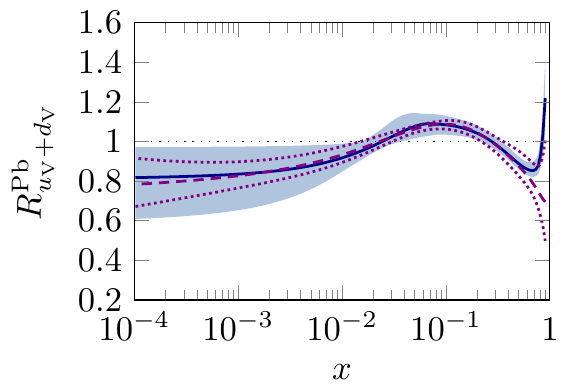}
  \hspace{-0.015\textwidth}
  \includegraphics[width=0.5\textwidth]{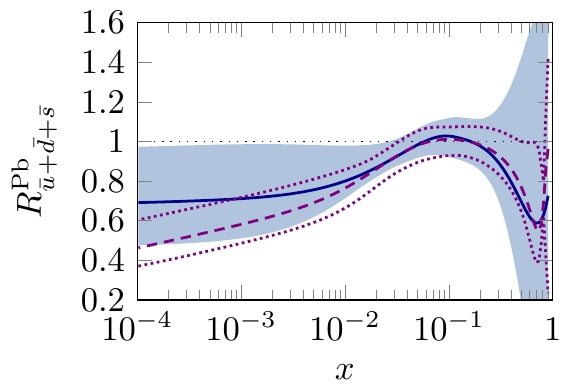}
  \\
  \vspace{-0.2cm}
  \includegraphics[width=0.5\textwidth]{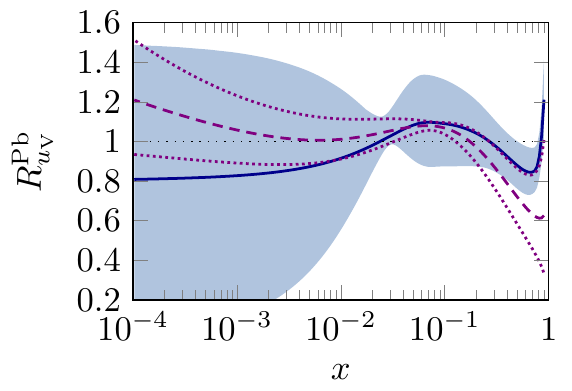}
  \hspace{-0.015\textwidth}
  \includegraphics[width=0.5\textwidth]{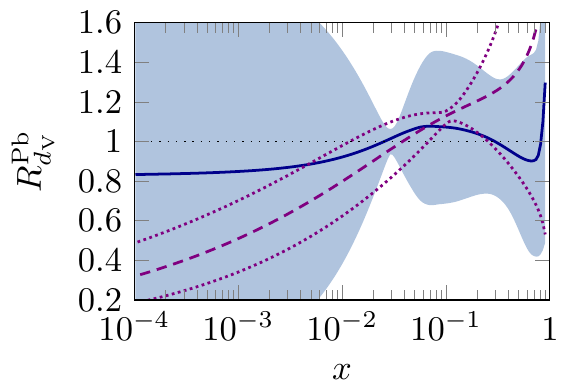}
  \\
  \vspace{-0.2cm}
  \includegraphics[width=0.5\textwidth]{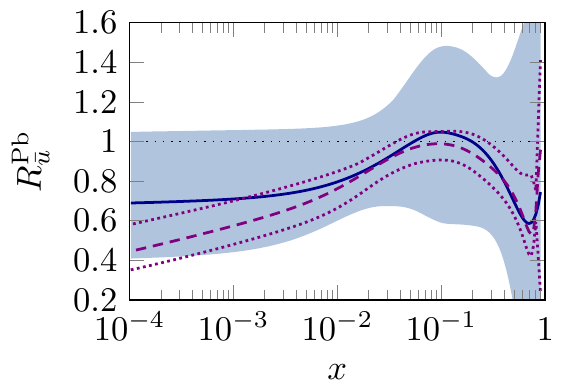}
  \hspace{-0.015\textwidth}
  \includegraphics[width=0.5\textwidth]{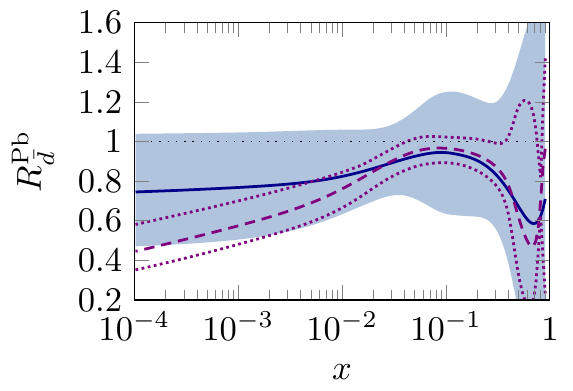}
  \\
  \vspace{-0.2cm}
  \includegraphics[width=0.5\textwidth]{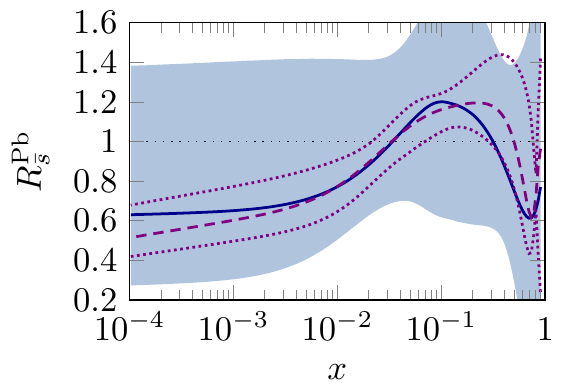}
  \hspace{-0.015\textwidth}
  \includegraphics[width=0.5\textwidth]{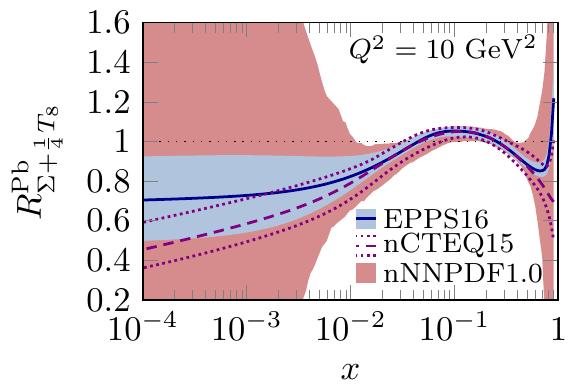}
  \caption{Comparison of quark nuclear modifications at the scale $Q^2 = 10\ {\rm GeV}^2$ in lead nucleus as found in the EPPS16, nCTEQ15 and nNNPDF1.0 analyses.}
  \label{fig:quarknmods}
\end{figure}

\fig{quarknmods} compares the quark nuclear modifications of the two analyses which allow flavour separation in the fits. The uppermost two panels show the average valence and light sea quark modifications in lead,
\begin{equation}
    R_{u_{\rm V}+d_{\rm V}}^{\rm Pb} = \frac{u^{p/{\rm Pb}}_{\rm V}+d^{p/{\rm Pb}}_{\rm V}}{u^{p}_{\rm V}+d^{p}_{\rm V}},
    \qquad
    R_{\bar{u}+\bar{d}+\bar{s}}^{\rm Pb} = \frac{\bar{u}^{p/{\rm Pb}}+\bar{d}^{p/{\rm Pb}}+\bar{s}^{p/{\rm Pb}}}{\bar{u}^{p}+\bar{d}^{p}+\bar{s}^{p}},
\end{equation}
at the scale $Q^2 = 10\ {\rm GeV}^2$. As should be expected, the EPPS16 and nCTEQ15 analyses are well constrained and agree nicely in the region $x \gtrsim 10^{-2}$ where data from fixed target DIS and DY are available. Within the uncertainties, we can clearly state that the valence quarks exhibit antishadowing and EMC effect and that shadowing for both valence and sea quarks seems to be preferred.

When we compare the modifications for individual valence quarks on the second row of \fig{quarknmods}, we find a large difference in the results. For EPPS16 the $u_{\rm V}$ and $d_{\rm V}$ are very similar, driven by the CHORUS data and also consistently with the pion--nucleus DY and CERN EW boson data, while for nCTEQ15, where no valence-quark constraints beyond neutral-current DIS and DY were included, the fit shows a large flavour asymmetry. For both EPPS16 and nCTEQ15 we find a narrow throat in the uncertainties, which is likely a fit-function artefact as in EPPS16 this happens at $x \approx 0.03$ where there are no data constraints from CHORUS. Even with the new constraints included, the EPPS16 uncertainties remain much larger for the individual flavours than for the average modification, simply reflecting the fact that the approximate isoscalarity of most nuclei makes it difficult to constrain the asymmetry. The same applies to the sea quarks, shown in the next three panels, where the nuclear modifications for all flavours are qualitatively similar and the EPPS16 and nCTEQ15 fits are in agreement. As flavour separation was allowed in EPPS16, the uncertainties of the individual flavours are larger than for the average sea-quark combination and also larger than in nCTEQ15 where the sea quarks are related to each other in a fixed way.

While the nNNPDF1.0 analysis uses only neutral-current DIS in the fit and is thus not yet in a fully global footing, the used methodology is somewhat different compared to other analyses, and hence it is interesting to compare EPPS16 and nCTEQ15 also with the results from this analysis.
The nNNPDF1.0 analysis uses Monte Carlo sampling of PDFs~\cite{Giele:1998gw,Giele:2001mr}, which allows for a more reliable uncertainty estimation than the Hessian method in regions poorly constrained by the data.
Since in lack of DY data no discrimination between valence and sea quarks was possible in the nNNPDF1.0 fit,
the only meaningfully constrained quark combination in their analysis, also bearing to the approximate isoscalarity of the targets, is, with the evolution basis variables $\Sigma = u^{p/A} + \bar{u}^{p/A} + d^{p/A} + \bar{d}^{p/A} + s^{p/A} + \bar{s}^{p/A}$ and $T_8 = u^{p/A} + \bar{u}^{p/A} + d^{p/A} + \bar{d}^{p/A} - 2(s^{p/A} + \bar{s}^{p/A})$ according to the nomenclature in Ref.~\cite{AbdulKhalek:2019mzd},
\begin{equation}
  \Sigma + \frac{1}{4}T_8 = \frac{5}{4}(u_{\rm V}^{p/A}+d_{\rm V}^{p/A}) + \frac{5}{2}(\bar{u}^{p/A}+\bar{d}^{p/A}) + \frac{1}{2}(s^{p/A} + \bar{s}^{p/A}).
\end{equation}
Hence, we compare nNNPDF1.0 with EPPS16 and nCTEQ15 only in this combination, shown in the bottom-right panel of \fig{quarknmods} for the lead nucleus. The three analyses agree nicely in the region constrained by the DIS data, but at small $x$ the nNNPDF1.0 uncertainties are vastly larger than those of EPPS16 and nCTEQ15. The EPPS16 and nCTEQ15 small-$x$ uncertainty bands should be understood as an extrapolation of those at higher $x$, through the assumed form of the fit functions motivated by low-$Q^2$ nuclear DIS data~\cite{Arneodo:1995cs} as well as requiring consistent $A$-systematics of nuclear effects. Studies with more flexible parametrizations within the EPPS16 framework, leading to similar inflation in small-$x$ uncertainties as seen with the nNNPDF1.0, can be found in Ref.~\cite{Aschenauer:2017oxs}.

\subsection{New observable for future pion--nucleus experiments}

\begin{figure}
  \centering
  \vspace{0.05cm}
  \includegraphics[width=0.9\textwidth]{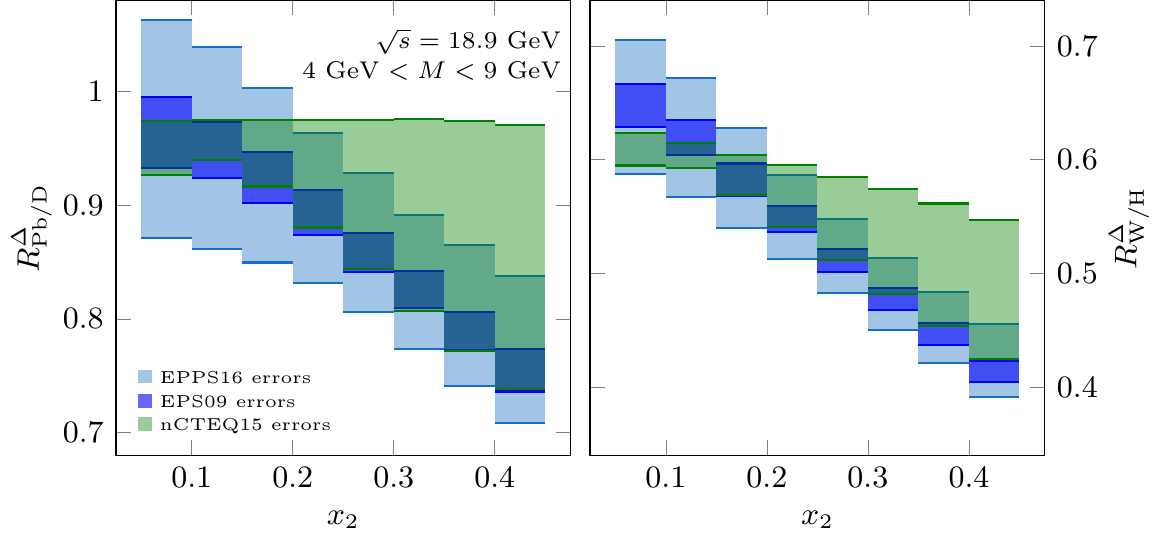}
  \caption{New pion--nucleus Drell--Yan observable for constraining the flavour asymmetry in valence quark nuclear modifications. Figure from Ref.~\cite{Paakkinen:2017eat}.}
  \label{fig:asympia}
\end{figure}

Even though the pion--nucleus DY data were not able to give stringent constraints in the EPPS16 fit, the increased sensitivity to the flavour separation makes these processes a potential probe in future experiments. To this end, we have proposed in Ref.~\cite{Paakkinen:2017eat} a new observable
\begin{equation}
  R^\Delta_{A_1/A_2}(x_2) = \frac{\frac{1}{A_1} (\mathrm{d}\sigma^{\pi^- + A_1}_\text{DY} / \mathrm{d}x_2 - \mathrm{d}\sigma^{\pi^+ + A_1}_\text{DY} / \mathrm{d}x_2)}{\frac{1}{A_2} (\mathrm{d}\sigma^{\pi^- + A_2}_\text{DY} / \mathrm{d}x_2 - \mathrm{d}\sigma^{\pi^+ + A_2}_\text{DY} / \mathrm{d}x_2)}.
\end{equation}
To leading-order accuracy, the contributions involving pion sea quarks cancel in the differences and we have
\begin{equation}
  R^\Delta_{A_1/\text{D}} \approx \frac{u_\mathrm{V}^{p/A} + d_\mathrm{V}^{p/A}}{u_\mathrm{V}^p + d_\mathrm{V}^p} + \frac{5}{3} \left(\frac{2Z}{A} - 1\right) \frac{u_\mathrm{V}^{p/A} - d_\mathrm{V}^{p/A}}{u_\mathrm{V}^p + d_\mathrm{V}^p}
\end{equation}
at \emph{all} $x_2$. Note that there is yet another factor $5/3$ increase in the sensitivity to the flavour separation compared to $R^{-}_{A/\text{D}}$ in \eq{piadyisospin}. \fig{asympia} shows this ratio for $\rm Pb/D$ and $\rm W/H$ computed in NLO. The errors calculated with EPS09, where no flavour separation was allowed, are rather small, whereas the EPPS16 and nCTEQ15 predictions have large uncertainties and somewhat different shape, showing that this observable could discern the differences in the nPDFs. There exist now plans to measure this ratio for $\rm W/C$ in a future experiment at the CERN M2 beam line~\cite{Denisov:2018unj}, with projections showing possible discriminating power over the nPDFs in the region $x \gtrsim 0.2$.

\section{New constraints for gluon nuclear modifications}
\label{sec:eppsgluonconstr}

Prior to EPPS16, direct gluon constraints were obtained only from RHIC inclusive pion-production data~\cite{Adler:2006wg,Abelev:2009hx} and indirect constraints mainly through the $Q^2$-dependence of DIS structure functions. The interpretation of the RHIC data is, however, not completely unambiguous. While the EPS08~\cite{Eskola:2008ca}, where these data were used for the first time, and the later EPS09 and nCTEQ15 analyses used these data under the assumption that the observed nuclear effects would only come from the nuclear modifications of the PDFs, the DSSZ analysis employed nuclear modified fragmentation functions. As a result, the DSSZ analysis finds very small gluon PDF nuclear modifications compared to the other analyses. New data were therefore needed to settle the issue.

\begin{figure}
  \centering
  \includegraphics[width=0.49\textwidth]{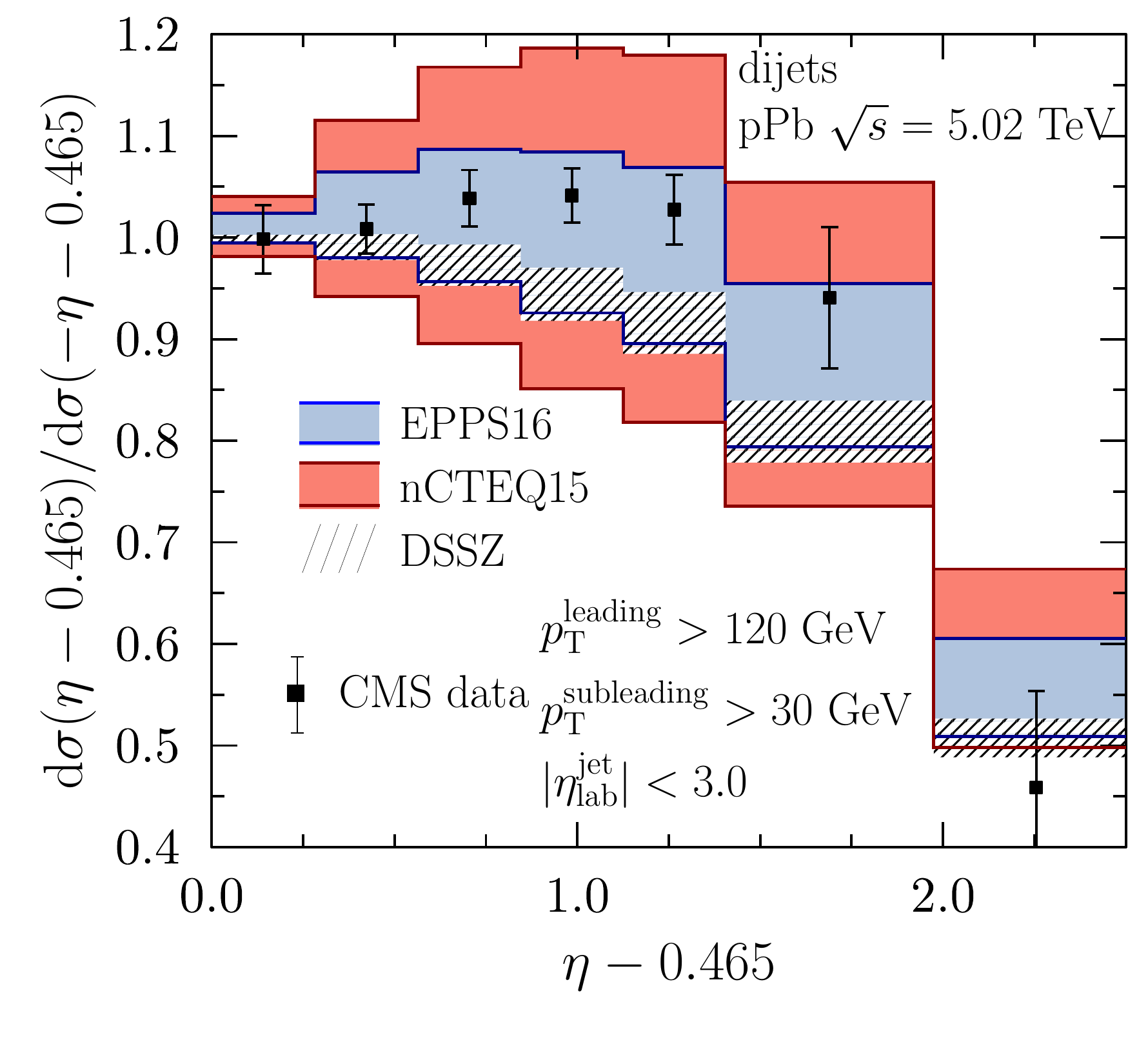}
  \includegraphics[width=0.49\textwidth]{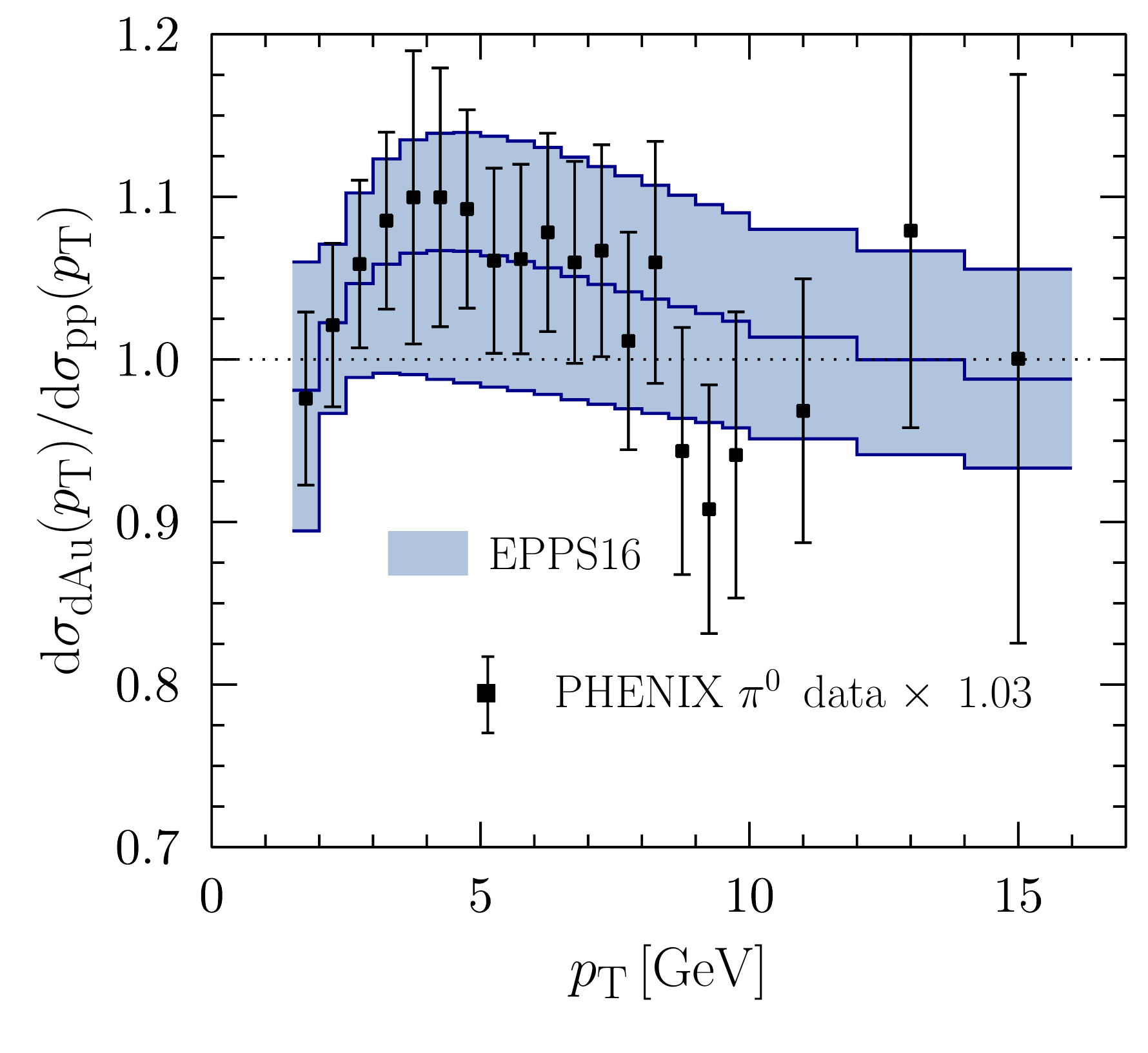}
  \caption{Left: Forward-to-backward ratio of the dijet production data measured by CMS compared with the EPPS16 fit results and predictions using the nCTEQ15 and DSSZ nPDFs. Right: PHENIX inclusive pion-production data and the EPPS16 fit. Figures from article~\cite{Eskola:2016oht}.}
  \label{fig:dijetrfb}
\end{figure}

In the EPPS16 analysis, described in the article~\cite{Eskola:2016oht} of this thesis, dijet data from CMS measurement at $5.02\ {\rm TeV}$ proton--lead collisions~\cite{Chatrchyan:2014hqa} were utilized for the first time. The EPPS16 fit results are compared with the data and predictions from nCTEQ15 and DSSZ in the left-hand-side panel of \fig{dijetrfb}. As is evident from the figure, the dijet data disagrees with DSSZ, whereas EPPS16 with nuclear modifications in the gluon PDF and no modifications on the fragmentation functions (with the KKP fragmentation functions~\cite{Kniehl:2000fe} used in the analysis) finds a good agreement with both the dijet data and the PHENIX pion data shown in the right-hand-side panel.

\begin{figure}
  \centering
  \includegraphics[width=0.5\textwidth]{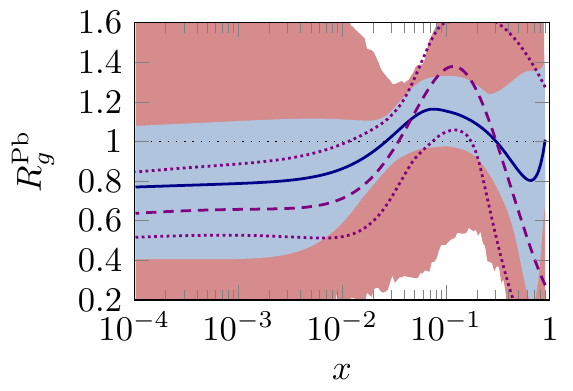}
  \caption{Comparison of the gluon nuclear modifications at the scale $Q^2 = 10\ {\rm GeV}^2$ in lead nucleus as found in the EPPS16, nCTEQ15 and nNNPDF1.0 analyses. Colours and line explanations are the same as in \fig{quarknmods}.}
  \label{fig:gluonmods}
\end{figure}

\fig{gluonmods} shows again the EPPS16, nCTEQ15 and nNNPDF1.0 nuclear mod\-i\-fi\-ca\-tions, now for comparison of the gluon PDFs. For the nNNPDF1.0 fit, where no direct gluon constraints were included, the uncertainties are large at all $x$ values. This emphasizes the importance of collider data in constraining the gluons. With the RHIC pion data included, the nCTEQ15 analysis was able to establish an antishadowing pattern, but the gluon modifications remained otherwise largely unconstrained, which also leads to the large uncertainties seen in \fig{dijetrfb}. In the EPPS16 analysis gluons are much better constrained due to the inclusion of the CMS dijet data. In particular, these data have a preference for an EMC-type slope for the gluon modification in lead. It should be also noted that the EW-boson data discussed in \sec{eppsflavoursep} are, to some extent, sensitive to the gluon PDF and seem to be consistent with the onsetting of small-$x$ gluon shadowing. At very small $x$, however, the EPPS16 uncertainties should again be understood as an extrapolation, fixed by the momentum sum rule and the assumed fit function form.

\subsection{Nuclear modification ratio of dijet spectra}
\label{sec:dijetnormalizedrpa}

In lack of a corresponding proton--proton baseline, the CMS dijet data were included in EPPS16 again as forward-to-backward ratios to reduce the sensitivity to the free-proton PDFs. Thus, again some information were lost and the full potential of a dijet measurement in proton--lead was not fully unleashed. The subsequent $5.02\ {\rm TeV}$ proton--proton data taking, allowing CMS to provide a measurement of the nuclear modification factor of dijet spectra~\cite{Sirunyan:2018qel}, was thus very fortunate for nPDF analyses.
The new CMS dijet data are provided as ratios of self-normalized rapidity distributions differential in $\eta_{\rm dijet}$ in bins of average transverse momentum $p_\mathrm{T}^{\rm ave}$ of the jet pair,
\begin{equation}
  R_{\rm pPb}^{\rm norm.} = \frac{\frac{1}{\mathrm{d}\sigma^{\rm p+Pb}/\mathrm{d}p_\mathrm{T}^{\rm ave}}\,\mathrm{d}^2\sigma^{\rm p+Pb}/\mathrm{d}p_\mathrm{T}^{\rm ave}\mathrm{d}\eta_{\rm dijet}}{\frac{1}{\mathrm{d}\sigma^{\rm p+p}/\mathrm{d}p_\mathrm{T}^{\rm ave}}\,\mathrm{d}^2\sigma^{\rm p+p}/\mathrm{d}p_\mathrm{T}^{\rm ave}\mathrm{d}\eta_{\rm dijet}}.
\end{equation}
This is advantageous due to the cancellation of the normalization uncertainty arising from imprecisions in the luminosity determination and also for the cancellation of hadronization effects, separately for proton--proton and proton--lead.

\afterpage{\begin{landscape}
\begin{figure}
  \centering
  \includegraphics[width=0.975\linewidth]{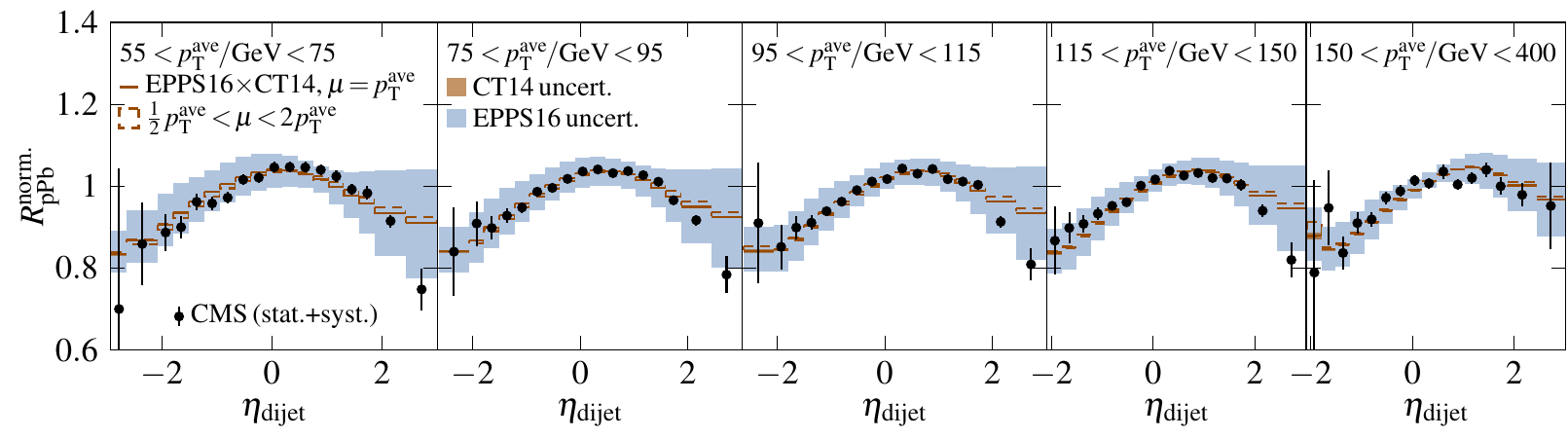}
  \includegraphics[width=0.975\linewidth]{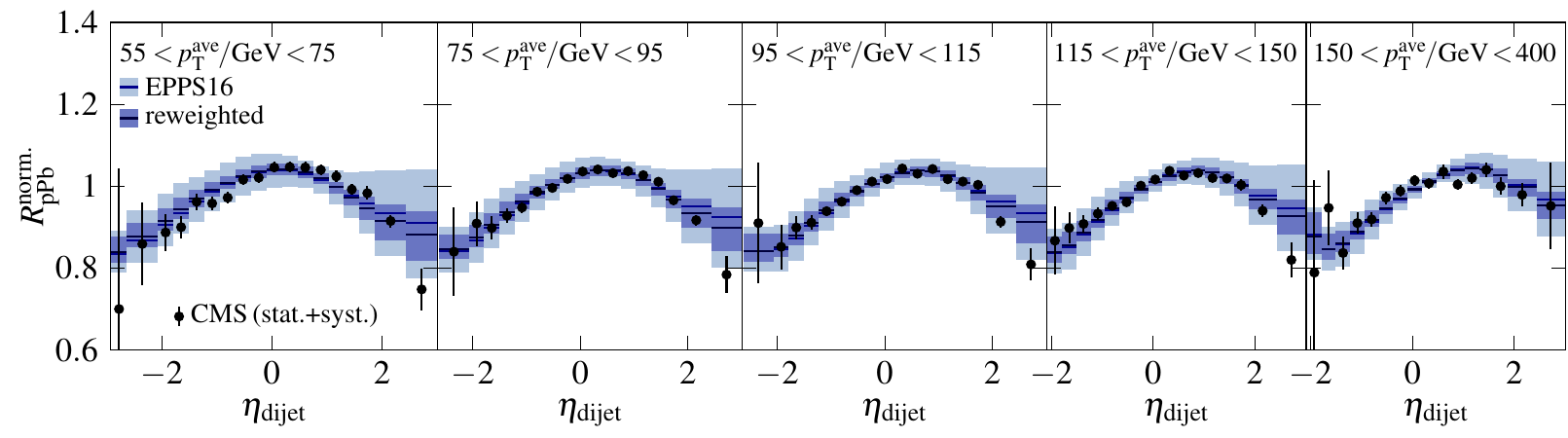}
  \caption{Top: Comparison of the CMS data on self-normalized dijet nuclear modification factor with NLO predictions using EPPS16 nPDFs. The hollow dashed boxes indicating the scale uncertainty partially overlap with the central predictions. Bottom: Results of reweighting the EPPS16 nPDFs with the CMS dijet data. Figures from article~\cite{Eskola:2019dui}.}
  \label{fig:dijetrpa}
  \label{fig:dijetrparw}
\end{figure}
\end{landscape}}

The expected impact of these data on the EPPS16 nPDFs were studied in article~\cite{Eskola:2019dui}
with the Hessian PDF reweighting outlined in \sec{hessianrw} including the higher order terms.
The original NLO predictions for this observable, produced using the NLOJet++~\cite{Nagy:2003tz} code, are shown in comparison with the data in the upper panels of \fig{dijetrpa}.
Compared to EPPS16, the data has much smaller uncertainties and strong additional constraints can be expected from including these data in a nPDF global analysis.
Further, as shown in the figure, the scale and proton-PDF uncertainties cancel very effectively in the ratio, making this observable an efficient probe of the PDF nuclear modifications. This is fortunate since, as studied extensively in article~\cite{Eskola:2019dui}, the proton--lead spectra before taking the ratio with proton--proton baseline have large uncertainties from the proton PDFs, preventing a clean extraction of nPDFs directly from the spectra.

The data are very precise and the uncertainties are systematics dominated, but unfortunately the correlations have not been published, and hence the statistical and systematical uncertainties had to be simply added in quadrature in the reweighting.
The impact on the predictions is quantified in lower panels of \fig{dijetrparw}, where the results with EPPS16 before and after the reweighting are shown. There is a substantial reduction in the uncertainties, showing that these data can place tight constraints on nPDFs. At forward rapidities, $\eta_{\rm dijet} \gtrsim 2$, where the data points lie systematically below the original EPPS16 central prediction, a downward pull is observed, indicating a preference for deeper gluon shadowing.

\begin{figure}
  \centering
  \includegraphics[width=0.95\textwidth]{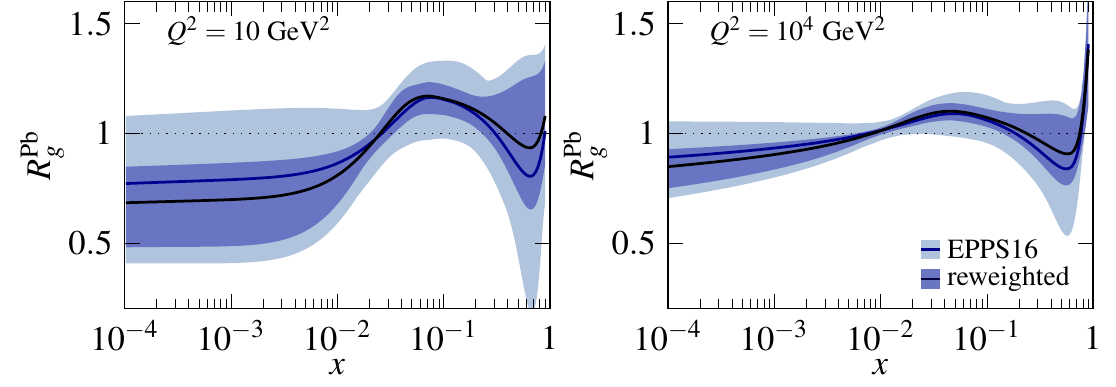}
  \caption{Impact on the EPPS16 gluon modifications at $Q^2 = 10\ {\rm GeV}^2$ and $Q^2 = 10^4\ {\rm GeV}^2$ upon reweighting with the CMS dijet $R_{\rm pPb}^{\rm norm.}$ measurement. Figure from article~\cite{Eskola:2019dui}.}
  \label{fig:EPPS16rwdijetrpa}
\end{figure}

The reweighting has a drastic effect on the EPPS16 gluon uncertainties, which shrink by a large factor throughout the probed $x$ range, as shown in \fig{EPPS16rwdijetrpa}. The most stringent constraints are put on the mid-$x$ region, where the resulting uncertainty is reduced to less than half of its original size. With the reweighted modifications exceeding unity in this region, we seem to be able to confirm the existence of gluon antishadowing. Similarly, at small $x$ the uncertainty band lies below one, supporting gluon shadowing.

Even with an enhanced shadowing in the central set compared to the original EPPS16, the fit has trouble in reproducing the most forward data points. Such a steep decrease as seen in the data in going from the second-to-most-forward data points to the most forward ones can be expected to be hard to come by in any global fit as the gluon modifications probed in this high-$Q^2$ region are smoothed by the scale evolution, as can be seen in \fig{EPPS16rwdijetrpa}. It is thus essential to have also other forward data to tell whether the drop in the data is a real physics effect, or perhaps caused by the systematic uncertainties. Further constraints are also needed in the high-$x$ region, where the uncertainties are are reduced compared to EPPS16 before reweighting, but still large enough such that we cannot yet confirm EMC effect for gluons. While we have here studied only the constraints on EPPS16, the impact on nCTEQ15 and nNNPDF1.0, with larger uncertainties to begin with, would be even more dramatic.

\subsection{Small-$x$ constraints with D-meson production}

While the normalized dijet nuclear modification factor discussed in the previous section proved to be an efficient probe of the gluon nuclear modifications, the $x$-reach of the CMS measurement goes only down to about $2 \cdot 10^{-3}$ in the lowest $\ptave$ bin, leaving the region of very small $x$ still unconstrained. To study the low-$x$ region, the use of inclusive forward production of $\D{}$ mesons has been proposed e.g.\ in Ref.~\cite{Gauld:2015lxa}. This process and its measurement at the LHCb experiment~\cite{Aaij:2017gcy} are discussed in the light of nPDFs in article~\cite{Eskola:2019bgf}. While these data have been studied previously~\cite{Kramer:2017gct,Kusina:2017gkz}, either a direct evaluation of the impact on nPDFs has not been given, or if done, then using a less rigorous theoretical framework. Thus, article~\cite{Eskola:2019bgf} provides the first fully QCD-based estimate of the impact of the LHCb $\D0$ data on nPDFs. More precisely, the analysis is performed in the SACOT-$m_{\rm T}$~\cite{Helenius:2018uul} scheme of GM-VFNS with KKKS08 fragmentation functions~\cite{Kneesch:2007ey} and using the Hessian PDF reweighting method.

\begin{figure}
  \centering
  \includegraphics[width=0.49\textwidth]{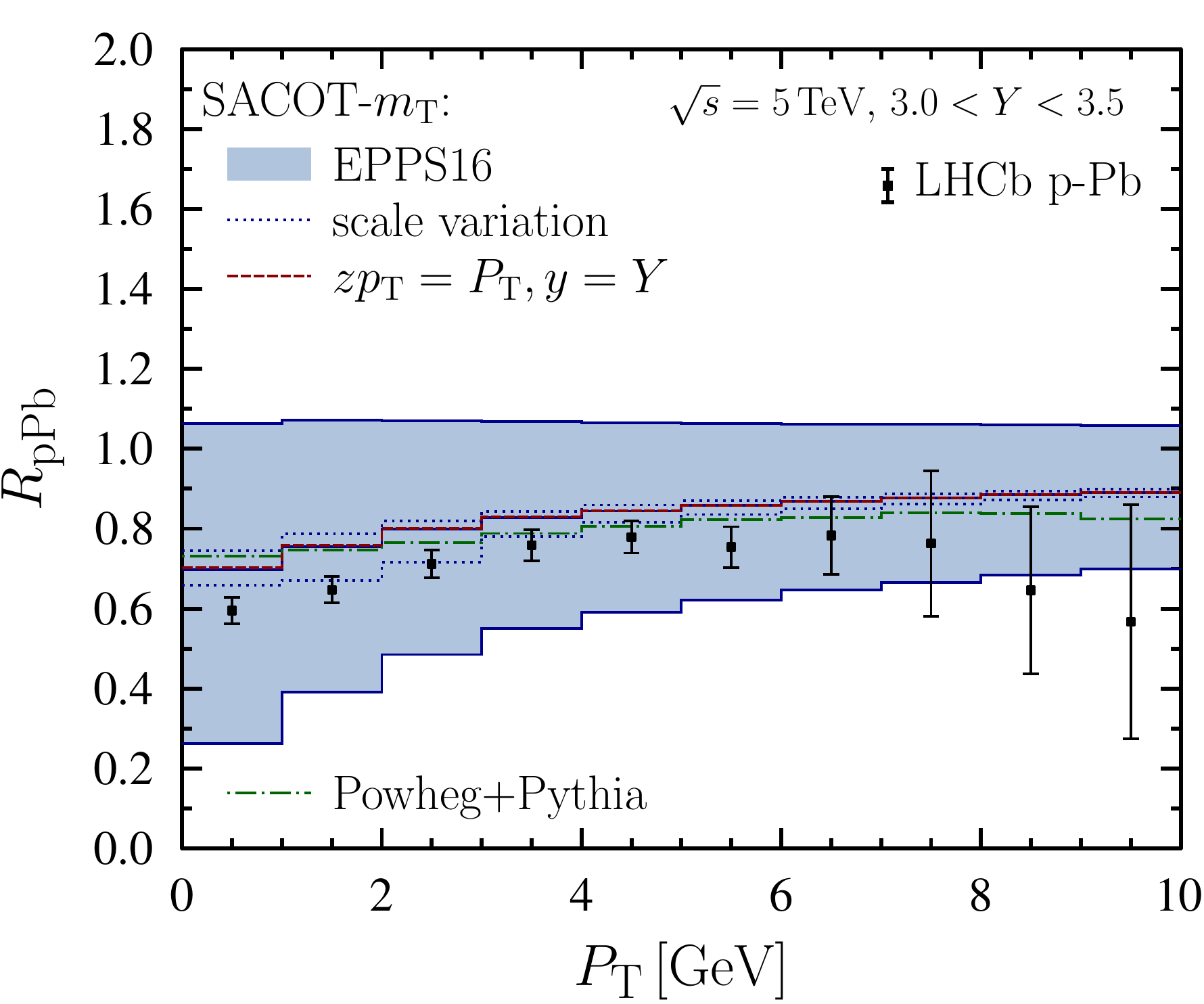}
  \includegraphics[width=0.49\textwidth]{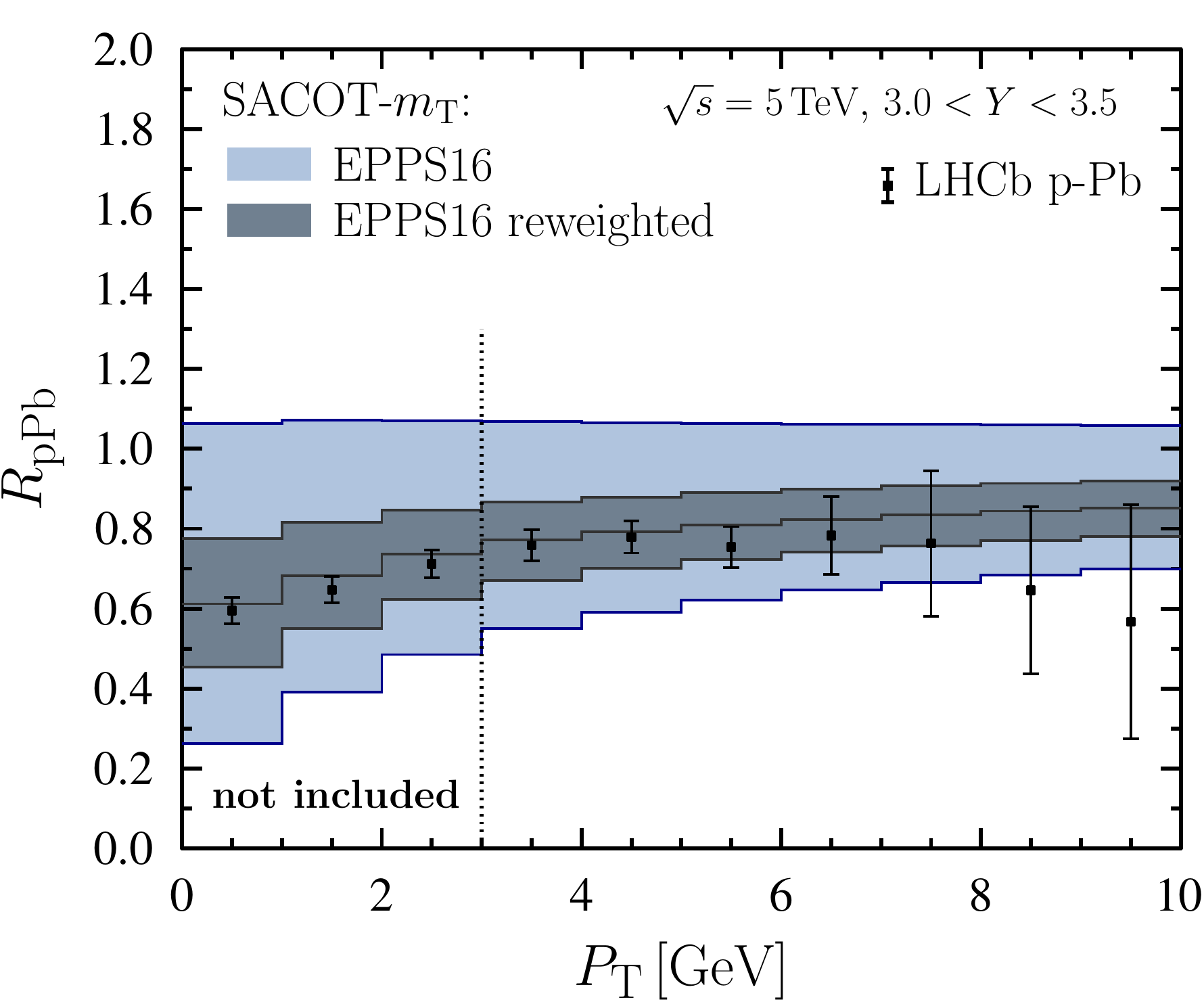}
  \caption{Data-to-theory comparison of inclusive $\D0$-meson production at forward rapidity (left) and the effect of reweighting on EPPS16 predictions in this bin (right). Figures from article~\cite{Eskola:2019bgf}.}
  \label{fig:dmeson}
\end{figure}

The LHCb measurements are given in terms of nuclear modification factors,
\begin{equation}
R_{\mathrm{pPb}}^{\D0}(P_{\mathrm{T}},Y) = \frac{\frac{1}{208}\mathrm{d}\sigma^{\mathrm{p+Pb}\rightarrow \D0 + X} / \mathrm{d}P_{\mathrm{T}} \mathrm{d}Y}{\mathrm{d}\sigma^{\mathrm{p+p}\rightarrow \D0 + X} / \mathrm{d}P_{\mathrm{T}} \mathrm{d}Y},
\label{eq:D0RpA}
\end{equation}
where $P_{\mathrm{T}}$ is the transverse momentum of the measured $\D0$ and $Y$ its rapidity. The left-hand-side panel of \fig{dmeson} shows this ratio in a bin of forward rapidity, $3.0 < Y < 3.5$. Again, the scale uncertainties are found to cancel to a large extent in the ratio. Still, at $P_{\mathrm{T}} < 3\ {\rm GeV}$ these uncertainties begin to grow and, due to the minimum scale $Q = 1.3\ {\rm GeV}$ in the EPPS16 PDFs, are potentially even underestimated in this region. To avoid possible bias, it is therefore safest not to include the $P_{\mathrm{T}} < 3\ {\rm GeV}$ data in a nPDF fit.

Even with this cut in place, the LHCb data are able to constrain nPDFs down to $x \approx 10^{-5}$. To study the possible impact on nPDFs in detail, we have performed a similar reweighting analysis as was done with the dijet data, here for both EPPS16 and nCTEQ15. The right-hand-side panel of \fig{dmeson} shows the resulting change in the EPPS16 predictions in the $3.0 < Y < 3.5$ bin, where a large reduction in the EPPS16 uncertainties are found.
At backward rapidities the reduction is not as large, but still significant. Interestingly, throughout the data range, also the data below the $3\ {\rm GeV}$ cut agree with the reweighted predictions, supporting the validity of collinear factorization down to $P_{\mathrm{T}} = 0\ {\rm GeV}$ in this process. No need for including nuclear modifications of fragmentation functions is found here either.

\begin{figure}
  \centering
  \includegraphics[width=0.9\textwidth]{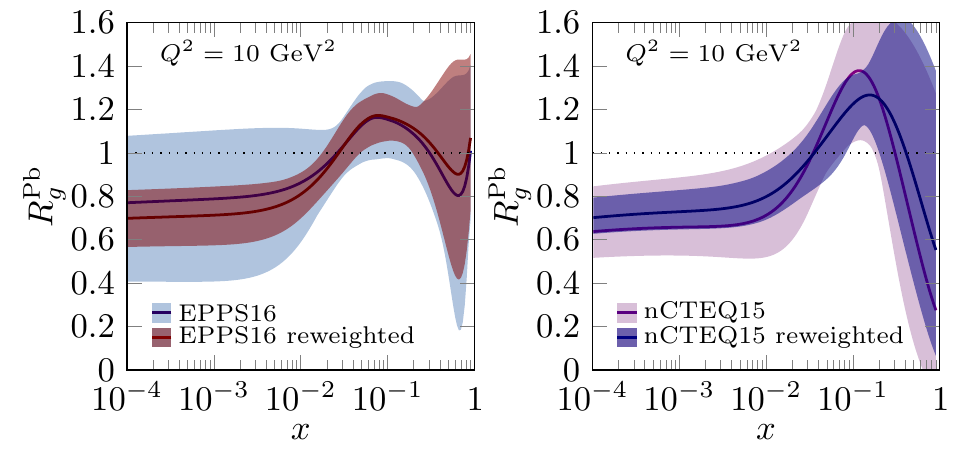}
  \caption{Impact of LHCb inclusive $\D0$-meson data on EPPS16 and nCTEQ15 nPDFs in Hessian reweighting. Figures from article~\cite{Eskola:2019bgf}.}
  \label{fig:dmesonrwgluons}
\end{figure}

\fig{dmesonrwgluons} shows the impact on the EPPS16 and nCTEQ15 gluon nuclear modifications.
The similarity with the results obtained in reweighting EPPS16 with the dijet data is striking (see \fig{EPPS16rwdijetrpa}), lending further support for the process independence of nuclear PDFs. We also find that the assumed parametrization in EPPS16 is not too restrictive and can describe both data simultaneously.
While the mid-$x$ constraints from $\D0$-meson data are somewhat less restrictive than those from dijets, at small $x$ significant further constraints are obtained, not only because the resulting uncertainty bands are smaller, but also more importantly since the data constraints extend to significantly lower $x$. The next generation nPDFs with both of these data included in the analysis can thus be expected to have the gluon modifications constrained with previously inaccessible precision.

\subsection{Multi-observable approach with RHIC}

While the BNL-RHIC provided the first direct constraints for the nuclear gluon PDFs, no further measurements have found their way to nPDF fits yet, even though the potential of the collider with its flexible beam line to provide nPDF constraints is indisputable. In article~\cite{Helenius:2019lop}, we have provided a systematic study on the prospects of a simultaneous analysis on multiple observables to constrain the nPDFs, revolving around the potential of the proposed forward upgrades of the STAR and sPHENIX experiments~\cite{Aschenauer:2016our,Adare:2015kwa}.

\begin{figure}
  \centering
  \begin{minipage}{0.5\textwidth}
    \includegraphics[width=\textwidth]{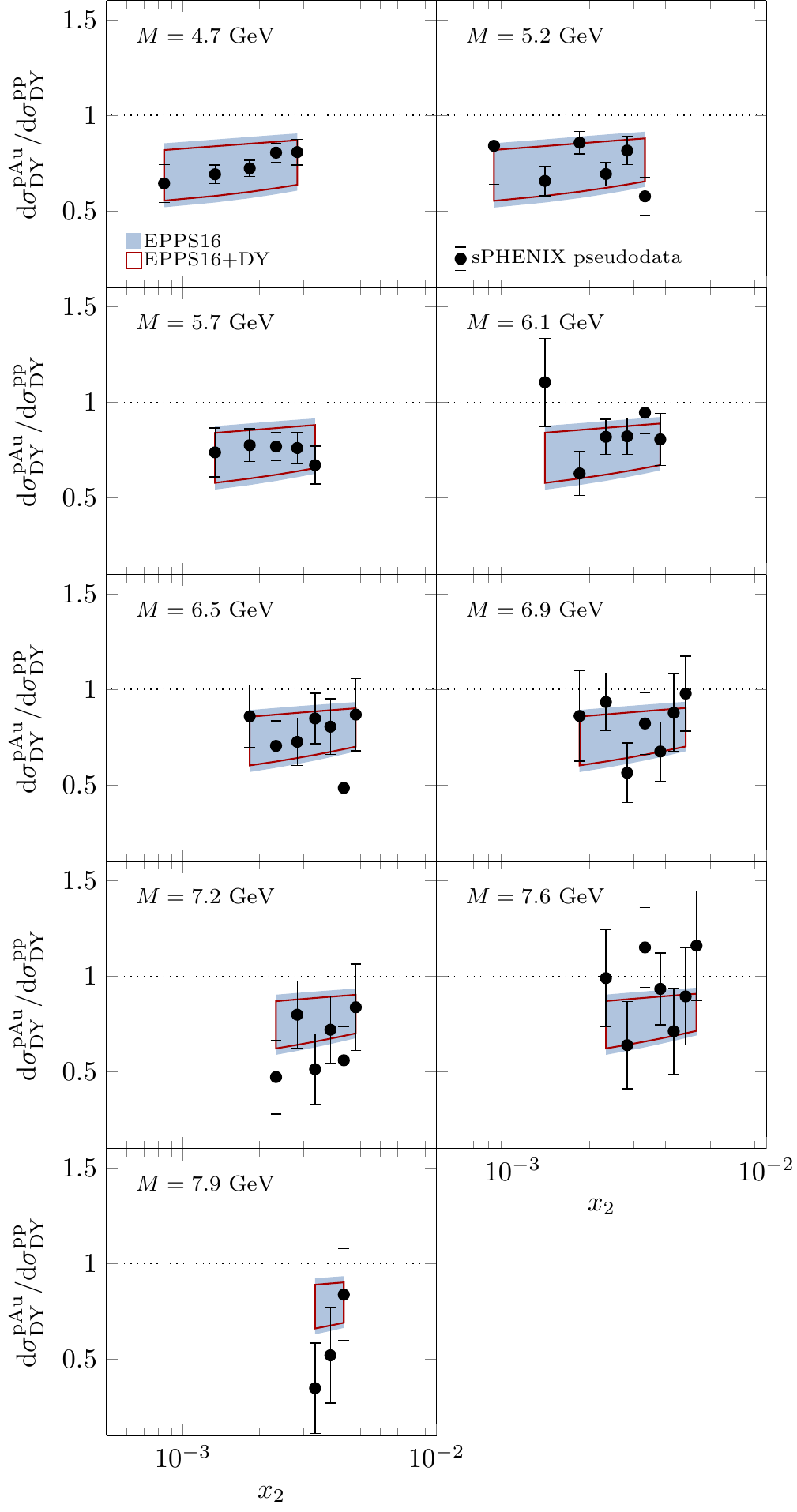}
  \end{minipage}
  \hspace{-0.015\textwidth}
  \begin{minipage}{0.5\textwidth}
    \includegraphics[width=\textwidth]{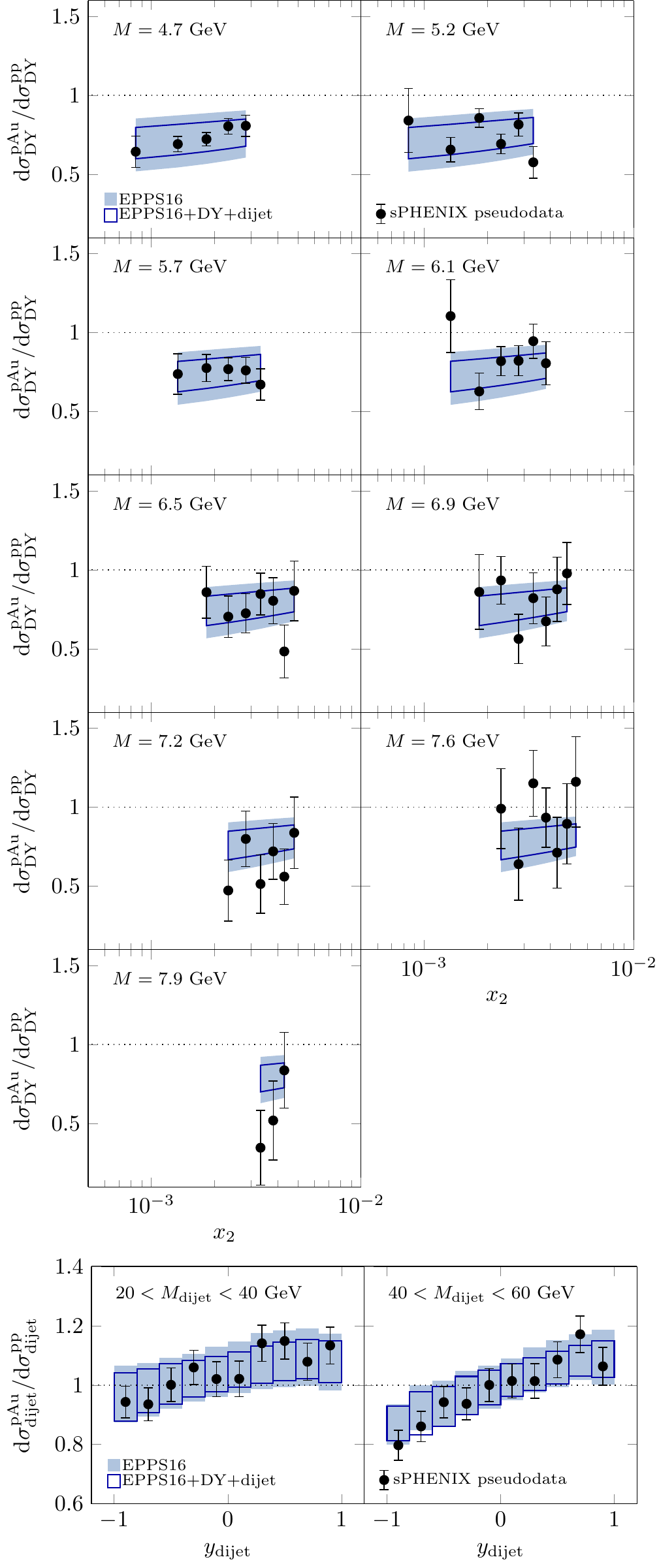}
  \end{minipage}
  \caption{The impact of sPHENIX pseudodata on EPPS16 predictions in reweighting with the forward Drell--Yan only (left) or including also the central dijets (right). Figures from article~\cite{Helenius:2019lop}.}
  \label{fig:corrnorm}
\end{figure}

The left-hand-side panels of \fig{corrnorm} show the pseudodata for DY nuclear modification factor generated with expected luminosities and efficiencies at the sPHENIX forward-arm upgrade. The data points at low dilepton invariant masses have smaller projected statistical uncertainties, shown as error bars, than the uncertainties from predictions with the EPPS16 nPDFs, which would promise new constraints on the nPDFs. However, on top of the statistical uncertainties, we are expecting a normalization uncertainty of the order of 4 percent, stemming from uncertanties in the luminosity determination. A reweighting performed with these pseudodata, ``EPPS16+DY'' in \fig{corrnorm}, thus finds barely any impact on the PDFs.

It is worth to note that this is also a situation where the d'Agostini bias can become potentially dangerous. As the EPPS16 nPDFs and the pseudodata generated from them have a rather flat $x$ dependence, any alteration in the predictions could be compensated by a respective change in the data normalization. Then, if the $\chi^2$ function from \eq{biasedchisq} were used, there would be a bias favoring smaller normalization (and thus enhanced shadowing). For this reason, we have used instead the unbiased $\chi^2$ function in \eq{unbiasedchisq}. The flatness of the data also prevents using a similar self-normalization trick as was used in \sec{dijetnormalizedrpa} to treat the normalization of dijets.

\begin{figure}
  \centering
  \includegraphics[width=0.95\textwidth]{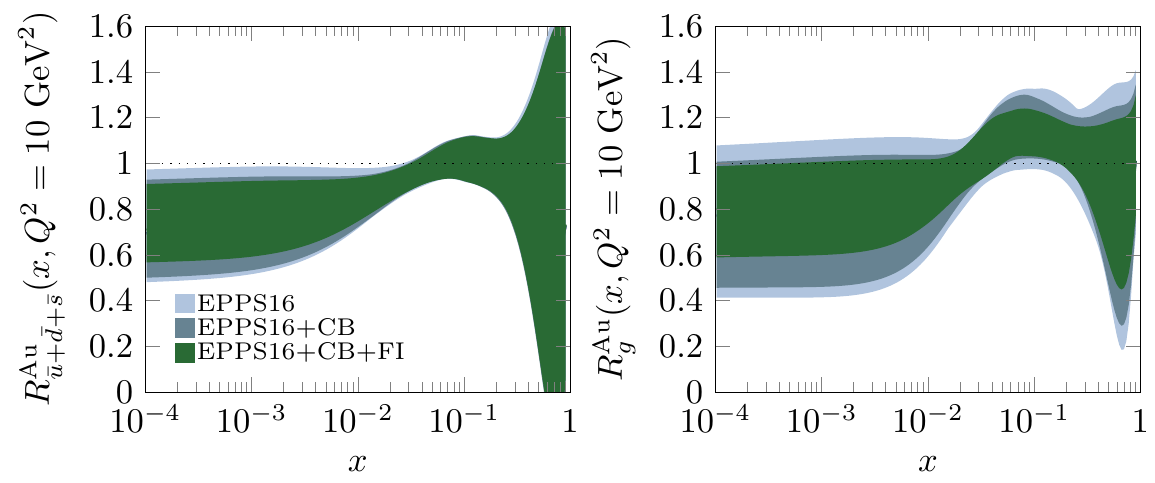}
  \caption{Projected impact of sPHENIX measurements of DY, dijet and gamma--jet nuclear modification ratios on EPPS16 in the central-barrel only (CB) and including forward instrumentation (CB+FI) scenarios. Figure from article~\cite{Helenius:2019lop}.}
  \label{fig:impactsphenix}
\end{figure}

Now, the idea of the multi-observable approach is that the luminosity uncertainty is correlated over all measurements using the same proton--gold and proton--proton run statistics. Thus, if one includes in the fit data from a better constrained region, say, dijets at central rapidity, this would constrain the normalization also in the less-constrained small-$x$ region. This is illustrated in the right-hand-side panels of \fig{corrnorm}, where now the combined ``EPPS16+DY+dijet'' fit, with normalization uncertainty correlated between the DY and dijet pseudodata, achieves a much larger impact at small $x$.

To further study the constraining potential in the multi-observable framework, we have performed reweighting analyses using combined sets of DY, dijet and photon--jet pseudodata. \fig{impactsphenix} shows the total expected impact on gluon modifications in gold in the central-barrel only (CB) and including forward instrumentation (CB+FI) scenarios. The constraints found in the CB scenario are rather modest, especially when acknowledging the fact that the small-$x$ constraints in this case are mostly due to momentum-conservation induced correlations.
The inclusion of forward instrumentation significantly increases the constraining power, particularly in the small-$x$ region. We have found that these additional small-$x$ constraints are driven at the present setup by the forward DY data. To leading order, the DY process happens through quark--antiquark annihilation, but since at small $x$ the level of sea quark distributions is set by the evolution from gluons and also the direct NLO contribution from quark--gluon scattering becomes increasingly important, the main impact at the scale shown in \fig{impactsphenix} is on the gluon modifications.

While the constraints appear to be smaller than what was observed with dijet and $\D0$ production data from the LHC, the proton--gold measurements at RHIC are important in checking that the results obtained for lead are still valid at slightly smaller nucleus and to guide our assumptions on how the nuclear effects will evolve towards smaller nuclei. In fact, while the LHC will keep on providing constraints mostly for the lead nucleus, the flexibility of the RHIC beam line would allow performing a proper $A$-scan to put constraints also on the mass-number dependence of the gluon modifications.

\chapter{Conclusions}
\label{chap:concl}

In this thesis, we have discussed the extraction of nuclear parton distribution functions (nPDFs), particularly in the light of new constraints derivable from various hadron--nucleus collision data which have not been previously included in nPDF global analyses. As a highlight, the article~\cite{Eskola:2016oht} of this thesis, with the EPPS16 nPDF set as output, presents the first nPDF global analysis including LHC data on electroweak gauge boson and dijet production. Summarizing further the main results of this thesis:

We have shown in the article~\cite{Paakkinen:2016wxk} that in certain ratios of pion--nucleus Drell--Yan cross sections the pion PDFs and thus also the uncertainties they come with efficiently cancel at the level of NLO cross sections. Due to contributions involving valence antiquarks of the pions, these observables show an enhanced sensitivity to the flavour separation of quark nuclear modifications, which, as discussed in \sec{flavourasym}, is hard to constrain. The existing data have somewhat large uncertainties and thus do not yield very strong constraints, but nonetheless indicate that valence quark modifications should not be too asymmetrical. As recognized also by the experimental community, there are interesting prospects in performing such measurements in future experiments.

The tensionless fit found in the article~\cite{Eskola:2016oht} gives evidence for the universality of nPDFs across a wide variety of different processes in the kinematical range $Q \geq 1.3$ GeV studied. In this analysis, we have found the most decisive new data to be those from the CHORUS neutrino--nucleus DIS and CMS proton--nucleus measurements, putting new constraints on the flavour separation and gluon nuclear modifications, respectively. This analysis is also the first one to allow for a full flavour separation in the quark nuclear modifications, thus significantly reducing the bias in predictions sensitive to such differences.

Using Hessian PDF reweighting tools, we have quantified in articles~\cite{Eskola:2019dui} and \cite{Eskola:2019bgf} the potential constraints on nPDFs from CMS dijet and LHCb $\D0$-production data. The impact on nPDFs is found to be dramatic, with the dijets putting stringent constraints on gluon modifications especially in mid-$x$ region and $\D0$ mesons respectively for gluons at small $x$. When used together in a global fit, these data can be expected to constrain the gluon modifications in lead to an unprecedented accuracy.

The field of nPDF analyses is currently evolving quickly, driven mainly by the constantly increasing amount of data constraints from the LHC proton--lead collisions. In addition to the data discussed above, newly-finalized measurements of ATLAS dijet conditional yields at 5.02 TeV~\cite{Aaboud:2019oop} and CMS W bosons at 8.16 TeV~\cite{Sirunyan:2019dox} can shed additional light on the nuclear modifications in lead. In the coming years, with the LHC turning into a high-luminosity mode, the precision of especially electroweak observables is expected to improve significantly~\cite{Citron:2018lsq}. On a further note, additional observables, such as direct photons~\cite{Helenius:2014qla} or photoproduction of dijets~\cite{Guzey:2019kik}, can also place new constraints on the gluons and test the universality of nPDFs.

All this makes the constraints for lighter nuclei to lag behind. While a fixed-target programme at the LHC~\cite{Hadjidakis:2018ifr} can help the situation significantly, the data from such measurements are bound to give constraints only in the large-$x$ region. The RHIC collider, with a history of successful studies in a wide class of different collision systems and forward upgrades in its experiments coming up, thus offers a unique opportunity to constrain the mass-number dependence of the nuclear modifications. Article~\cite{Helenius:2019lop} discusses the prospects of constraining the gluon nuclear modifications with RHIC. There, we found that a simultaneous analysis on multiple observables can help in reducing luminosity-related normalization uncertainties and thus improve the impact of forward DY measurements. Ultimately still, an electron--ion collider would be needed to truly pin down the PDF nuclear modifications~\cite{Aschenauer:2017oxs,Accardi:2012qut,AbelleiraFernandez:2012cc,Paukkunen:2017phq}.

\printbibliography[notcategory=contribution,title=References,heading=bibintoc]

\end{document}